\documentclass[floatfix,showpacs,amsmath,amsfonts,amssymb,aps,twocolumn,prb,superscriptaddress]{revtex4-1}

\usepackage[table,dvipsnames]{xcolor}
\usepackage{graphicx}
\usepackage{dcolumn}
\usepackage{amsthm}
\usepackage{bm}
\usepackage{siunitx}
\usepackage{nicefrac}
\usepackage{mathtools}
\usepackage{float}
\usepackage{braket}
\usepackage{makecell} 
\usepackage{multirow}
\usepackage{wasysym}
\usepackage{titletoc}
\usepackage{multirow}
\usepackage{mwe}
\usepackage{diagbox}
\usepackage{pifont}
\newcommand{\cmark}{\ding{51}}
\newcommand{\gcmark}{{\color{Green}\ding{51}}}

\definecolor{darkred}{rgb}{0.6,0.,0.}
\definecolor{darkgreen}{rgb}{0.,0.5,0.}
\definecolor{darkblue}{rgb}{0.,0.,0.6}
\usepackage[breaklinks, colorlinks=true, linkcolor=darkred, citecolor=darkgreen, urlcolor=darkblue]{hyperref}

\expandafter\def\csname ver@etex.sty\endcsname{3000/12/31}

\usepackage{autonum}

\newcommand{\includegraphicsgood}{\includegraphics}

\usepackage[inactive,tightpage]{preview}
\PreviewMacro[*[[!]{\input{}}
\PreviewMacro[*[[!]{\includegraphicsgood}
\newcommand{\normord}[1]{:\mathrel{#1}:}

\begin{document}

\title{Stability, phase transitions, and numerical breakdown of fractional Chern insulators in higher Chern bands of the Hofstadter model}
\author{Bartholomew Andrews}
\affiliation{Department of Physics, University of Zurich, Winterthurerstrasse 190, 8057 Zurich, Switzerland}
\author{Titus Neupert}
\affiliation{Department of Physics, University of Zurich, Winterthurerstrasse 190, 8057 Zurich, Switzerland}
\author{Gunnar M{\"o}ller}
\affiliation{Physics  of  Quantum  Materials Group, School of Physical Sciences, University of Kent, Canterbury CT2 7NZ, United Kingdom}
\date{\today}

\begin{abstract}

The Hofstadter model is a popular choice for theorists investigating the fractional quantum Hall effect on lattices, due to its simplicity, infinite selection of topological flat bands, and increasing applicability to real materials. In particular, fractional Chern insulators in bands with Chern number $|C|>1$ can demonstrate richer physical properties than continuum Landau level states and have recently been detected in experiments. Motivated by this, we examine the stability of fractional Chern insulators with higher Chern number in the Hofstadter model, using large-scale infinite density matrix renormalization group simulations on a thin cylinder. We confirm the existence of fractional states in bands with Chern numbers $C=1,2,3,4,5$ at the filling fractions predicted by the generalized Jain series [\href{https://journals.aps.org/prl/abstract/10.1103/PhysRevLett.115.126401}{Phys.~Rev.~Lett.~\textbf{115}, 126401 (2015)}]. Moreover, we discuss their metal-to-insulator phase transitions, as well as the subtleties in distinguishing between physical and numerical stability. Finally, we comment on the relative suitability of fractional Chern insulators in higher Chern number bands for proposed modern applications.      

\end{abstract}

\maketitle         

\section{Introduction}

Lattice generalizations of the fractional quantum Hall effect, also known as fractional Chern insulators (FCIs)~\cite{Regnault11}, have received substantial interest in the past decade, primarily due to their low magnetic field~\cite{Neupert11}/high temperature realizations~\cite{Tang11}, shorter characteristic length scales, and richer physical phenomena, compared to their continuum counterparts~\cite{Bergholtz13, Parameswaran13}. In early works, the analog of the Laughlin state was demonstrated~\cite{Sorensen05}, and extensions to the Jain hierarchy of states and more general fractional Chern insulators were found in Hofstadter-type~models~\cite{Hafezi07, Moller09}, based on small-scale exact diagonalization (ED) calculations. Since then our understanding of FCIs has been deepened with respect to: the adiabatic continuity to fractional quantum Hall states~\cite{Scaffidi12, Zhang16}, the role of band geometry~\cite{Parameswaran12, Roy14, Jackson15}, as well as optimal methods for band engineering~\cite{Lee17}. In particular, research efforts have focused on FCIs in bands with Chern number $|C|>1$, which cannot be continuously connected to the Landau level continuum limit and therefore can host lattice-specific fractional states~\cite{Kol93, Moller09, Liu12, Udagawa14, Wu15, Moller15, Andrews18}, which have potential applications to topological quantum computing~\cite{Nayak08, Barkeshli12, Barkeshli13}. Consequently, such FCIs have been the focus for a large proportion of numerical studies~\cite{Andrews18} covering quasi-charge excitations~\cite{Jaworowski19}, non-Abelian states~\cite{Sterdyniak13, Bergholtz15}, non-Abelian twist defects \cite{Liu17}, and the bosonic integer quantum Hall effect~\cite{Moller09, He15, Zeng16, He17, Andrews18}. Most significantly, $|C|>1$ FCIs have now been realized experimentally in van der Waals heterostructures with an external magnetic field~\cite{Spanton18} and, in the case of $|C|>1$ CIs, also without a magnetic field~\cite{Chen20}, which provides strong motivation to revisit the topic.        

Building on the theoretical foundation of Refs.~\onlinecite{Moller09,Moller15} and inspired by recent experimental advances, we compare the stability of $|C|>1$ FCIs in the Hofstadter model in order to identify promising candidates for imminent experimental investigations. To this end, we employ large-scale infinite density matrix renormalization group (iDMRG) simulations on a thin cylinder geometry. We present a direct follow-up to the ED study on a torus by Andrews and M{\"o}ller~\cite{Andrews18} and leverage the DMRG algorithm on an infinite cylinder to stabilize a larger set of FCIs predicted by the generalized Jain series~\cite{Moller15}. Furthermore, through the application of a modern tensor network method, we are able to access larger (semi-infinite) system sizes, Hamiltonians without a band projection, and illuminating entanglement properties, which we use in conjunction with the previous study to more precisely analyze the physical stability of $|C|>1$ FCIs in the Hofstadter model, as well as the numerical stability of the two methods employed. In this paper, we diagnose FCIs based on their charge pumping and two-point correlation functions and quantify stability with respect to the interaction strength and single-particle gap-to-width ratio. We confirm the existence of FCIs in bands with Chern number $C=1,2,3,4,5$ in accordance to the generalized Jain series~\cite{Moller15}, going beyond our previous ED results~\cite{Andrews18}, and we present case studies of metal-to-FCI phase transitions, showing their relation to the single-particle band structure. Moreover, we expose the inherent limitations of theoretical studies, including misleading numerical breakdowns in charge pumping computations due to an insufficient system size. In all cases, we analyze our results in light of current experiments on moir{\'e} superstructures~\cite{Spanton18, Chen20, Xie21}, as well as potential realizations in optical flux lattices~\cite{Cooper13, Aidelsburger15, Cooper19}, Floquet systems~\cite{Xiong16}, and quantum spin liquids~\cite{Yao13, Trescher12, Cook14}.        

The structure of the paper is as follows. In Sec.~\ref{sec:model}, we introduce the Hofstadter model and the many-body Hamiltonian. In Sec.~\ref{sec:method}, we outline our method, including an explanation of our lattice configurations and an overview of the iDMRG algorithm. In Sec.~\ref{sec:results}, we present our numerical results, showing stabilized FCIs as well as their metal-to-FCI transitions. Finally, in Sec.~\ref{sec:discussion}, we discuss the implications of our findings and outline avenues for future research. 

\section{Model}
\label{sec:model}

We consider spinless fermions hopping on a square lattice with lattice constant $a$, taken to lie in the $xy$-plane, in the presence of a perpendicular magnetic field $\mathbf{B}=B\hat{\mathbf{e}}_z$. We select a square lattice since it yields comparable FCI stability to general single-component lattices~\cite{Wu12_2}, such as the triangular lattice~\cite{Yang12, Kourtis12}, and it is applicable to leading cold-atom experiments~\cite{Aidelsburger15, Motruk17, Motruk20}. The particles interact with each other via nearest-neighbor density-density interactions, such that the many-body Hamiltonian may be written as
\begin{equation}
\label{eq:ham}
H=\sum_{\braket{i,j}} \mathrm{e}^{\mathrm{i}\theta_{ij}}c^\dagger_i c_j + V \sum_{\braket{i,j}} \rho_i \rho_j,
\end{equation} 
where $\mathrm{e}^{\mathrm{i}\theta_{ij}}$ is the Peierls phase factor, $c^\dagger_i/c_i$ are the fermionic creation/annihilation operators, $V$ is the interaction strength and $\rho_i=c^\dagger_i c_i$ is the density operator. Unless specified otherwise, we use $V=10$, and all energy scales are measured in the nearest-neighbor hopping strength, which we have set to unity. The perpendicular magnetic field is incorporated via the Peierls substitution, such that the Peierls phase is defined as $\theta_{ij}=(2\pi/\phi_0)\int_i^j \mathbf{A}\cdot\mathrm{d}\mathbf{l}$, where $\phi_0=h/e$ is the flux quantum, $\mathbf{A}$ is the vector potential, and $\mathrm{d}\mathbf{l}$ is the infinitesimal line element connecting sites $i$ and $j$~\cite{Peierls33}. The effect of this substitution is to accommodate the larger magnetic translation group~\cite{Zak64}. In this paper we use the Landau gauge with a conserved $k_y$ momentum, choosing $\mathbf{A}=Bx\hat{\mathbf{e}}_y$, which corresponds to enlarging the unit cell in the $x$ direction. We define $p'$ as the magnetic flux passing though each magnetic unit cell, yielding the flux density $n_\phi \equiv BA_\text{UC}/\phi_0 \equiv p'/q$ per lattice plaquette, where $A_\text{UC}=a^2$ is the area of the unit cell and $p', q$ are coprime integers. In the selected Landau gauge for the square-lattice Hofstadter model, the unit cell is enlarged from $1\times1\to q\times 1$, known as the magnetic unit cell in the presence of a perpendicular magnetic field, and so $q$ directly corresponds to the number of bands in the single-particle energy spectrum~\cite{Harper55}. The frustration between the magnetic unit cell area and the irreducible area occupied by one flux quantum results in the famous Hofstadter butterfly --- a fractal spectrum of eigenenergies $E$ as a function of $n_\phi$~\cite{Hofstadter76}. The interactions between the particles are modeled using a nearest-neighbor density-density term, which has been shown to be favorable for stabilizing composite fermion states~\cite{Liu13, Sheng11, Wu12_2}, experimentally relevant due to screening lengths in common electronic devices~\cite{Rosner15, Pizarro19, Kim17}, and also computationally tractable~\cite{Moller09, Regnault11, Sterdyniak13, Moller15, Bauer16, Andrews18, Andrews20, Andrews21}.           

\section{Method}
\label{sec:method}

In this section we describe the method employed to compare the stability of FCIs in higher Chern bands. In Sec.~\ref{subsec:lat_geom} we outline the selection of lattice geometries and in Sec.~\ref{subsec:idmrg} we summarize the iDMRG algorithm. 

\subsection{Lattice geometries}
\label{subsec:lat_geom}

\begin{figure}
	\includegraphics[width=\linewidth]{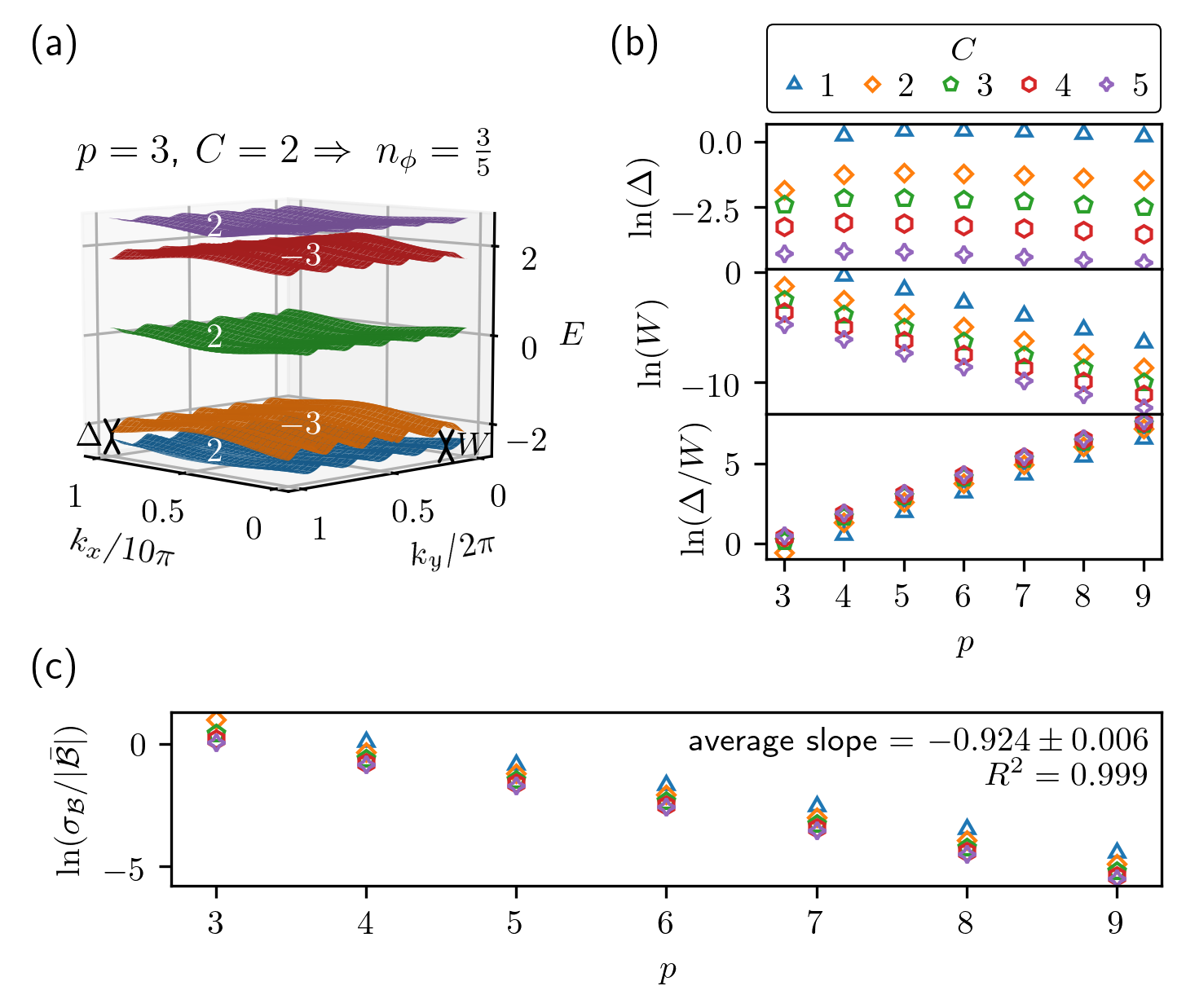}
	\caption{\label{fig:bandstructure_analysis}\textbf{Band flatness in the Hofstadter model.} (a)~Single-particle band structure for the Hofstadter model at $n_\phi=3/5$, as a function of momentum $k$. The Chern number of each band, $C$, as well as the width, $W$, and gap, $\Delta$, of the lowest band are labeled. (b)~Band gaps, widths, and gap-to-width ratios against $p<10$ for the first five Chern numbers. (c)~Standard deviation of the Berry curvature for the lowest band, $\sigma_\mathcal{B}$, in units of the mean Berry curvature, $\bar{\mathcal{B}}$.}
\end{figure}

In order to accurately compare the stability of FCIs in higher Chern bands, we first need to systematically generate topological flat bands of arbitrary Chern number. To this end, we employ the Hofstadter model, as introduced in the previous section, due to its simplicity, versatility, and experimental relevance~\cite{Miyake13, Aidelsburger13, Hafezi13, Aidelsburger15, Dean13, Roushan17, Ni19, Dutt20}. Here, we build on work by M{\"o}ller and Cooper~\cite{Moller09, Moller15}, who showed evidence for novel fractional Chern insulators in higher Chern bands of the Hofstadter model~\cite{Moller09} and later recast this result in the language of composite fermions~\cite{Moller15}, showing how the Jain series~\cite{Jain89} generalizes to higher Chern bands. 

Following Ref.~\onlinecite{Moller15}, we will exploit their insight that a system with flux density
\begin{equation}
\label{eq:nphi}
n_\phi = \frac{p}{|C|p-\mathrm{sgn}(C)}\bmod 1 \equiv \frac{p'}{q}
\end{equation}
has a lowest band of Chern number $C$ with a significant single-particle gap, provided that $p$ is large enough such that the lowest band is distinct. Hence, these flux densities are particularly appropriate for stabilizing Chern insulator states, as well as facilitating numerical studies~\cite{Moller15}. We note that, due to the symmetry of the Hofstadter butterfly, we define the flux density modulo $1$~\footnote{This implies that $p$ on the left-hand side of Eq.~\eqref{eq:nphi} does not always equal $p'$.}. For example, if we require a distinct lowest band with Chern number $C=2$, this is first achieved at $p=3$, as illustrated in Fig.~\ref{fig:bandstructure_analysis}(a). As we increase the value of $p$, the bands in the Hofstadter model become exponentially flatter while there is a polynomial decrease of the band gaps~\cite{Harper14}. Hence, the gap-to-width ratio exponentially increases, whereas the fluctuations of the Berry curvature exponentially decrease, as shown in Figs.~\ref{fig:bandstructure_analysis}(b,c) (cf.~Supplementary Material in Ref.~\onlinecite{Moller15}). We note that we require $p\geq4$ to obtain a distant lowest band with $C=1$, whereas for $C>1$ this is achieved with $p\geq 3$. For a systematic comparison, we consider all distinct lowest bands with $p<10$ for each Chern number. For Chern number $|C|=1$, increasing $p$ corresponds to taking the continuum limit $n_\phi\to0$, whereas for $|C|>1$, this corresponds to the effective continuum limit $n_\phi\to1/|C|$~\cite{Andrews18}.

Once we have tuned to the appropriate flux density to obtain a topological flat band of the required Chern number, the next step is to fractionally fill this lowest band. According to M{\"o}ller and Cooper~\cite{Moller15}, building on the notion of flux attachment~\cite{Kol93, Moller09}, FCIs are predicted to be stabilized at filling fractions of the generalized Jain series
\begin{equation}
\label{eq:nu}
\nu = \frac{r}{|kC|r+1} \equiv \frac{r}{s},
\end{equation}
which provides a framework to explain observations of FCI states in higher-$|C|$ bands~\cite{Wang12, Liu12, Sterdyniak13}. Here $r\in\mathbb{Z}$ is the number of filled composite fermion bands, $s$ corresponds to the ground-state degeneracy, $k\in\mathbb{Z}$ is the number of flux attached per composite fermion, and hence $k=1$ for bosons and $k=2$ for fermions. When filling the Chern band however, the geometry of the lattice is restricted by a number of constraints. For example, the flux density fixes the size of the magnetic unit cell to $q\times 1$ and the desired filling fraction, together with the numerical cost of large system sizes and the need for an integer number of particles, limits the numerically accessible finite-size lattice configurations which may stabilize a given FCI. As a result of these constraints, and the more fragile nature of higher-order fractional quantum Hall plateaus, the number of FCIs that we are able to stabilize decreases with Chern number.

In the limit of $|C|=1$ and $r\in\mathbb{Z}^+$, we see that Eq.~\eqref{eq:nu} reproduces the celebrated Haldane hierarchy $\nu=1/3, 2/5, 3/7, \dots$~\cite{Haldane83}, whereas for $|C|=1$ and $r\in\mathbb{Z}$, we reproduce the Jain series $\nu=r/(|k|r+1)$~\cite{Jain89} with either positive or negative flux attachment. Consequently, we refer to FCIs with $|r|=1$ as primary composite fermion states, FCIs with $|r|=2$ as secondary composite fermion states, and so on~\cite{Andrews18}. The cases where $r=+1$ are also referred to as `Laughlin-like'. For a systematic comparison of FCIs in this paper, we consider filling fractions with increasing $|r|$ for each Chern number.        

\subsection{iDMRG algorithm}
\label{subsec:idmrg}

In order to find and analyze the ground state of this 2D many-body problem, we employ the iDMRG algorithm on a thin cylinder~\cite{White92, Schollwock11, Stoudenmire12}. This is a well-established method that can be formulated in the tensor network framework and has been used to successfully model fractional quantum Hall states in recent years~\cite{Zaletel13, Zaletel15, Grushin15, Schoonderwoerd19, Andrews20, Andrews21}. The method is defined by transcribing the Hamiltonian to a matrix product operator (MPO) and serves to optimize an ansatz wavefunction in the form of a matrix product state (MPS). In this paper, the MPO and MPS refer to a 1D chain that zig-zags to cover the cylinder, and the MPS is defined in canonical form, such that a bipartition at any bond on the chain corresponds to a Schmidt decomposition. Specifically, we consider the bipartition at a bond such that the system is spatially split into two semi-infinite cylinders (denoted ``left'' and ``right'' by convention), with a Schmidt decomposition $\ket{\Psi}=\sum_{i=1}^\chi \lambda^2_i \ket{\psi_i}_\text{L}\otimes\ket{\psi_i}_\text{R}$, where $\chi$ is the bond dimension, $\lambda_i>0$ are the Schmidt values, and $\ket{\psi_i}_\text{L/R}$ are the left/right Schmidt states. We start with a finite MPS unit cell, minimize the energy using a Lanczos algorithm, and then perform the Schmidt decomposition up to a given bond dimension. Subsequently, we symmetrically enlarge the MPS unit cell and iterate the process until we reach a convergence of the relevant observables. Crucially, the Schmidt eigenbasis is directly related to the reduced density matrix eigenbasis, which allows us to conveniently extract entanglement properties, such as the von Neumann entanglement entropy $S_\text{vN}=-\sum_i \lambda_i^2 \ln \lambda_i^2$ and the entanglement spectrum $\{\epsilon_i\}$, defined through $\lambda_i^2=\mathrm{e}^{-\epsilon_i}$. Moreover, since the $U(1)$ symmetry of the Hamiltonian is also a symmetry of the reduced density matrix, we can label the Schmidt states according to their $U(1)$ charges $Q_{\text{L/R},i}\in\mathbb{Z}$~\cite{Grushin15, Zaletel13}.

For the systems considered in this paper we define an MPS unit cell composed of $L_x\times L_y$ magnetic unit cells, where $x$ is the direction of the cylinder axis. Since our MPS unit cell is translationally invariant in the $x$ direction due to the infinite cylinder ansatz of the iDMRG algorithm, we set $L_x=1$ and tile magnetic unit cells in the $y$ direction.  Hence, the number of lattice sites in our simulation cell is $q\times L_y$, which defines the finite-size simulation cell for our calculations.

Compared to the more traditional ED studies on a torus~\cite{Regnault11}, iDMRG on an infinite cylinder offers a number of advantages. First, since iDMRG operates with a real-space representation and a truncated Hamiltonian corresponding to the dominant $\chi$ Schmidt values, no band projection needs to be taken. Typically in ED computations, the interaction term is projected to the lowest band to make the problem computationally tractable in cases where Landau level mixing is not expected to play a significant role ($W\ll V \ll \Delta$)~\cite{Andrews18, AndrewsThesis}. Conversely, iDMRG does not require such a band projection, since the many-body problem is instead made tractable via a bond dimension truncation. Second, iDMRG can model larger (semi-infinite) system sizes, which are competitive with ED even in their finite $L_y$ dimension~\cite{Grushin15}. Although ED simulations are limited combinatorially by the number of particles and sites (or the number of magnetic unit cells, if a projection to the lowest band is used), in iDMRG simulations the number of sites does not play a direct role. Instead, iDMRG simulations scale exponentially with the chosen cylinder circumference, which typically translates to an improved, albeit sharp, system size cut-off. Furthermore, it was demonstrated that the relevant circumference for quantum Hall states should be measured in terms of the magnetic length $l_B$~\cite{Schoonderwoerd19, SchoonderwoerdThesis}. Finally, the geometry of the infinite cylinder DMRG lends itself to a complementary set of numerical tests. Compared to ED, which focuses primarily on the many-body energy spectrum, DMRG focuses on the ground states~\footnote{We note that variants of the DMRG algorithm geared towards excited states are currently being explored~\cite{Khemani16}.}. Coupled with the infinite cylinder geometry, this enables alternative ways to analyze these states, e.g., via the correlation length, which is related to the many-body gap through $\xi\sim 1/\Delta_\text{m.b.}$.         

\section{Results}
\label{sec:results}

In this section we present the results of our many-body numerics. In Sec.~\ref{subsec:FCIs} we stabilize FCIs with higher Chern number and in Sec.~\ref{subsec:phase_trans} we analyze their metal-to-FCI phase transitions.

\subsection{FCIs in higher Chern bands}
\label{subsec:FCIs}

In order to comment on the relative stability of FCIs, we first need a reliable way to demonstrate their existence. One of the original, most common, and numerically cheapest methods to diagnose fractional quantum Hall states is via Laughlin's charge pumping argument~\cite{Laughlin81}. In our system, this corresponds to adiabatically inserting a flux $\Phi_x$ through the center of the cylinder and measuring the expected charge pumped on the left half of the surface $\braket{Q_\text{L}}=\sum_{i=1}^{\chi}\lambda_{i}^2 Q_{\text{L},i}$~\cite{Cincio13, Zaletel13, Grushin15}. This expected charge is directly related to the Hall conductivity, $\sigma_\text{H}=(e/h^2) C\nu$, and therefore is proportional to the product of the Chern number of the band and the filling fraction. Provided the numerics are reliable, this is sufficient to show the existence of a fractional quantum Hall state. However, since it is often a subtle issue as to whether the charge pumping is accurate (discussed later), we also provide an analysis of the corresponding two-point correlation functions in Appendix~\ref{sec:corr_func}.

\begin{figure}
	\includegraphics[width=\linewidth]{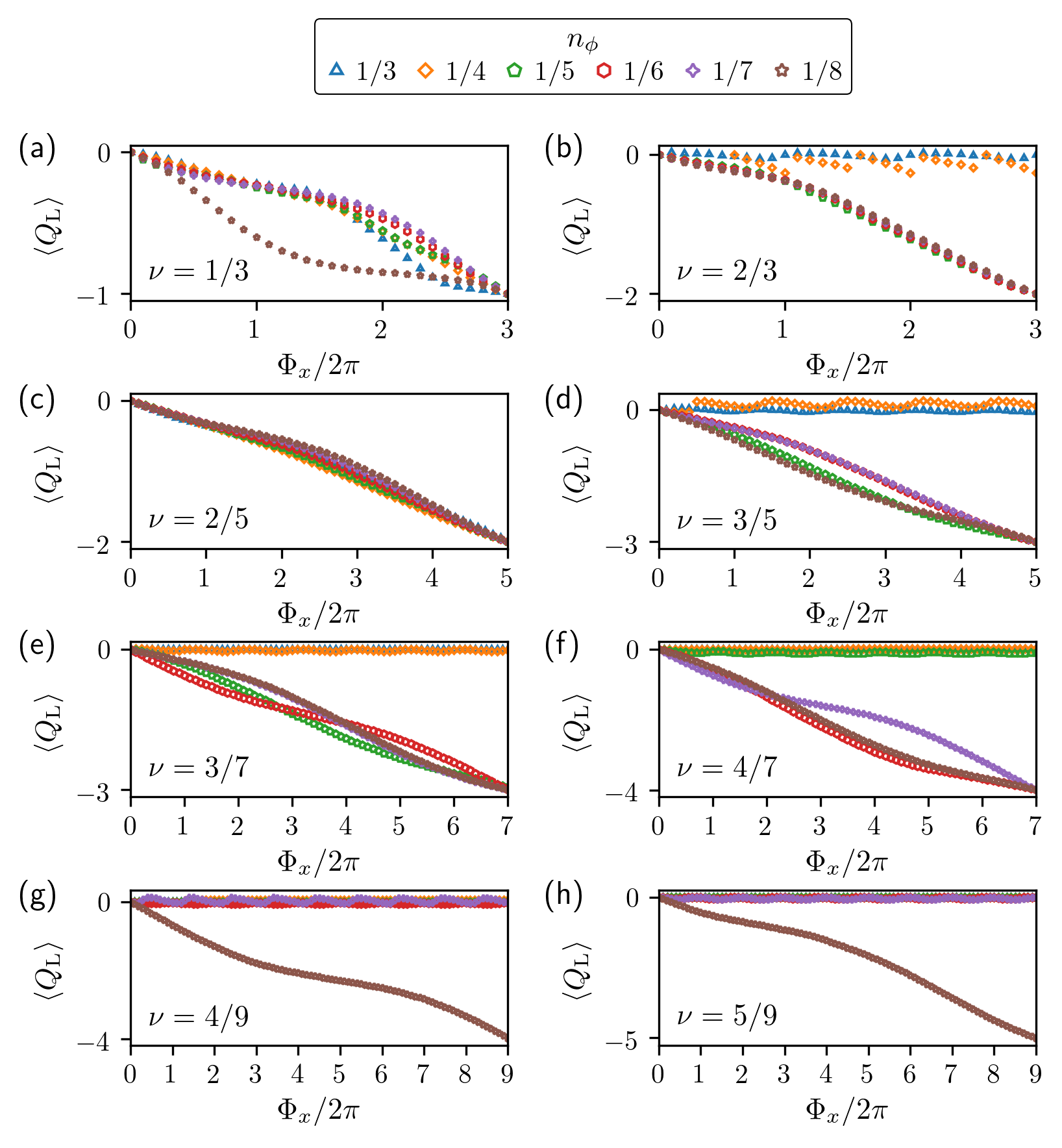}
	\caption{\label{fig:phiflow_analysis}\textbf{FCIs in $C=1$ bands.} Expectation value of charge pumped on the left half of the cylinder, $\braket{Q_\text{L}}$, as a function of adiabatic flux insertion through the cylinder, $\Phi_x$. All of the computations are performed with cylinder circumference $L_y=2s$, interaction strength $V=10$, bond dimension $\chi=250$, and flux interval $\Delta\Phi_x/2\pi=0.1$. The flux densities corresponding to $p<10$ are considered, where the lowest band is gapped.}
\end{figure}

We start our investigation by examining the charge pumping for FCIs in bands with $C=1$, as shown in Fig.~\ref{fig:phiflow_analysis}. As stated before, we systematically consider all distinct $C=1$ bands generated with $p<10$, by setting the flux density according to Eq.~\eqref{eq:nphi}, and we consider all computationally tractable filling fractions by incrementing $|r|$ in Eq.~\eqref{eq:nu}. With these constraints, we are able to demonstrate FCIs for the first four values of $|r|$ with our DMRG parameters. The charge pumping for the first three hierarchy states ($r=1,2,3$) in Figs.~\ref{fig:phiflow_analysis}(a,c,e) have been demonstrated previously (with certain values of $n_\phi$)~\cite{Grushin15, Schoonderwoerd19, Andrews20, Andrews21}, and so these results may be used to benchmark our numerics. In contrast, the charge pumping for the fourth-order hierarchy state ($r=4$) in Fig.~\ref{fig:phiflow_analysis}(g), as well as the states with negative $r$ in Figs.~\ref{fig:phiflow_analysis}(b,d,f,h), have not been previously demonstrated and therefore represent our first set of original results. 

There are several important points that can be learned already from the examination of $C=1$ FCIs. First, we note that FCIs are not demonstrated for every value of $n_\phi$ that generates an isolated band (as per Eq.~\eqref{eq:nphi}). For example, for the Laughlin state in Fig.~\ref{fig:phiflow_analysis}(a), we observe that an FCI is stabilized for every value of $p<10$ that generates a distinct lowest band, whereas for its particle-hole conjugate in Fig.~\ref{fig:phiflow_analysis}(b), FCIs are not found for $n_\phi=1/3$ ($\Delta/W\approx1.73$) and $n_\phi=1/4$ ($\Delta/W\approx3.57$). Since the interaction strength is much larger than the scale of the band structure, this is likely due to the increasing gap-to-width ratio with decreasing $n_\phi$, coupled with the fact that some fractional quantum Hall plateaus are more difficult to stabilize. It is known that the $\nu=2/3$ state, for example, is more difficult to stabilize than the $\nu=1/3$ state in numerical simulations due to the higher density of particles exacerbating finite-size effects. Moreover, as we increase the flatness ratio, the conditions for physically stabilizing FCIs improve. This effect is also reflected in the fact that higher-order states generally require a smaller $n_\phi$ to be demonstrated. For example, for the states with $|r|=4$ in Figs.~\ref{fig:phiflow_analysis}(g,h), we were only able to demonstrate FCIs with $n_\phi=1/8$ ($\Delta/W\approx741$). Second, we observe the existence of FCIs with an interaction strength that far exceeds the band gap. This is an effect that has been noted before in other FCIs~\cite{Moller09, Regnault11, Kourtis14} and we confirm it here, since $\Delta\ll 10$ in all cases. Finally, we note that the dependency of charge pumping on inserted flux can take drastically different forms for a given filling fraction, however the direction of the respective curves always corresponds to $\text{sign}(C)$, which is best illustrated in Fig.~\ref{fig:phiflow_analysis}(a). Generally, we find that discontinuities in the charge pumping curve, such as the $n_\phi=1/4$ curve for $\nu=2/3$ in Fig.~\ref{fig:phiflow_analysis}(b), are a hint of a numerical instability. However, in this figure we present all of our results with the same bond dimension ($\chi=250$) and relative cylinder circumference ($L_y=2s$, where $s$ is the denominator of $\nu$) for a fair comparison of states that is readily reproducible.

In order to select the $C=1$ FCIs presented in Fig.~\ref{fig:phiflow_analysis}, we systematically computed the charge pumping for $|r|\in[1,2,3,4,5]$, $p\in[4,5,6,7,8,9]$, $\chi\in[50,100,\dots,500]$, and $L_y\in[s,2s]$, which is a total of 1200 computations. In the process, we noticed a couple of numerical instabilities that may be useful to know for practitioners of the algorithm. Most importantly, we emphasize that the two widely held assumptions that (i) a smaller $n_\phi$ will improve the stability of an FCI and (ii) a larger $\chi$ will improve the precision of a numerical result for a legitimate FCI configuration, both come with the caveat that the system size $L_y$ is large enough. This can be particularly misleading in the case of charge pumping computations, since we found several examples where an accurate charge pumping result breaks down when the flux density is \emph{decreased} or the bond dimension is \emph{increased}, due to an insufficient system size. We discuss these examples in further detail in Appendix~\ref{sec:stability}. In summary, for numerical computations, statements (i) and (ii) hold only if the system size is sufficiently large. We consequently confirm that all of the claimed states are indeed FCIs by additionally analyzing their two-point correlation functions in Appendix~\ref{sec:corr_func}.                      

\begin{figure}
	\centering{$C=2$ bands}\\
	\vspace{0.5em}
	\includegraphics[width=\linewidth]{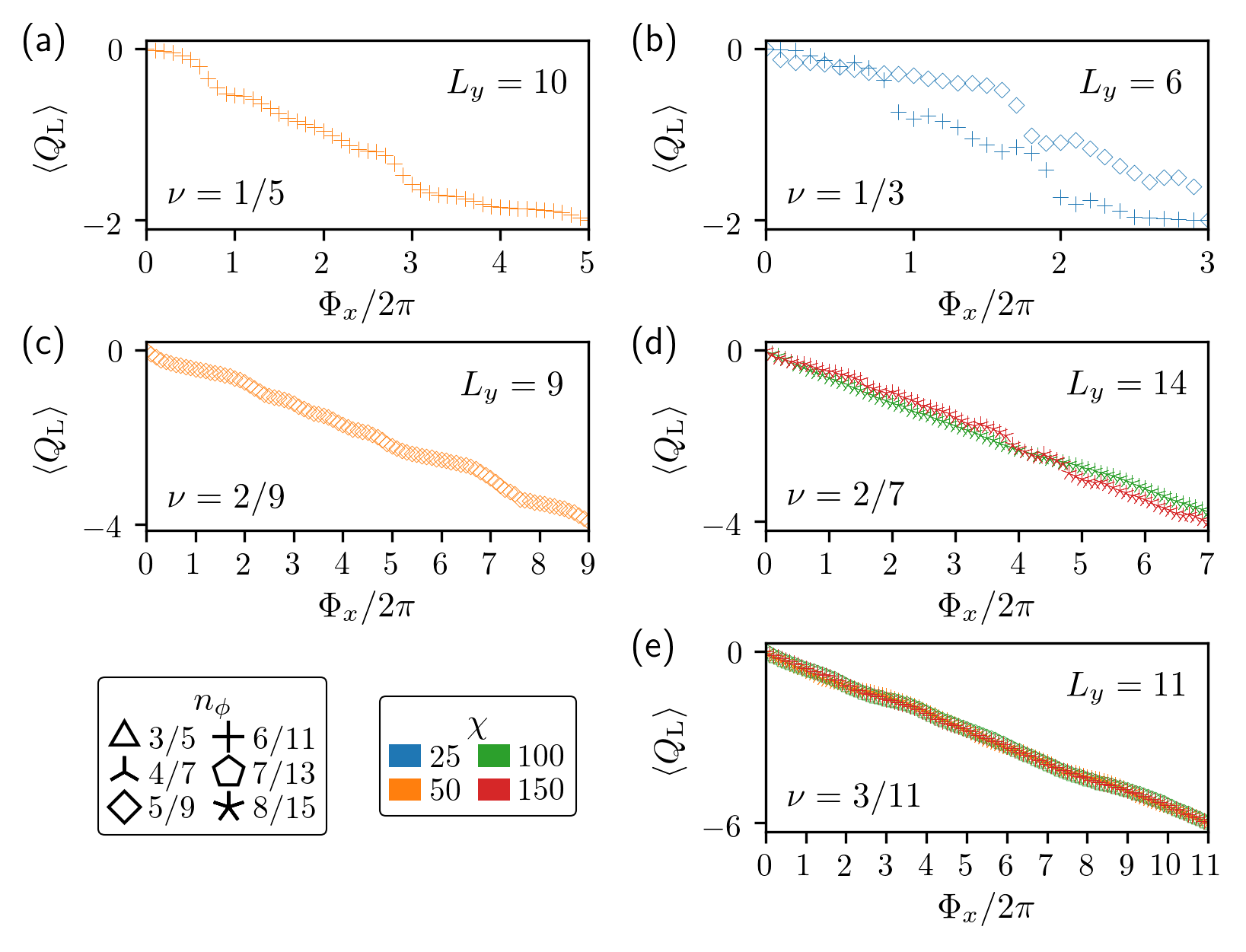}
	\centering{$C=3$ bands}\\
	\vspace{0.5em}
	\includegraphics[width=\linewidth]{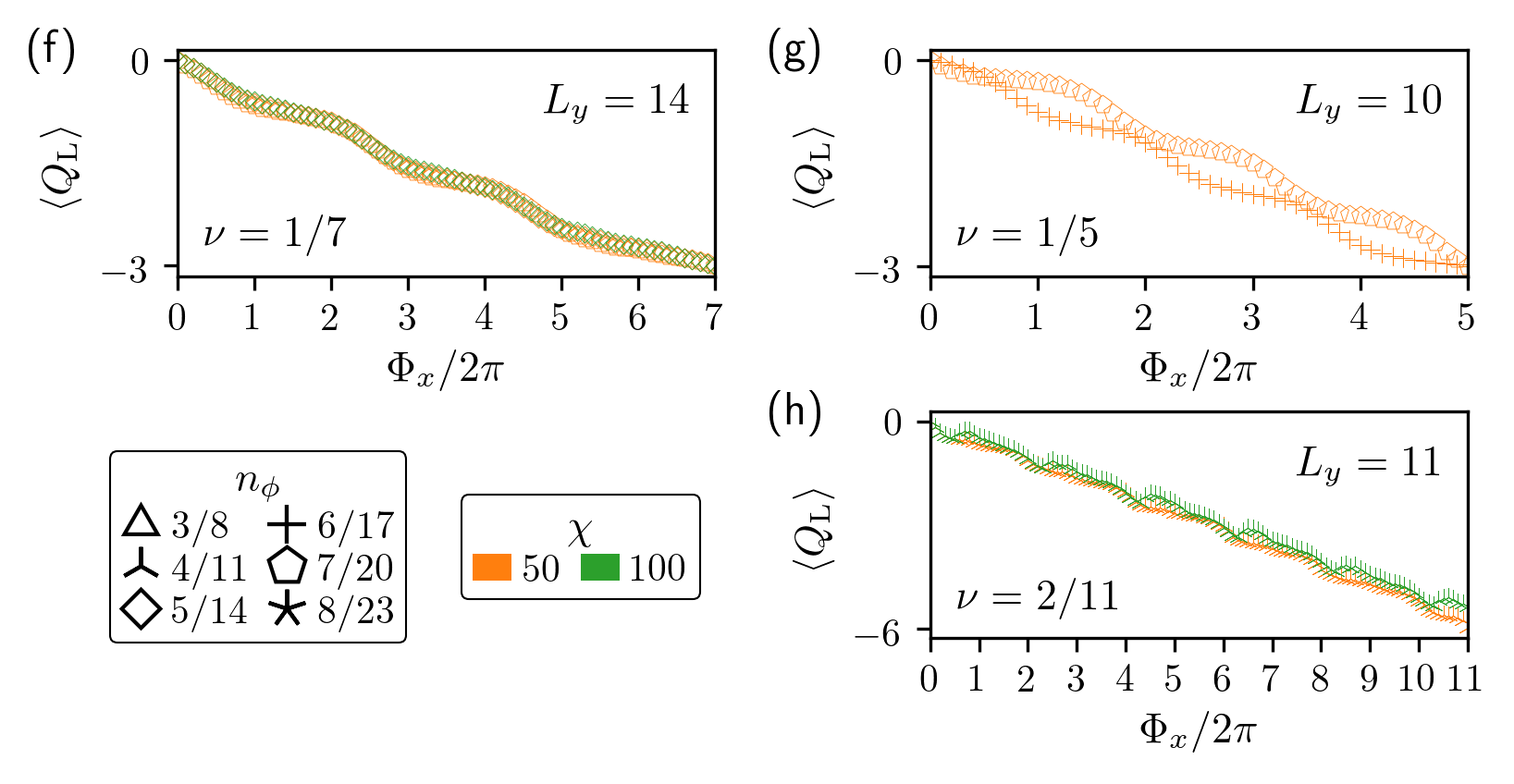}
	\caption{\label{fig:phiflow_c23_analysis}\textbf{FCIs in $C=2,3$ bands.} Charge pumping for FCIs in [(a)--(e)] $C=2$ and [(f)--(h)] $C=3$ bands. All of the computations are performed with interaction strength $V=10$ and flux interval $\Delta\Phi_x/2\pi=0.1$. The flux densities corresponding to $p<10$ are considered, where the lowest band is gapped.}
\end{figure}

Applying the lessons learned from $C=1$ FCIs allows us to efficiently explore the parameter space for FCIs with higher Chern number. In Fig.~\ref{fig:phiflow_c23_analysis}, we present the corresponding charge pumping results from a study of FCIs in $C=2$ and $3$ bands. The computational expense of the higher Chern number FCIs precluded a systematic comparison with respect to flux density and bond dimension, as in Fig.~\ref{fig:phiflow_analysis}, and so instead, we show only the verified $C>1$ FCIs. We notice both of the previously discussed numerical instabilities at play, whereby a decrease in $n_\phi$, as well as an increase in $\chi$, are capable of numerically destabilizing the charge pumping. This is particularly apparent for $\nu=1/3$ in Fig.~\ref{fig:phiflow_c23_analysis}(b), where the bond dimension had to be reduced drastically to achieve the charge pumping result, albeit with noisy, non-monotonic curves. Moreover, the corresponding charge pumping is unsuccessful with smaller values of $n_\phi<6/11$, i.e., closer to the corresponding flat band limit $n_\phi\to1/2$. As previously mentioned, these effects are indicative of the fact that the computations are restricted by system size, which unfortunately cannot be further increased here due to the computational expense. We note that a lack of charge pumping results due to numerical limitations does not exclude the possibility of stabilizing FCIs with those parameters, since we can only verify the existence of FCIs and not the converse. Furthermore, we observe that as we increase the Chern number, the number of accessible $r$ values decreases: where we can verify FCIs with five different $r$ values for $C=2$ but only three different $r$ values for $C=3$. This is due to the decreasing physical stability of these higher-order fractional quantum Hall plateaus, coupled with the increasing numerical expense due to more demanding system size requirements. Interestingly, we find that FCIs with negative $r$ are easier to verify, likely due to their smaller values of $s$, which is physically associated to more robust fractional quantum Hall states and results in more numerically-favorable system sizes.           

\begin{figure}
	\centering{$C=4$ bands}\\
	\vspace{0.5em}
	\includegraphics[width=\linewidth]{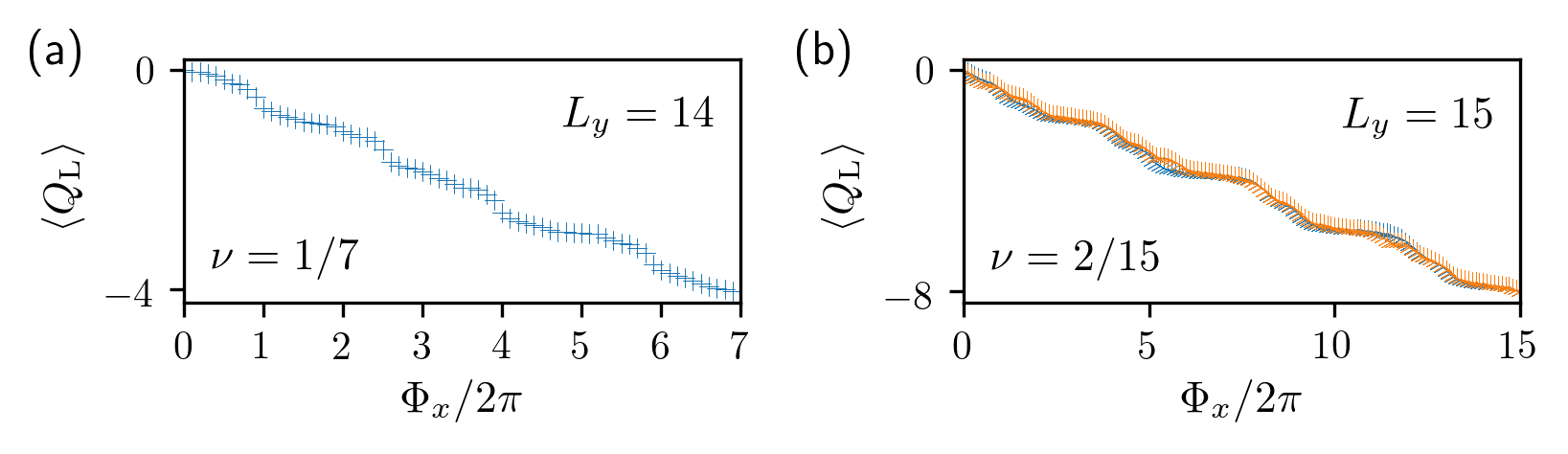}
	\centering{$C=5$ bands}\\
	\vspace{0.5em}
	\includegraphics[width=\linewidth]{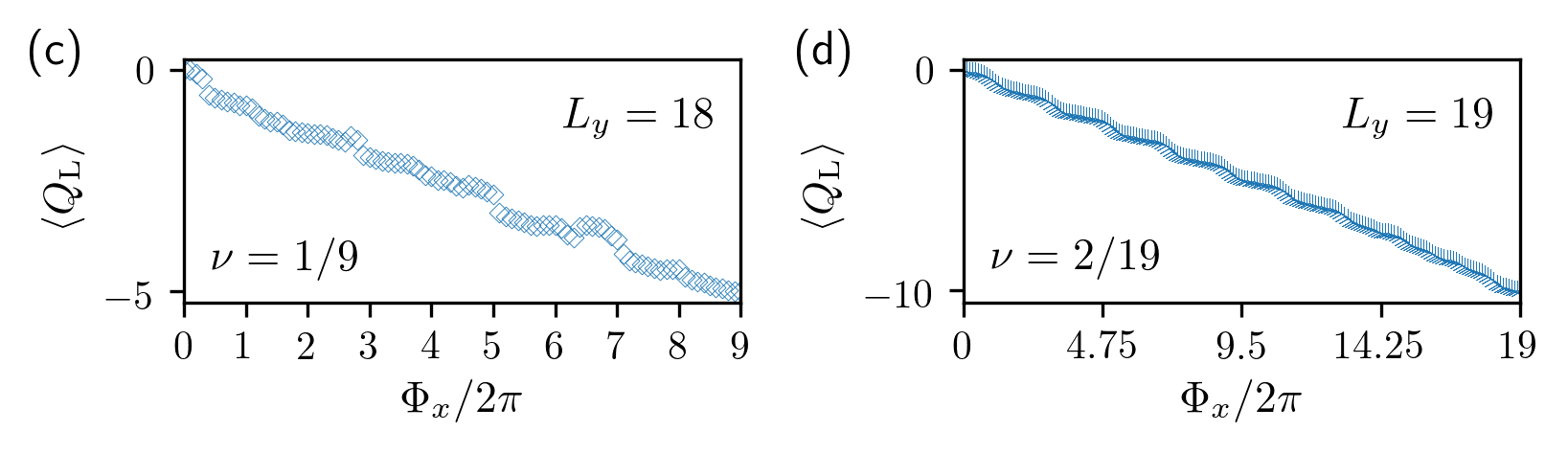}
	\caption{\label{fig:phiflow_c45_analysis}\textbf{FCIs in $C=4,5$ bands.} Charge pumping for FCIs in [(a),(b)] $C=4$ bands with (a)~$n_\phi=6/23$, (b)~$n_\phi=4/15$, and [(c),(d)] $C=5$ bands with (c)~$n_\phi=5/24$, (d)~$n_\phi=4/19$. The bond dimension coloring is the same as in Fig.~\ref{fig:phiflow_c23_analysis}. All of the computations are performed with interaction strength $V=10$ and flux interval $\Delta\Phi_x/2\pi=0.1$. The flux densities corresponding to $p<10$ are considered, where the lowest band is gapped. Note that, unlike in Figs.~\ref{fig:phiflow_analysis} and~\ref{fig:phiflow_c23_analysis}, both columns correspond to negative~$r$.}
\end{figure}

Finally, we extend our analysis to FCIs in Chern bands with $C=4,5$, as shown in Fig.~\ref{fig:phiflow_c45_analysis}. For these values of the Chern number, it is a considerable computational effort to verify the existence of any FCIs, and so we present simply the specific configurations for which we were able to demonstrate flux pumping. In accordance with the previously noted trends, we are able to verify FCIs for a further reduced set of $r$ and, specifically in this case, only for the negative values $r=-1,-2$. As with the most challenging $\nu=1/3$ state in Fig.~\ref{fig:phiflow_c23_analysis}(b), we have to reduce the bond dimension in order to obtain the charge pumping results in most cases. This shows that the system size is strongly restricting the numerics and should be increased. However, as discussed in Appendix~\ref{sec:stability}, increasing the system size typically requires a corresponding increase in $q$ and $\chi$ and so comes at a compound numerical cost. Since the fact that charge pumping can be seen at all is due to its robust topological nature that can persist at such low bond dimensions, this type of scaling analysis may be adapted in the future as a technique to quantify the suitability of a given system size to stabilize an FCI.             

\subsection{Metal-to-FCI phase transitions}
\label{subsec:phase_trans}

Now that we have detected a selection of FCIs in higher Chern bands of the Hofstadter model, we proceed to quantify their stability. A common method for quantifying the stability of fractional quantum Hall states is with respect to their band flatness and interaction strength. Historically, it was conjectured that $W\ll V\ll \Delta$ is required for FCI phases, so that the interaction strength is larger than the band width, such that particles are strongly interacting, but smaller than the band gap, such that band-mixing remains small. It is now known that this is not a necessary condition, since many FCIs have been demonstrated with these inequalities relaxed~\cite{Moller09, Regnault11, Kourtis14} (including the FCIs in this paper). Moreover, the flat-band criterion has since been extended in terms of quantum geometry, where it was demonstrated that it is, in fact, a suppression of Berry curvature fluctuations that improves FCI conditions~\cite{Wu12_2, Roy14, Jackson15, Bauer16}, as long as interactions are suitably tied by locality to the lattice geometry~\cite{Simon20}\footnote{It has recently been shown that it is impossible to engineer an ideal flat band with constant Berry curvature from a lattice model with a finite number of sites per unit cell~\cite{Varjas21, Mera21}.}. For the Hofstadter model, these two criteria directly coincide and are governed by the flux density, as shown in Figs.~\ref{fig:bandstructure_analysis}(b,c). In this section, we use the robustness of an FCI with respect to its interaction strength as a quantifier of stability. To this end, we tune the verified FCIs from Sec.~\ref{subsec:FCIs} from $V=0$ to $10$, such that they undergo a phase transition. Since $\Delta\ll 10$ in all cases, we focus on the metal-to-FCI phase transitions and compare values of $V_\text{crit}$~\footnote{We did not observe any FCI breakdown transitions as the interaction strength is increased to $V=10$}. In the interests of brevity, we present two illustrative case studies: one for $C=1$ and another for $C=2$.

\begin{figure}
	\includegraphics[width=\linewidth]{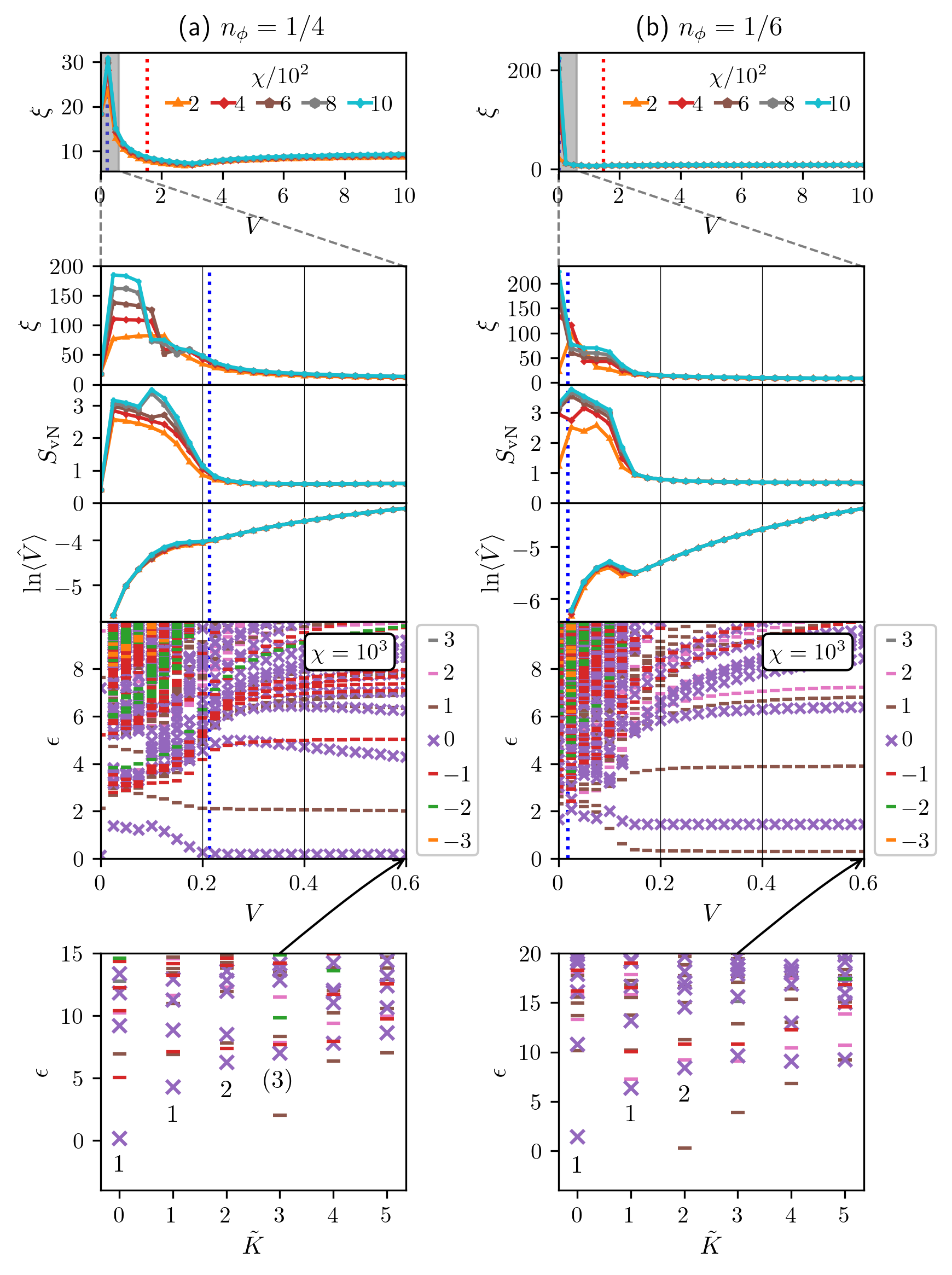}
	\caption{\label{fig:phase_trans}\textbf{Metal-to-FCI phase transitions in $C=1$ bands.} (top panels)~Correlation length $\xi$, von Neumann entanglement entropy $S_\mathrm{vN}$, interaction energy $\braket{\hat{V}}$, and entanglement spectrum $\{\epsilon_i\}$, as a function of interaction strength $V$, for $\nu=1/3$, $L_y=6$ FCIs in $C=1$ bands with (a)~$n_\phi=1/4$ and (b)~$n_\phi=1/6$. The band widths ($W$) and gaps ($\Delta$) are marked with blue- and red-dotted lines, respectively. The entanglement energies are additionally colored corresponding to their $U(1)$ charge eigenvalues $Q_{\text{L},i}$. (bottom panel)~Momentum-resolved entanglement spectrum at $V=0.6$. We select the energy scale and rotate the momentum eigenvalues, such that $\tilde{K}=(K+c)\bmod L_y$ where $c$ is a constant, to emphasize the edge states. The counting of the edge states in the zeroth charge sector is also labeled.}
\end{figure}

In Fig.~\ref{fig:phase_trans}, we demonstrate the metal-to-FCI phase transitions for two $C=1$, $\nu=1/3$ states, with $n_\phi=1/4$ and $1/6$, corresponding to the states shown in Fig.~\ref{fig:phiflow_analysis}(a). We plot the correlation length $\xi$ and von Neumann entanglement entropy $S_\text{vN}$ as a function of $V$, which are both expected to diverge at the transition. Moreover, we examine the interaction energy $\braket{\hat{V}}$~\footnote{The interaction energy $\braket{\hat{V}}=V\braket{\sum_{\braket{i,j}}\rho_i\rho_j}$ is computed for the ground-state wavefunction on the MPS unit cell with periodic boundary conditions.} and entanglement spectrum $\{\epsilon_i\}$ for further hallmarks of metallic and FCI phases. For the $n_\phi=1/4$ configuration in Fig.~\ref{fig:phase_trans}(a), we observe a metallic phase at $V \lesssim 0.2$ and an FCI at all larger values considered. Although $\xi$ and $S_\text{vN}$ diverge with bond dimension over a finite $V$ interval, precluding a precise identification of $V_\text{crit}$, the interaction energy has a clear point of inflection at $V_\text{crit}=0.2\pm0.025$, after which the gradient $\partial \ln\braket{\hat{V}}/\partial V$ is constant\footnote{The kinetic energy $\braket{\hat{T}}=E-\braket{\hat{V}}$ provides similar insight.}. This is characteristic of the transition point, since in the compressible phase the liquid can react to changes in the interaction strength by reconfiguring itself, which leads to a non-linear dependence on $V$, whereas in the incompressible phase the liquid cannot, and so the $\braket{\hat{V}}$ matrix elements are simply scaled by the interaction strength. From the entanglement spectrum, it can also be seen that in the metallic phase the energies are highly sensitive to changes in $V$, whereas in the FCI phase they converge continuously to fixed values. Moreover, since $\ket{\Psi}$ is invariant under rotations about the cylinder axis, the Schmidt states simply acquire a phase factor $e^{-\mathrm{i}2\pi K/L_y}$ under such a transformation and therefore, may be labeled by their momentum quantum number $K=0, 1, \dots, L_y-1$~\cite{Cincio13, Zaletel13}. The momentum-resolved entanglement spectrum at $V=0.6$ shows one branch for the edge states with a $1,1,2,(3),\dots$ counting, consistent with the $\nu=1/3$ FCI state~\cite{Wen92, Li08}. This agrees with analogous studies of the Laughlin state in the Haldane model~\cite{Grushin15}, albeit now also with $V\gg\Delta$. In this example, we can see from the top panel of Fig.~\ref{fig:phase_trans}(a) that the phase transition roughly accords with the energy scale of the band width (blue-dotted line). As the interaction strength surpasses the band width, we stabilize an FCI, which persists even as $V$ exceeds the band gap (red-dotted line). We notice a similar picture for the $n_\phi=1/6$ flux density in Fig.~\ref{fig:phase_trans}(b). We observe the same divergence of $\xi$ and $S_\text{vN}$ with $\chi$ over a finite $V$ interval, as well as a point of inflection of $\braket{\hat{V}}$ directly preceding a steady growth in the FCI phase and following a hallmark weakly-correlated metallic energy spectrum. From this, we deduce that there is a metallic phase at $V \lesssim 0.15$ and an FCI at all greater values. Moreover, the edge states from the momentum-resolved entanglement spectrum at $V=0.6$ are structured in a single branch with a counting of $1, 1, 2, \dots$, again confirming the expected FCI state. However, in contrast to $n_\phi=1/4$, we notice that the transition occurs at $V_\mathrm{crit}=0.15\pm0.025$, which is significantly above the band gap. This means that the $\nu=1/3$ state at $n_\phi=1/6$ is less stable than at $n_\phi=1/4$ in the relative sense that we require a larger $V/W$, but more stable than at $n_\phi=1/4$ in the absolute sense that we require a smaller $V$. 


\begin{figure}
	\includegraphics[width=\linewidth]{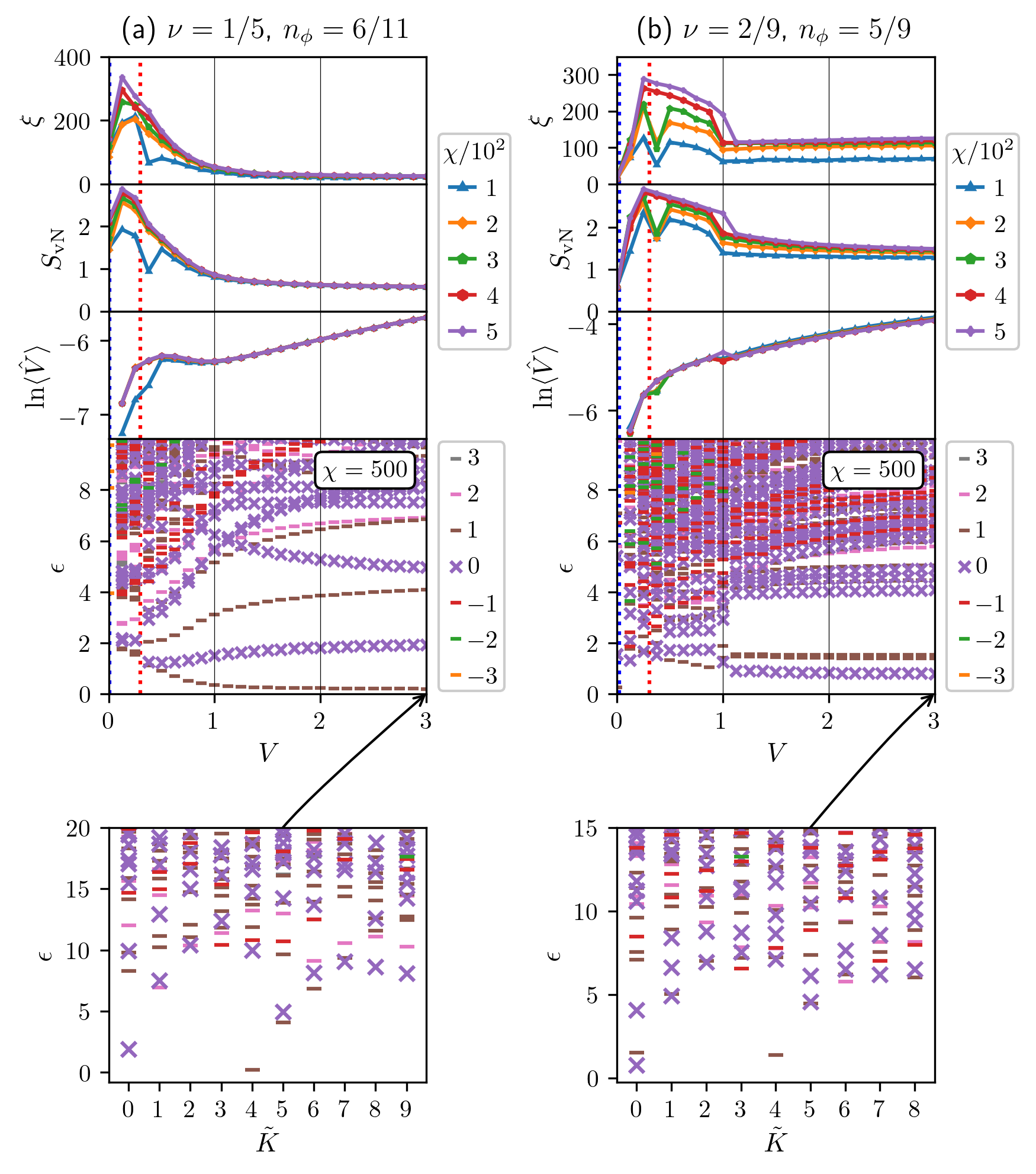}
	\caption{\label{fig:phase_trans_c2}\textbf{Metal-to-FCI phase transitions in $C=2$ bands.} (top panels)~Correlation length $\xi$, von Neumann entanglement entropy $S_\mathrm{vN}$, interaction energy $\braket{\hat{V}}$, and entanglement spectrum $\{\epsilon_i\}$, as a function of interaction strength $V$, for FCIs in $C=2$ bands with (a)~$\nu=1/5$, $L_y=10$, $n_\phi=6/11$, and (b)~$\nu=2/9$, $L_y=9$, $n_\phi=5/9$. The band widths ($W$) and gaps ($\Delta$) are marked with blue- and red-dotted lines, respectively. The entanglement energies are additionally colored corresponding to their $U(1)$ charge eigenvalues $Q_{\text{L}, i}$. (bottom panel)~Momentum-resolved entanglement spectrum at $V=3$. We select the energy scale and rotate the momentum eigenvalues, such that $\tilde{K}=(K+c)\bmod L_y$ where $c$ is a constant, to emphasize the edge states.}
\end{figure}

In Fig.~\ref{fig:phase_trans_c2}, we demonstrate the metal-to-FCI phase transitions for two $C=2$ states, with $\nu=1/5$, $n_\phi=1/4$ and $\nu=2/9$, $n_\phi=1/6$, corresponding to the states shown in Figs.~\ref{fig:phiflow_c23_analysis}(a,c). Perhaps the most striking difference compared to the $C=1$ states in Fig.~\ref{fig:phase_trans}, is the significantly larger interaction strength required to stabilize the FCI phases. In Fig.~\ref{fig:phase_trans_c2}(a), we examine the primary composite fermion state $\nu=1/5$ with $n_\phi=6/11$. Here we observe a metallic phase at $V \lesssim 1$ and an FCI at all larger values. Interestingly in this case, we note that $V_\text{crit}$ is closer to the energy scale of the band gap (red-dotted line) than the band width (blue-dotted line). This means that not only does the FCI phase persist with $V>\Delta$ for this configuration, but this is actually a \emph{required} condition. This behavior has parallels with the secondary composite fermion state $\nu=2/9$ with $n_\phi=5/9$, shown in Fig.~\ref{fig:phase_trans_c2}(b). As before, we record a metallic phase at $V\lesssim 1$ and an FCI phase at all greater values. Naively, a similar value of $V_\mathrm{crit}$ may be expected since the flux densities of the two configurations, and hence also the values of the band flatness, are of the same order of magnitude. However, in general, the values of $V_\mathrm{crit}$ can vary significantly as a function of $n_\phi$ (discussed later). Again, we observe that the transition point is closer to the value of the band gap than the band width in this case. In both examples, the momentum-resolved entanglement spectra at $V=3$ show a two-branch structure for $C=2$~\footnote{The edge-state counting cannot be resolved.}. We note that for Hofstadter bands with increasing Chern number, the band gaps and band widths decrease at different rates, as shown in Fig.~\ref{fig:bandstructure_analysis}(b), and so the comparisons drawn to $W$ and $\Delta$ are specific to these case studies.

\begin{figure}
	\includegraphics[width=\linewidth]{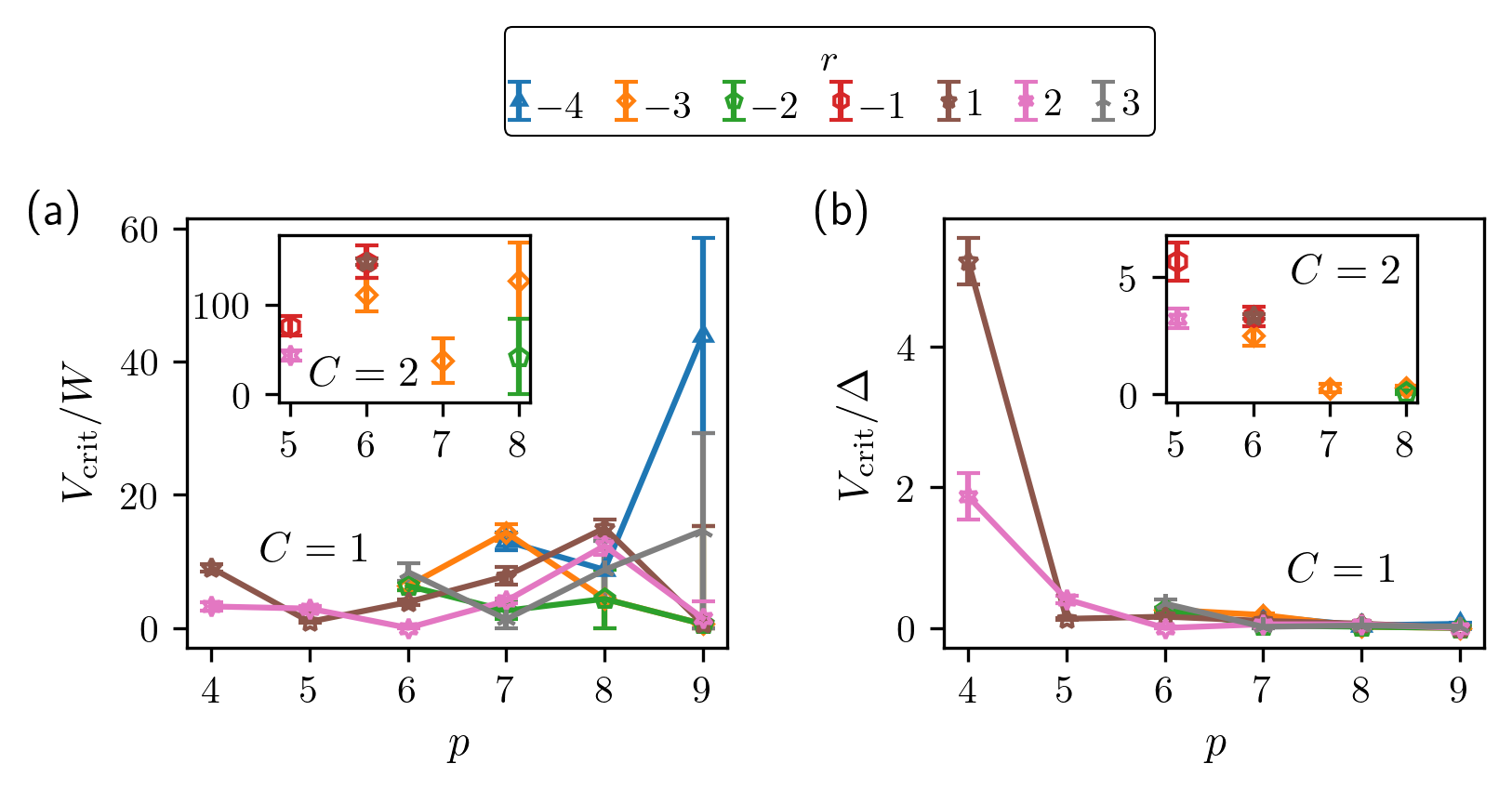}
	\caption{\label{fig:Vcritical}\textbf{Scaling of $V_\mathrm{crit}$ in $C=1,2$ bands.} Critical interaction strength $V_\mathrm{crit}$ in units of (a) band width $W$ and (b) band gap $\Delta$, as a function of $p$. Values are shown for the $C=1$ FCIs from Fig.~\ref{fig:phiflow_analysis} in the main plots and for the $C=2$ FCIs from Fig.~\ref{fig:phiflow_c23_analysis} in the insets.}
\end{figure}

In order to develop a broader understanding of the $V_\mathrm{crit}$ scaling, we present the transition points for a variety of states and flux densities in $C=1,2$ bands in Fig.~\ref{fig:Vcritical}. In Fig.~\ref{fig:Vcritical}(a), we show the dependence of $V_\mathrm{crit}/W$ on the flux density. Here, we can see that the value of $V_\mathrm{crit}$ can vary significantly, even for different states with the same flux density or different flux density configurations of the same state. There is also no general trend that $V_\mathrm{crit}\sim W$, as our case studies in Fig.~\ref{fig:phase_trans} and the original theory may suggest. Instead, we observe from the $C=1$ data that $V_\mathrm{crit}/W$ is, on average, approximately independent of changes to flux density. Since the band width decreases exponentially with $p$ (as shown in Fig.~\ref{fig:bandstructure_analysis}(b)), the error and imprecision of our $V_\mathrm{crit}$ estimates is prohibitively amplified for $p>8$. However, outside of this region, we see that the values of $V_\mathrm{crit}/W$ within an $r$-series, as well as the values of $V_\mathrm{crit}/W$ among different $r$-series, are of a comparable order of magnitude $\braket{V_\mathrm{crit}/W}\sim 10$. From the $C=2$ data we see a similar picture, albeit with fewer points, larger error bars, and an increased average order of magnitude $\braket{V_\mathrm{crit}/W}\sim 10^2$. For comparison, in Fig.~\ref{fig:Vcritical}(b) we show the dependence of $V_\mathrm{crit}/\Delta$ on the flux density. In this case, since the band gap is approximately constant with flux density and not exponentially dependent like the band width, the $V_\mathrm{crit}/\Delta$ values rapidly approach zero with increasing $p$. From the band structure scaling in Fig.~\ref{fig:bandstructure_analysis}(b), we expect the convergence to be asymptotic. This rapid decay is also reflected for the $C=2$ data. Note that there is no general relationship $V_\mathrm{crit}\sim\Delta$, which our case studies in Fig.~\ref{fig:phase_trans_c2} may suggest. The analysis shows that $V_\mathrm{crit}/W$ is the relevant quantity to compare the stability of FCIs. Although it is roughly constant with flux density for FCIs in bands of the same Chern number, there are significant fluctuations present, which prevents a simple universal scaling relation being established. Generally, it is important to be aware that FCIs may be stabilized via different mechanisms and hence, should be studied on a case-by-case basis.        

\section{Discussion and conclusions}
\label{sec:discussion}

\begin{table}
	\caption{\label{tab:summary}\textbf{Summary of stabilized FCIs.} List of FCIs presented in this paper with respect to $r$ and $C$. FCIs that have been demonstrated using ED calculations are colored red~\cite{Andrews18}, FCIs that have been demonstrated in this paper (using iDMRG) are colored green, and the intersection is colored black. Note that the $r=-1$ state is excluded for $C=1$, since this corresponds to integer filling.}
	\begin{ruledtabular}
		\begin{tabular}{c|ccccccccc}
			\diagbox[innerwidth=0.8cm, height=0.6cm]{$C$}{$r$} & $-5$ & $-4$ & $-3$ & $-2$ & $-1$ & $1$ & $2$ & $3$ & $4$ \\
			\hline
			$1$ & \gcmark & \gcmark & \cmark & \cmark & --- & \cmark & \cmark & \cmark & \gcmark \\
			$2$ & & & \cmark & \cmark & \cmark & \cmark & \cmark & {\color{Red}\cmark} & \\
			$3$ & & & & \gcmark & \gcmark & \gcmark & & & \\
			$4$ & & & & \gcmark & \gcmark & & & & \\
			$5$ & & & & \gcmark & \gcmark & & & & \\
		\end{tabular}
	\end{ruledtabular}
\end{table}

In this paper, we have analyzed the stability of FCIs in higher Chern bands of the Hofstadter model. Using a combination of charge pumping and correlation function analysis, we have numerically demonstrated the existence of FCIs in Chern number $C=1,2,3,4,5$ bands at the filling fractions predicted by the generalized Jain series~\cite{Moller15}, summarized in Table~\ref{tab:summary}. Moreover, we studied their metal-to-FCI phase transitions with respect to interaction strength. We found that $V_\mathrm{crit}/W$ is the relevant quantity to compare the stability of FCIs and showed that it is, on average, of the same order of magnitude for FCIs in bands of the same Chern number, as a function of flux density, despite appreciable fluctuations among different configurations. Moreover, we showed that the values of $V_\mathrm{crit}/W$ for $C=2$ FCIs are typically an order of magnitude larger than for $C=1$. We did not observe any FCI breakdown transitions with increasing interaction strength in the interval $V\in[0,10]$. 

In order to comment on the stability of FCIs in such a theoretical study, it is crucial to distinguish between numerical and physical stability. In terms of numerical stability, we found that, just like for ED computations~\cite{Andrews18}, FCIs in higher Chern bands are more challenging to stabilize. In iDMRG studies, this is due to the more demanding lattice geometries resulting in larger required system sizes and a greater flux insertion needed to demonstrate charge pumping. Moreover, FCIs with smaller and/or negative $r$ are easier to stabilize, for analogous reasons. Although charge pumping is often used to demonstrate FCIs because it can persist at extremely low bond dimensions, it also comes with its own notable disadvantages. In particular, it is difficult to maintain an adiabatic flux insertion at insufficient system sizes, which can result in misleading breakdowns with decreasing flux density or increasing bond dimension.

In terms of physical stability, our results accord with, and extend, the findings of Andrews and M{\"o}ller~\cite{Andrews18}. We find that FCIs in $C=1$ bands are more stable than $C>1$ FCIs, since they require a smaller $V_\mathrm{crit}/W$ value for the metal-to-FCI transition, and hence are stable for a larger range of $V$ in units of the band width. Out of the FCIs that we stabilized numerically, we found that physical stability decreases with increasing $|r|$, with only a few exceptions. For example, the state at $\nu=1/3$ for $C=2$ is significantly more fragile than the corresponding $r=1$ filling, and the $\nu=2/19$ state for $C=5$ is more robust. These outliers are potentially due to competing states of similar energy detracting from the stability of the overall ground states~\cite{Andrews18}. In our previous ED study, we concluded that $C=2$ FCIs with $\nu=1/5$ and $3/11$ are particularly stable owing to their large gaps in the particle entanglement spectra~\cite{Andrews18}. Using iDMRG, we can support this claim and also add the $\nu=1/7$ and $\nu=1/5$ FCIs with $C=3$ as prime candidates for future investigations of FCIs in higher Chern bands.   

There are several promising experimental realizations and applications that motivate this work. For example, CIs with $C=2$ have recently been demonstrated in van der Waals heterostructures without a magnetic field~\cite{Chen20}, which naturally leads investigations in the direction of fractional quantum Hall states under similar conditions. Moreover, twisted double bilayer graphene has been shown to be a suitable candidate to host FCIs in $C=2$ bands with the application of an external electric field~\cite{Liu21}, and it is known that such FCIs can also be demonstrated using small-scale cold-atom experiments with existing detection methods~\cite{Repellin20}. Coupled with these experimental advancements, there are already a number of proposed applications. Most saliently, FCIs with $|C|>1$ can be mapped to $|C|$-layer fractional quantum Hall systems, up to boundary conditions~\cite{Qi11, Wu12, Wu13, Wu14}. Aside from the potential insights this can bring to multi-layer fractional quantum Hall research, this also implies that lattice dislocations in a $|C|>1$ FCI can be mapped to layer permutations, which can increase the genus of the ground-state manifold~\cite{Barkeshli12, Liu17}. Convenient methods for manipulating the genus in this way open the door to a new type of topological quantum computing utilizing extrinsic defects~\cite{Barkeshli12, Barkeshli13, Knapp19}.           

Since FCIs are stabilized via different mechanisms, some of which are still unknown~\cite{Zaletel15, Fu16}, it is unwise to make sweeping generalizations about their stability with respect to particular parameters. Instead, the priority in the short term is to establish a few robust candidates for $|C|>1$ FCIs to guide experiments. Future work in this area may involve, for example, an analysis of the role of interaction range. There is currently a wealth of numerical evidence to suggest that Abelian Jain states favor short-range interactions~\cite{Liu13, Sheng11, Wu12_2, Andrews21}, whereas exotic fractional quantum Hall states may be stabilized exclusively via long-range interactions~\cite{Liu13_2, Yang19}. It is important to establish where $|C|>1$ FCIs fall on this spectrum to facilitate reliable device engineering. Computationally, this is challenging using iDMRG due to the growth of the Hamiltonian MPO dimension, the scaling of 2D interaction range in a 1D MPS representation, and the larger system sizes required to alleviate finite-size effects. Other avenues for future investigation include: a re-evaluation of these states using complementary computational methods, such as projected entangled pair states~\cite{Chen20_2}; a direct comparison between the stability of $|C|>1$ FCIs and $C$-component fractional quantum Hall states; as well as a tailored analysis for the FCIs recently observed in moir{\'e} materials, such as magic-angle twisted bilayer graphene~\cite{Xie21}. We hope that this study will help focus research efforts to promote the widely-accessible realization of this promising phase of matter.

\begin{acknowledgments}
We thank Madhav Mohan for helping to gather the data in Figs.~\ref{fig:phiflow_analysis},~\ref{fig:phiflow_c23_analysis},~\ref{fig:phiflow_c45_analysis}, and Johannes Hauschild for useful discussions. Calculations were performed using the \textsc{TeNPy} Library (version 0.5.0)\cite{tenpy} and \textsc{GNU Parallel}~\cite{Tange11}. We thank S3IT (\href{www.s3it.uzh.ch}{www.s3it.uzh.ch}), the University of Zurich's Service and Support for Science IT team, and in particular Darren Reed, for help with optimizing the computations. B.A. and T.N. acknowledge support from the Swiss National Science Foundation under Grant No.~PP00P2\_176877, and G.M. acknowledges support from the Royal Society under University Research Fellowship URF\textbackslash R\textbackslash180004. 
\end{acknowledgments}

\appendix

\section{Correlation function analysis}
\label{sec:corr_func}

To complement our charge pumping computations, we additionally diagnose our quantum Hall states using a correlation-function-based approach~\cite{Pu17, He17, Andrews18, Andrews21}. To this end, we study the connected two-point density correlation functions $g(x)=\braket{\normord{\rho_{0,0}\rho_{x,0}}}-\braket{\rho_{0,0}\rho_{\infty,0}}$ for each of the FCIs presented in Sec.~\ref{subsec:FCIs}. This quantity represents the probability of finding two fermions at a certain separation relative to a uniform uncorrelated gas. One fermion is kept fixed at the origin $(0,0)$, whereas the other fermion is at a position $(x,0)$, in units of lattice sites, which corresponds to positions on a straight line along the (infinite) cylinder axis. Due to the radial symmetry of the correlation function (at short distances $x\ll L_y/2$), analogous results can also be obtained along the circumference. However, we choose to study a path along the cylinder axis, so that we are able to analyze the correlation functions at the correlation length scale.

One of the defining features of a quantum Hall state are its gapped bulk and gapless edge, which give rise to exponential ($\sim\mathrm{e}^{-x/\xi}$) and algebraic ($\sim x^{-\alpha}$) correlation functions at long distances, respectively~\cite{He17}. By studying a path along the cylinder axis, we are probing the bulk and therefore expect to record an exponential convergence for an FCI ground state. Specifically, we anticipate the microscopic physics to be reflected at short distances, with potential fluctuations, followed by an exponential decay in the asymptotic $x\to\infty$ limit. Since this exponential decay is absent in competing phases, such as superfluid phases and charge density waves, it is often used as a hallmark of a quantum Hall state~\cite{He17}.

\begin{figure}
	\includegraphics[width=\linewidth]{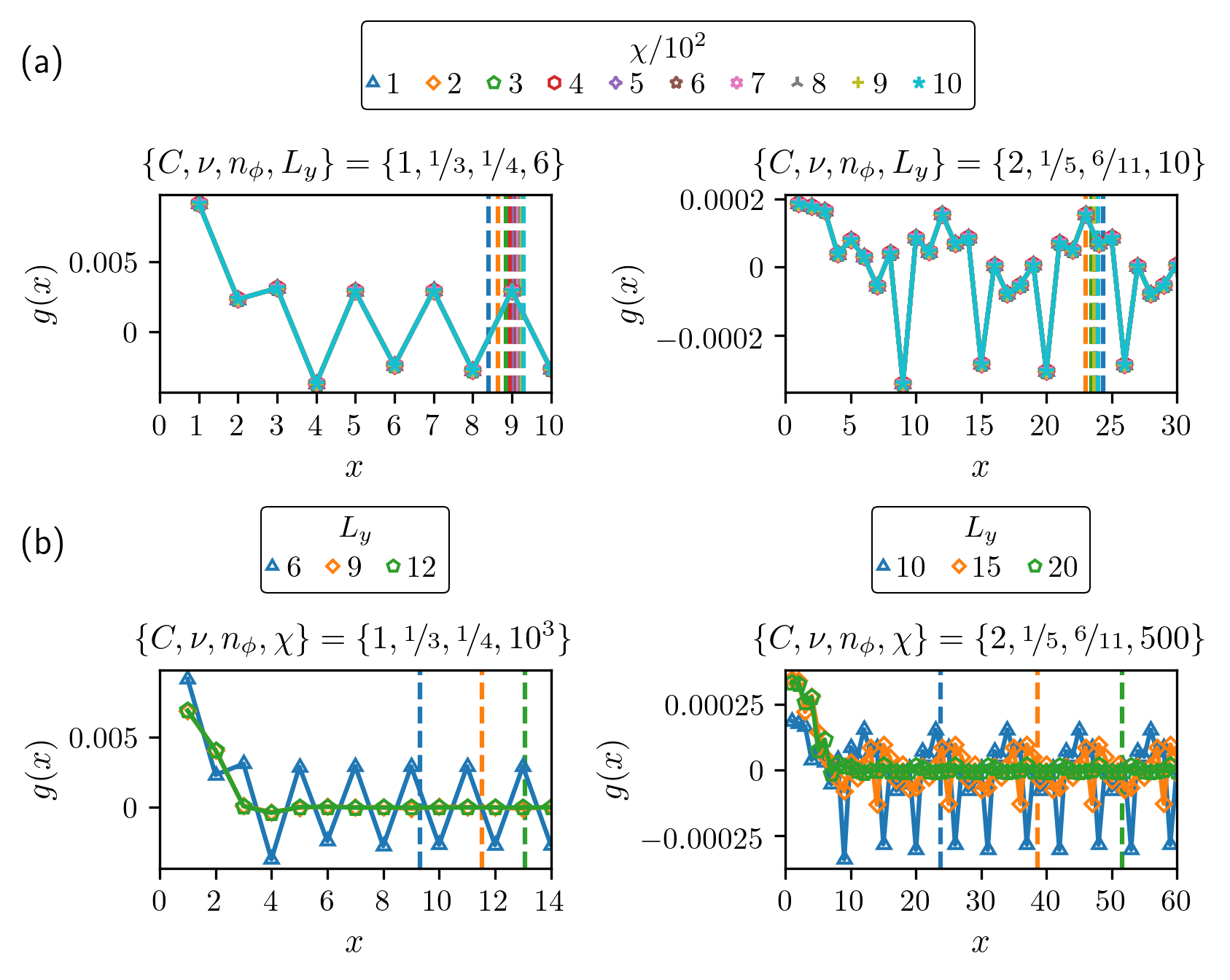}
	\caption{\label{fig:corrfunc_chi_Ly}\textbf{Scaling of correlation functions with bond dimension and system size.} Scaling of density-density correlation functions $g(x)=\braket{\normord{\rho_{0,0}\rho_{x,0}}}-\braket{\rho_{0,0}\rho_{\infty,0}}$, as a function of $x$, the number of sites in the $x$ direction, for the case studies analyzed in (left) Fig.~\ref{fig:phase_trans}(a) and (right) Fig.~\ref{fig:phase_trans_c2}(a), with respect to (a) bond dimension and (b) system size. The $x=0$ point has been excluded since trivially $\braket{\normord{\rho_{0,0}\rho_{0,0}}}=0$ by the Pauli exclusion principle. The corresponding correlation lengths are marked with dashed lines.}
\end{figure}

Before we analyze the correlation functions in depth, it is important to review the numerical scaling of this observable. Since many of the FCIs demonstrated via charge pumping in Sec.~\ref{subsec:FCIs} used minimal values for the bond dimension and system size, it is crucial to understand the effect that this can have on the correlation functions, in order to accurately interpret the results. In Fig.~\ref{fig:corrfunc_chi_Ly}, we present the scaling of correlation functions with bond dimension $\chi$ and system size $L_y$, for the case studies shown in Figs.~\ref{fig:phase_trans}(a) and~\ref{fig:phase_trans_c2}(a). From the scaling with bond dimension, shown in Fig.~\ref{fig:corrfunc_chi_Ly}(a), it can be seen that $\chi$ has a negligible impact on the correlation functions, above a certain threshold. In these examples, the threshold is $\chi\lesssim100$, however for larger cylinder circumferences this can be higher. In both cases, the correlation length converges to a fixed value with $\chi$, at values of $\chi$ well above the threshold required to obtain the converged form of the correlation function. In the $C=1$ example, $\xi$ smoothly converges from below, whereas in the $C=2$ example, it undergoes an oscillatory convergence. From the scaling with system size, shown in Fig.~\ref{fig:corrfunc_chi_Ly}(b), it is clear that $L_y$ has a drastic impact on the form of the correlation functions. In both cases, it can be seen that the high-frequency oscillations of the correlation function are an artifact of the quasi-1D nature of the system. Once the value of $L_y$ is increased, these oscillations diminish. In the $C=1$ example, we see that for $L_y\geq9$ the correlation functions take the quintessential form for a Laughlin state: smooth oscillations exponentially decaying to a fixed value~\cite{Chakraborty}. In the $C=2$ example, we observe that the high-frequency oscillations are again filtered out; however, on this occasion a slight low-frequency undulation remains. On average, we still observe a smooth exponential decay to a finite value. However, it should be noted that owing to the reduced particle density in the $C=2$ example, the scale on the plot is more than an order of magnitude smaller than for $C=1$. Moreover, minor oscillations corresponding to competing charge density wave phases have been shown to be more prevalent in higher-$|C|$ FCIs~\cite{Andrews18}. For these example configurations, $\xi$ converges from below in both cases. This analysis shows that the correlation functions in this section are predominantly restricted by the system size. Above a modest $\chi$ threshold, high-frequency oscillations are removed in the limit of large $L_y$.    

\begin{figure}
	\includegraphics[width=\linewidth]{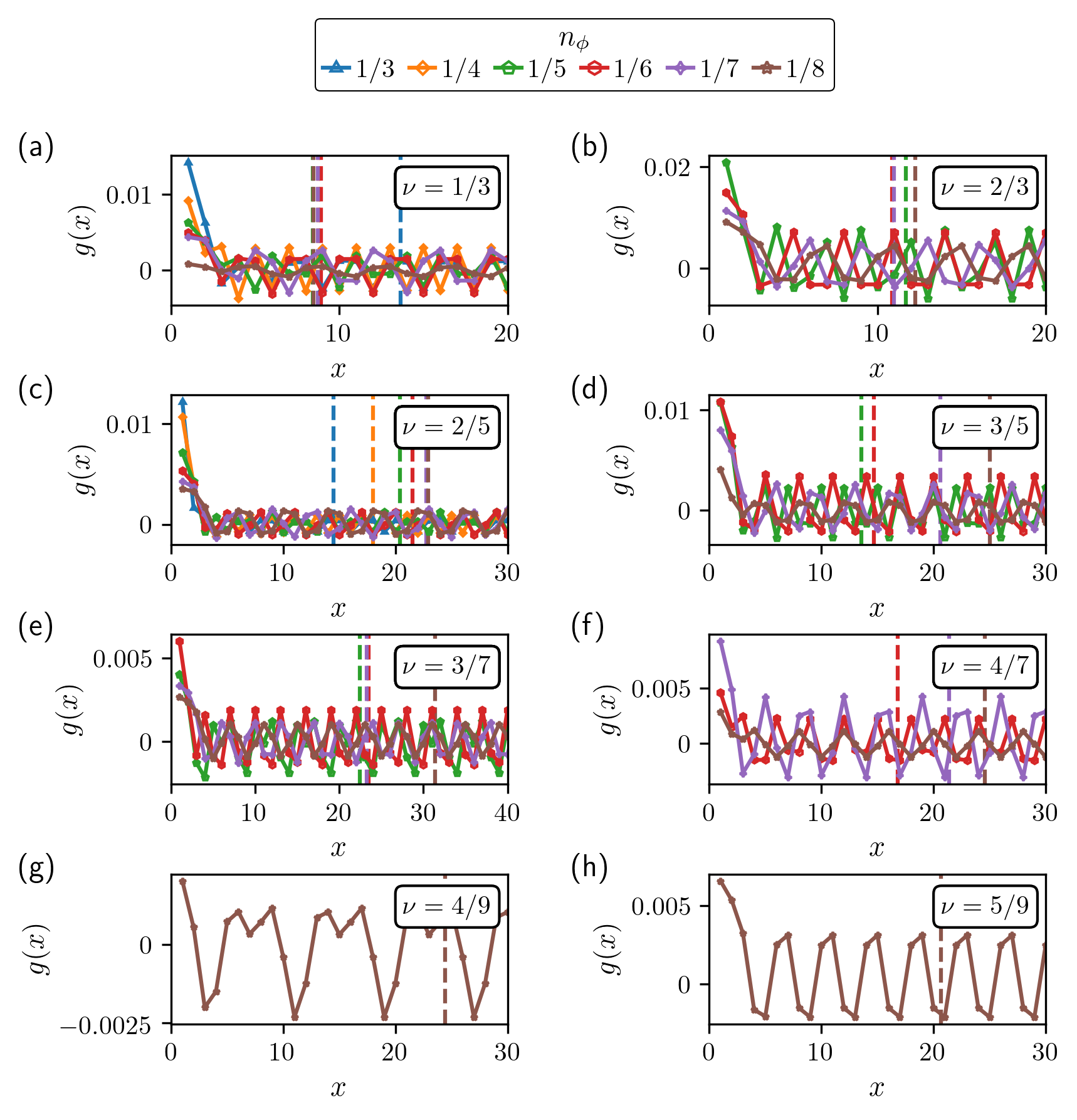}
	\caption{\label{fig:corrfunc_analysis}\textbf{Connected two-point correlation functions for FCIs in $C=1$ bands.} Density-density correlation functions for the FCIs in Fig.~\ref{fig:phiflow_analysis}, presented as in Fig.~\ref{fig:corrfunc_chi_Ly}. The parameters are the same as those used for the charge pumping. In this case, $\chi=250$ and $L_y=2s$. The corresponding correlation lengths are marked with dashed lines.}
\end{figure}

With this in mind, we now examine the correlation functions for the $C=1$ FCIs, in Fig.~\ref{fig:corrfunc_analysis}. In all cases, the $x=0$ point is omitted since $\braket{\normord{\rho_{0,0}\rho_{0,0}}}=0$, as demanded by the Pauli exclusion principle. There are several points to note here, which highlight both the numerical and physical features of our simulations. With respect to the numerics, we see that the correlation functions appear to start from their maximum values and fluctuate at short distances before exhibiting a steady high-frequency oscillation in the $x\to\infty$ limit. In fact, the correlation functions converge to their maximum values at small $x<a$ before decaying rapidly (cf., Supplementary Material of Ref.~\onlinecite{Andrews18}), which is not resolved on the lattice scale. Moreover, the high-frequency oscillations in the large-$x$ limit are directly akin to the $C=1$ example in Fig.~\ref{fig:corrfunc_chi_Ly}(b) and likewise diminish as the system size is increased. With respect to the physics, we observe the typical irregular fluctuations at short distances, which reflects the microscopic details of the Hamiltonian, followed by an exponential decay to a fixed value, on average, in the asymptotic limit, which confirms the quantum Hall state. The simplest examples of this are the first three hierarchy states $\nu=1/3,2/5,3/7$ together with their first two particle-hole conjugates $\nu=2/3,3/5$, particularly at larger $n_\phi$. In these cases, the oscillations induced by the 1D nature of the system are pronounced with a single frequency component and so are easy to identify. For the other configurations, with smaller $n_\phi$ and higher-order $\nu$, the systems are less one-dimensional and so the oscillations are smaller, cf., the diminishing oscillations with decreasing $n_\phi$ in the $\nu=1/3$ example. Moreover, they are occasionally composed of more than one frequency component, such as for $\nu=4/9$, which indicates the potential influence of a minor charge density wave contribution. Overall, however, the correlation functions for the $C=1$ FCIs clearly support their identification as quantum Hall states, once the numerical artifacts have been taken into account.       

\begin{figure}
	\centering{$C=2$ bands}\\
	\vspace{0.5em}
	\includegraphics[width=\linewidth]{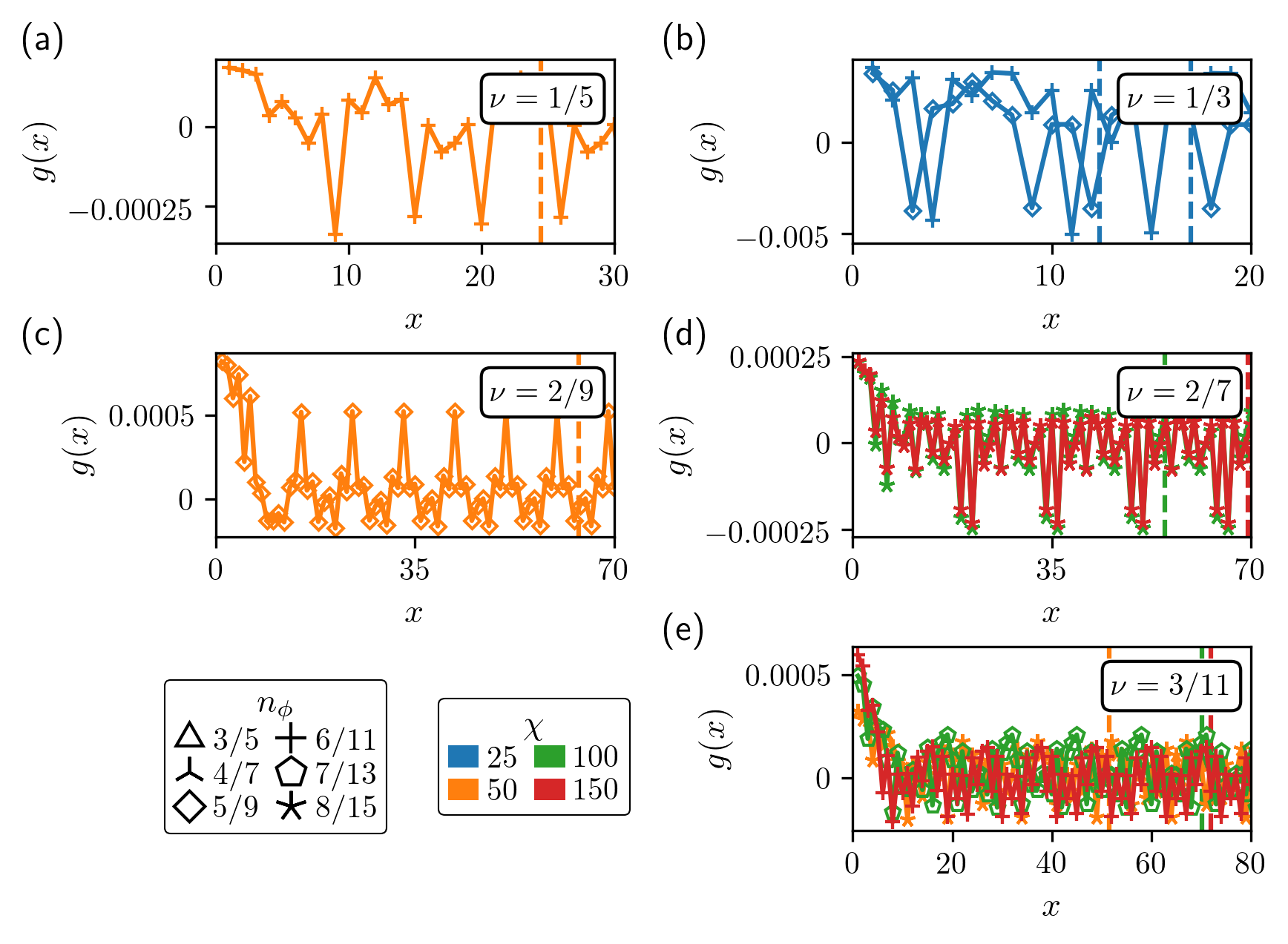}
	\centering{$C=3$ bands}\\
	\vspace{0.5em}
	\includegraphics[width=\linewidth]{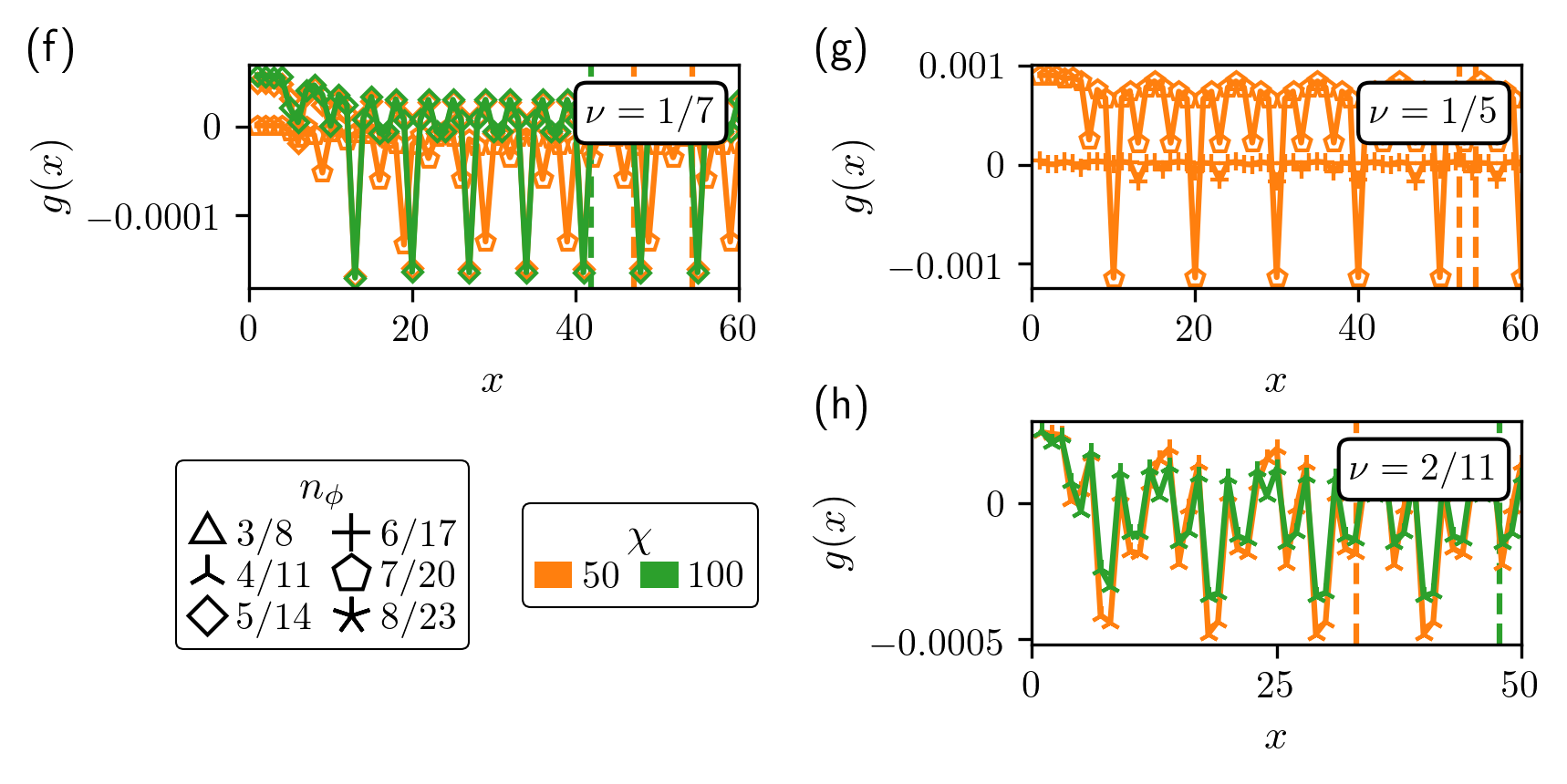}
	\caption{\label{fig:corrfunc_c23_analysis}\textbf{Connected two-point correlation functions for FCIs in $C=2,3$ bands.} Density-density correlation functions for the FCIs in Fig.~\ref{fig:phiflow_c23_analysis}, presented as in Fig.~\ref{fig:corrfunc_chi_Ly}. The parameters are the same as those used for the charge pumping. In this case, (a)~$L_y=10$, (b)~$L_y=6$, (c)~$L_y=9$, (d)~$L_y=14$, (e)~$L_y=11$, (f)~$L_y=14$, (g)~$L_y=10$, and (h)~$L_y=11$. The corresponding correlation lengths are marked with dashed lines.}
\end{figure}

We progress by examining the correlation functions for the $C=2,3$ FCIs, in Fig.~\ref{fig:corrfunc_c23_analysis}. As before, we omit the $\braket{\normord{\rho_{0,0}\rho_{0,0}}}=0$ point in all cases, and present the correlations functions that directly correspond to the FCIs identified via charge pumping in Sec.~\ref{subsec:FCIs}. In this case, there are a number of notable differences between the $C=1$ examples in Fig.~\ref{fig:corrfunc_analysis} from the numerical perspective. In particular, we now consistently observe oscillations in the asymptotic limit that consist of multiple frequency components, akin to the $C=2$ example in Fig.~\ref{fig:corrfunc_chi_Ly}(b). We have identified three reasons for this discrepancy. First, the smaller particle density in these systems reduces the scale of the correlation functions and makes the oscillations more visible; second, the larger Chern number makes charge density wave competition more prevalent~\cite{Andrews18}; and finally, the severe numerical constraints of these systems exacerbates any finite-size effects. In addition to this, since some configurations for $C=2,3$ are shown for various bond dimensions, we can now comment on their scaling. Complementing Fig.~\ref{fig:corrfunc_chi_Ly}(a), where we showed that the bond dimension has negligible effect on the correlation functions above a certain threshold, here we identify a couple of cases where, for small $\chi$, the scaling does have an appreciable effect, e.g., for $\nu=2/7$ at $n_\phi=8/15$ and $\nu=2/11$ at $n_\phi=4/11$. In both examples, an increasing bond dimension works similarly to an increasing system size, in that it diminishes high-frequency oscillations. This observation accords with Fig.~\ref{fig:corrfunc_chi_Ly}(a) in that the effect is only seen for comparatively small bond dimensions, which are determined by the cylinder circumference. From the physical perspective, we can take the average of these oscillations and recover an exponential decay in the asymptotic limit, as before. However, in many cases, low-frequency undulations remain in the $x\to\infty$ limit signaling a charge density wave competition of physical origin. Notwithstanding the significant numerical effects that partially obscure the underlying structure, the correlation functions for the $C=2,3$ FCIs support their identification as quantum Hall states, albeit with a clear charge density wave contribution in several cases.

\begin{figure}
	\centering{$C=4$ bands}\\
	\vspace{0.5em}
	\includegraphics[width=\linewidth]{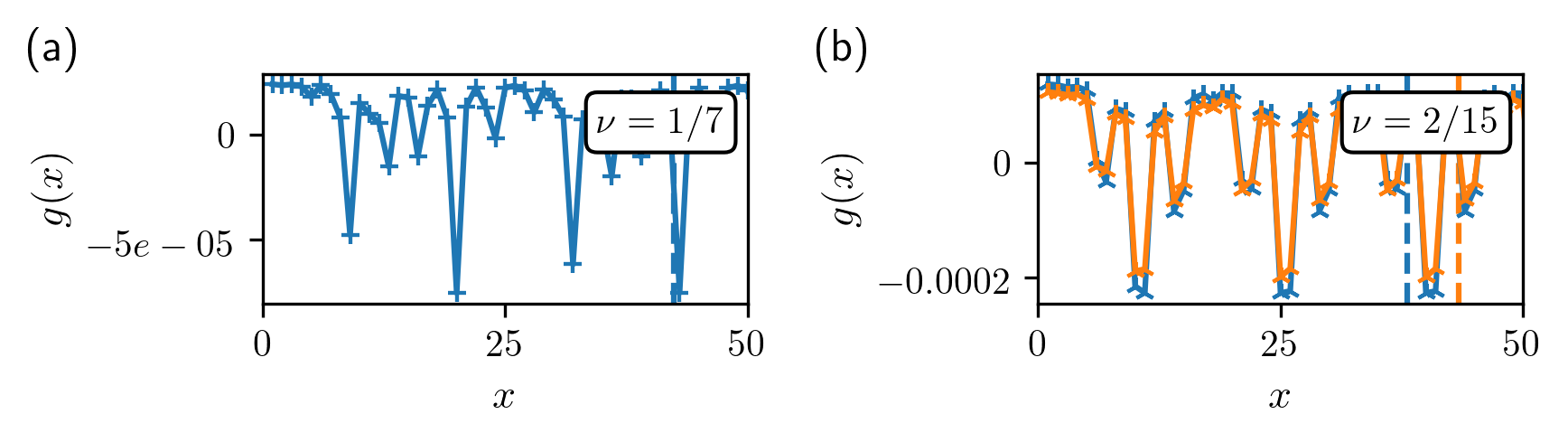}
	\centering{$C=5$ bands}\\
	\vspace{0.5em}
	\includegraphics[width=\linewidth]{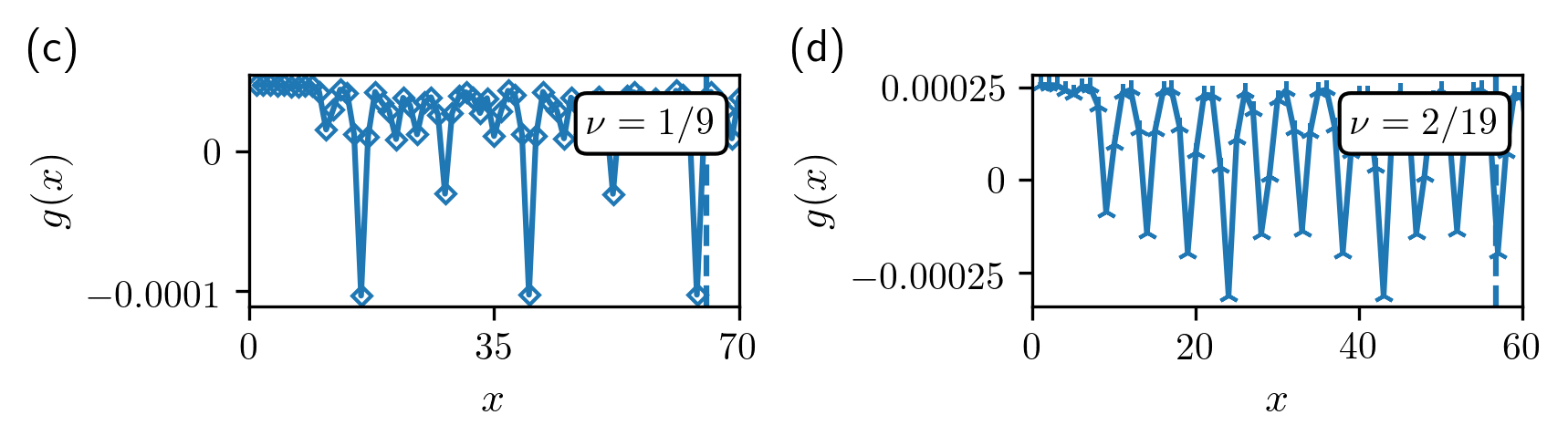}
	\caption{\label{fig:corrfunc_c45_analysis}\textbf{Connected two-point correlation functions for FCIs in $C=4,5$ bands.} Density-density correlation functions for the FCIs in Fig.~\ref{fig:phiflow_c45_analysis}, presented as in Fig.~\ref{fig:corrfunc_chi_Ly}. The parameters are the same as those used for the charge pumping. In this case, (a)~$L_y=14$, (b)~$L_y=15$, (c)~$L_y=18$, and (d)~$L_y=19$. The bond dimensions and $p$ values are presented as in Fig.~\ref{fig:corrfunc_c23_analysis}. The corresponding correlation lengths are marked with dashed lines.}
\end{figure}

Finally, we examine the correlation functions for the $C=4,5$ FCIs, in Fig.~\ref{fig:corrfunc_c45_analysis}. We again omit the $\braket{\normord{\rho_{0,0}\rho_{0,0}}}=0$ points and present the correlation functions for exactly the same set of parameters as the FCIs demonstrated in Sec.~\ref{subsec:FCIs}. Here, we notice a continuation of the trends established in Figs.~\ref{fig:corrfunc_analysis} and~\ref{fig:corrfunc_c23_analysis}. In particular, the particle density of the systems is further decreased, e.g., in Fig.~\ref{fig:corrfunc_c45_analysis}(a), which reduces the scale of the plots and exposes spurious fluctuations, and the long-distance oscillations have yet more frequency components, e.g., in Fig.~\ref{fig:corrfunc_c45_analysis}(b), for the reasons outlined above. Moreover, we can see that for the $\nu=2/15$ state at $n_\phi=4/15$, the high-frequency oscillations are diminished with increasing bond dimension for these small values of $\chi$, which indicates that the configurations are restricted by bond dimension as well as system size. Even after taking the distance average for these correlation functions, it is difficult to definitively confirm the exponential decay in the asymptotic limit due to large and persistent low-frequency oscillations. In this case, the correlation functions for the $C=4,5$ FCIs may accord with their identification as quantum Hall states, however it is not clear from these example configurations due to the fluctuations at large $x$, which may be physical or numerical. In all cases, the correlation functions shown in this section are a testament to the robustness of charge pumping as an indicator of the quantum Hall effect. In the majority of cases, the correlation functions corroborate the charge pumping data in Sec.~\ref{subsec:FCIs}. For FCIs in higher Chern number bands, however, the identification from correlation functions alone is often obstructed by persistent oscillations of a physical and/or numerical origin.

\section{Effect of system size on numerical stability}
\label{sec:stability}

In this section, we discuss two counter-intuitive examples where (i) an decrease in flux density, and (ii) an increase in bond dimension, may lead to numerical instabilities of an FCI.  

\begin{figure}
	\includegraphics[width=\linewidth]{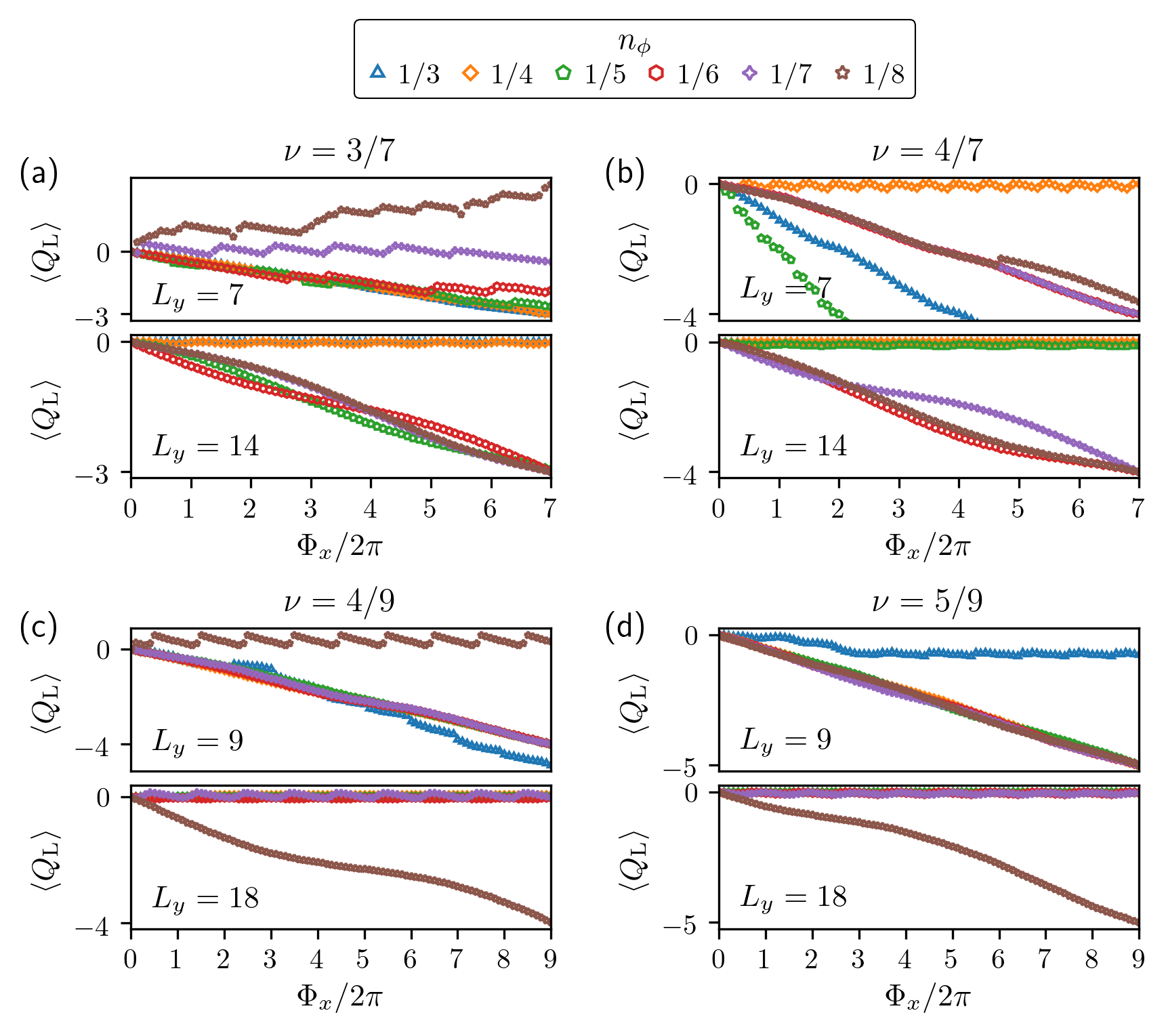}
	\caption{\label{fig:flatness_scaling}\textbf{Numerical instability with decreasing flux density.} Comparison of the charge pumping for the $C=1$ FCIs from Figs.~\ref{fig:phiflow_analysis}[(e)--(h)] at the first two allowed cylinder circumferences. The charge pumping is shown for (a)~$\nu=3/7$, $L_y=7,14$, (b)~$\nu=4/7$, $L_y=7,14$, (c)~$\nu=4/9$, $L_y=9,18$, and (d)~$\nu=5/9$, $L_y=9,18$.}
\end{figure}

In Fig.~\ref{fig:flatness_scaling}, we demonstrate a numerical instability caused by a decrease in flux density. For this example, we study the $|r|=3,4$ FCIs in Figs.~\ref{fig:phiflow_analysis}(e--h) at two different system sizes, $L_y=s$ and $L_y=2s$, which are the two smallest possible values of $L_y$. For the $\nu=3/7$ state with $L_y=7$ in Fig.~\ref{fig:flatness_scaling}(a), we observe a continuous, monotonic, and correct charge pumping result for $n_\phi=1/3$ and $1/4$, whereas when the flux density is decreased, such that $n_\phi<1/4$, the charge pumping curve diverges and exhibits discontinuities, non-monotonicity, and non-quantized $\braket{Q_\text{L}}$, which are all signs of a numerical instability. Conversely, when we examine identical configurations with $L_y=14$, we no longer observe charge pumping for $n_\phi=1/3$ and $1/4$, however a decrease in flux density improves the response, such that we obtain continuous, monotonic, and correct values for $n_\phi<1/4$. Moreover, the charge pumping for the larger system size does not show signs of discontinuities, non-monotonicity, or non-quantized values of $\braket{Q_\text{L}}$, for any value of $n_\phi$ shown. We note that this does not rule out the existence of FCIs at $\nu=3/7$ with $n_\phi=1/3,1/4$, rather it shows that with $L_y=14$, configurations with larger flux densities are more difficult to stabilize and configurations with smaller flux densities are no longer numerically constrained. Similar behavior is also observed at $\nu=4/7$, $4/9$, and $5/9$, shown in Figs.~\ref{fig:flatness_scaling}(b--d). 

To better understand this phenomenon, we examine the implications of decreasing the flux density. One of the most apparent consequences is an increase in band flatness, as illustrated in Fig.~\ref{fig:bandstructure_analysis}(b). Since an increased gap-to-width ratio is typically associated with more favorable FCI conditions, and $n_\phi\to0$ corresponds to the Landau level limit, a breakdown in this regime is unexpected on physical grounds. However, there are also important numerical side effects of tuning the flux density, owing to the finite system size. For example, a decrease in the flux density corresponds to an increase in the magnetic length, the relevant length scale in the problem ($n_\phi\sim l_B^{-1/2}$). Hence, for the infinite cylinder geometry, decreasing $n_\phi$ will exacerbate finite-size effects at fixed $L_y$ by making the system more one-dimensional. Moreover, by incrementing $p$, we are also increasing $q$: the width of the MPS unit cell. This means that a decreasing $n_\phi$ implies an increasing system size ($L_x/a\propto q$) and decreasing many-body gap ($\Delta_\text{m.b.}\propto q^{-2}$)~\cite{Bauer16}, which can have a detrimental impact on charge pumping at fixed $\chi$. Since, in this example, the corresponding charge pumping works for configurations with a larger $L_y$, which has an exponentially larger impact on the numerical cost and a comparable many-body gap, we attribute this numerical breakdown to an increased magnetic length. For practitioners of the algorithm, we recommend choosing a system size, such that any discontinuities, non-monotonicity, and non-quantized values of $\braket{Q_\text{L}}$ are mitigated, as in Fig.~\ref{fig:phiflow_analysis}. However, for more demanding systems, such as in Figs.~\ref{fig:phiflow_c23_analysis} and~\ref{fig:phiflow_c45_analysis}, we recognize that this is not always possible and so we advise caution.

\begin{figure}
	\includegraphics[width=\linewidth]{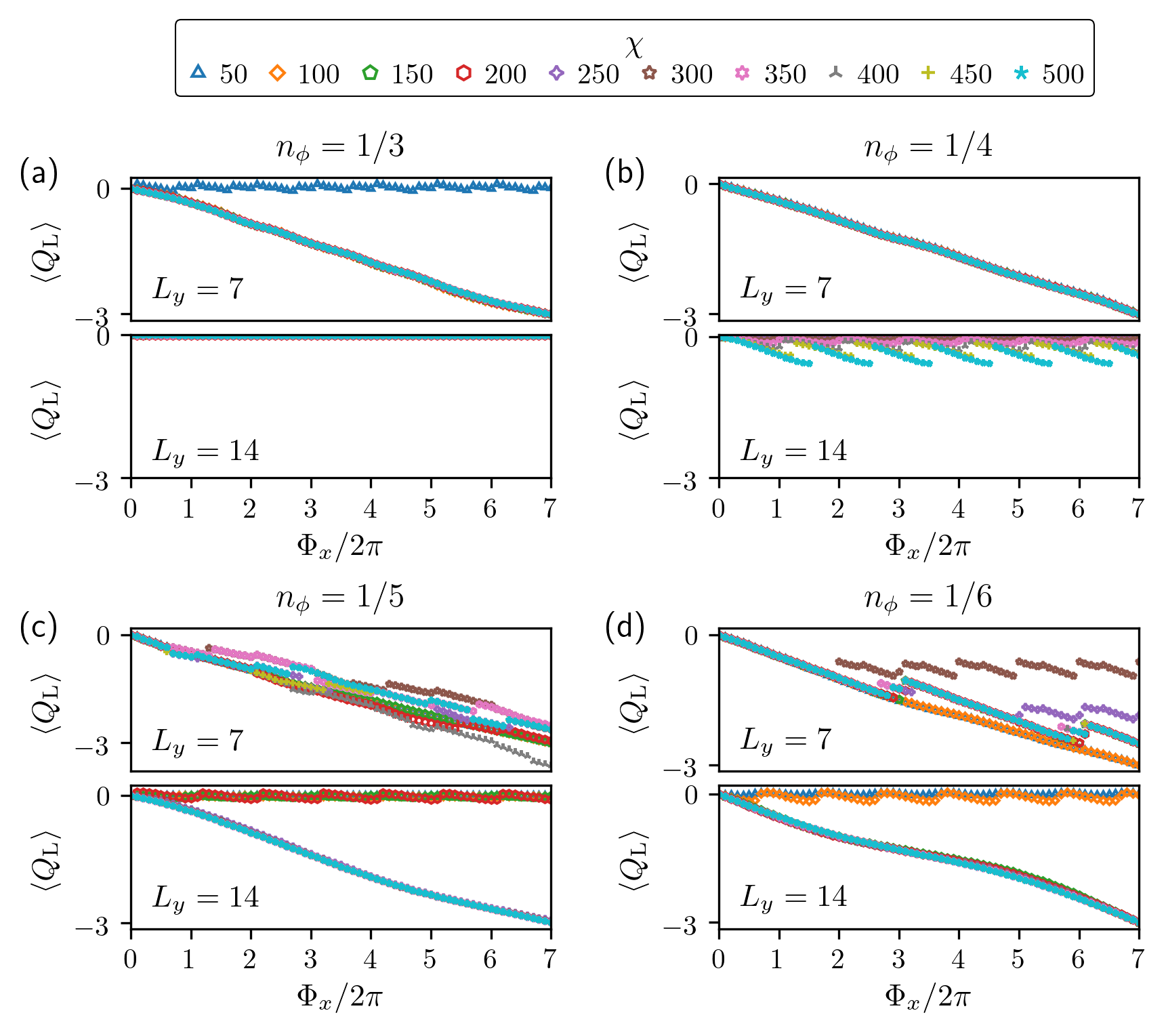}
	\caption{\label{fig:bond_scaling}\textbf{Numerical instability with increasing bond dimension.} Comparison of the charge pumping for the $\nu=3/7$ FCI from Fig.~\ref{fig:flatness_scaling}(a) as a function of bond dimension. The charge pumping is shown for the flux densities (a)~$n_\phi=1/3$, (b)~$n_\phi=1/4$, (c)~$n_\phi=1/5$, and (d)~$n_\phi=1/6$.}
\end{figure}

In Fig.~\ref{fig:bond_scaling}, we demonstrate a numerical instability caused by an increase in the bond dimension. For this example, we study the flux densities $n_\phi=1/3$, $1/4$, $1/5$, and $1/6$, for the $\nu=3/7$ state presented in Fig.~\ref{fig:flatness_scaling}(a). Here, we see many similar features to the numerical instabilities with increasing flux density. In Fig.~\ref{fig:bond_scaling}(a), we show the charge pumping for the $n_\phi=1/3$ configuration for varying values of $\chi\in[50,100,\dots, 500]$. We notice that for $L_y=7$, we observe the correct charge pumping for $\chi>50$, whereas for $L_y=14$, $\braket{Q_\text{L}}=0$ for all values of $\chi$. This reiterates the previous point, that smaller system sizes are more likely to stabilize configurations with a larger $n_\phi$. This is also reflected for $n_\phi=1/4$ in Fig.~\ref{fig:bond_scaling}(b). Note that the charge pumping in Fig.~\ref{fig:flatness_scaling} was performed at $\chi=250$. As we move to smaller values of $n_\phi$, however, we observe a more unusual phenomenon. For example, for the $n_\phi=1/5$ configuration with $L_y=7$ in Fig.~\ref{fig:bond_scaling}(c), we observe the correct charge pumping up to $\chi=150$, above which the curve becomes discontinuous at $\chi=200$, and non-quantized at $\chi\geq250$. Similar behavior is observed for the $n_\phi=1/6$ configuration with $L_y=7$ in Fig.~\ref{fig:bond_scaling}(d). In both cases, this pathology is remedied by increasing the system size to $L_y=14$. We verified that this behavior is not a result of poorly-converged DMRG by drastically: increasing the number of sweeps without bond optimization to update the environment (\texttt{update\_env} in \textsc{TeNPy}~\cite{tenpy}), increasing the number of sweeps to decrease the norm error of the wavefunction below the defined threshold (\texttt{norm\_tol\_iter}~\cite{tenpy}), decreasing the flux interval ($\Delta \Phi_x$), and decreasing the norm error threshold (\texttt{norm\_tol}~\cite{tenpy}). This shows that, when the system size is insufficient, there may be both lower and upper bounds on the bond dimensions that yield the correct charge pumping.

To gain insight into this effect, we examine the implications of increasing the bond dimension. By increasing $\chi$, we are effectively increasing the dimension of the matrices in our MPS representation of the ground state, and so this is expected to improve the precision and accuracy of the result. However, as before, there may be undesired side effects of increasing the bond dimension at finite system sizes. For example, over-parameterizing the MPS at an insufficient system size can increase the probability of converging to spurious competing ground states. Moreover, since the correlation length increases with bond dimension in the vicinity of a transition ($\xi\sim\ln\chi$), and elsewhere for unconverged numerics, this can exacerbate finite-size effects for strongly-correlated ground states, such as FCIs. In this example, the correlation length increase is marginal across the breakdown, and so we attribute this numerical instability to competing ground states due to an over-parameterization at inadequate system sizes. The consequence of these two numerical instabilities is that, when simulating a demanding FCI, there is a window of optimal $n_\phi$ and $\chi$ to observe valid charge pumping. For extremely demanding states, where the system size is especially inadequate, the optimal $\chi$ is likely to be low.

\bibliographystyle{apsrev4-1}
\bibliography{fci}

\begin{thebibliography}{108}%
\makeatletter
\providecommand \@ifxundefined [1]{%
 \@ifx{#1\undefined}
}%
\providecommand \@ifnum [1]{%
 \ifnum #1\expandafter \@firstoftwo
 \else \expandafter \@secondoftwo
 \fi
}%
\providecommand \@ifx [1]{%
 \ifx #1\expandafter \@firstoftwo
 \else \expandafter \@secondoftwo
 \fi
}%
\providecommand \natexlab [1]{#1}%
\providecommand \enquote  [1]{``#1''}%
\providecommand \bibnamefont  [1]{#1}%
\providecommand \bibfnamefont [1]{#1}%
\providecommand \citenamefont [1]{#1}%
\providecommand \href@noop [0]{\@secondoftwo}%
\providecommand \href [0]{\begingroup \@sanitize@url \@href}%
\providecommand \@href[1]{\@@startlink{#1}\@@href}%
\providecommand \@@href[1]{\endgroup#1\@@endlink}%
\providecommand \@sanitize@url [0]{\catcode `\\12\catcode `\$12\catcode
  `\&12\catcode `\#12\catcode `\^12\catcode `\_12\catcode `\%12\relax}%
\providecommand \@@startlink[1]{}%
\providecommand \@@endlink[0]{}%
\providecommand \url  [0]{\begingroup\@sanitize@url \@url }%
\providecommand \@url [1]{\endgroup\@href {#1}{\urlprefix }}%
\providecommand \urlprefix  [0]{URL }%
\providecommand \Eprint [0]{\href }%
\providecommand \doibase [0]{http://dx.doi.org/}%
\providecommand \selectlanguage [0]{\@gobble}%
\providecommand \bibinfo  [0]{\@secondoftwo}%
\providecommand \bibfield  [0]{\@secondoftwo}%
\providecommand \translation [1]{[#1]}%
\providecommand \BibitemOpen [0]{}%
\providecommand \bibitemStop [0]{}%
\providecommand \bibitemNoStop [0]{.\EOS\space}%
\providecommand \EOS [0]{\spacefactor3000\relax}%
\providecommand \BibitemShut  [1]{\csname bibitem#1\endcsname}%
\let\auto@bib@innerbib\@empty
\bibitem [{\citenamefont {Regnault}\ and\ \citenamefont
  {Bernevig}(2011)}]{Regnault11}%
  \BibitemOpen
  \bibfield  {author} {\bibinfo {author} {\bibfnamefont {N.}~\bibnamefont
  {Regnault}}\ and\ \bibinfo {author} {\bibfnamefont {B.~A.}\ \bibnamefont
  {Bernevig}},\ }\href {\doibase 10.1103/PhysRevX.1.021014} {\bibfield
  {journal} {\bibinfo  {journal} {Phys. Rev. X}\ }\textbf {\bibinfo {volume}
  {1}},\ \bibinfo {pages} {021014} (\bibinfo {year} {2011})}\BibitemShut
  {NoStop}%
\bibitem [{\citenamefont {Neupert}\ \emph {et~al.}(2011)\citenamefont
  {Neupert}, \citenamefont {Santos}, \citenamefont {Chamon},\ and\
  \citenamefont {Mudry}}]{Neupert11}%
  \BibitemOpen
  \bibfield  {author} {\bibinfo {author} {\bibfnamefont {T.}~\bibnamefont
  {Neupert}}, \bibinfo {author} {\bibfnamefont {L.}~\bibnamefont {Santos}},
  \bibinfo {author} {\bibfnamefont {C.}~\bibnamefont {Chamon}}, \ and\ \bibinfo
  {author} {\bibfnamefont {C.}~\bibnamefont {Mudry}},\ }\href {\doibase
  10.1103/PhysRevLett.106.236804} {\bibfield  {journal} {\bibinfo  {journal}
  {Phys. Rev. Lett.}\ }\textbf {\bibinfo {volume} {106}},\ \bibinfo {pages}
  {236804} (\bibinfo {year} {2011})}\BibitemShut {NoStop}%
\bibitem [{\citenamefont {Tang}\ \emph {et~al.}(2011)\citenamefont {Tang},
  \citenamefont {Mei},\ and\ \citenamefont {Wen}}]{Tang11}%
  \BibitemOpen
  \bibfield  {author} {\bibinfo {author} {\bibfnamefont {E.}~\bibnamefont
  {Tang}}, \bibinfo {author} {\bibfnamefont {J.-W.}\ \bibnamefont {Mei}}, \
  and\ \bibinfo {author} {\bibfnamefont {X.-G.}\ \bibnamefont {Wen}},\ }\href
  {\doibase 10.1103/PhysRevLett.106.236802} {\bibfield  {journal} {\bibinfo
  {journal} {Phys. Rev. Lett.}\ }\textbf {\bibinfo {volume} {106}},\ \bibinfo
  {pages} {236802} (\bibinfo {year} {2011})}\BibitemShut {NoStop}%
\bibitem [{\citenamefont {Bergholtz}\ and\ \citenamefont
  {Liu}(2013)}]{Bergholtz13}%
  \BibitemOpen
  \bibfield  {author} {\bibinfo {author} {\bibfnamefont {E.~J.}\ \bibnamefont
  {Bergholtz}}\ and\ \bibinfo {author} {\bibfnamefont {Z.}~\bibnamefont
  {Liu}},\ }\href {\doibase 10.1142/S021797921330017X} {\bibfield  {journal}
  {\bibinfo  {journal} {International Journal of Modern Physics B}\ }\textbf
  {\bibinfo {volume} {27}},\ \bibinfo {pages} {1330017} (\bibinfo {year}
  {2013})},\ \Eprint
  {http://arxiv.org/abs/https://doi.org/10.1142/S021797921330017X}
  {https://doi.org/10.1142/S021797921330017X} \BibitemShut {NoStop}%
\bibitem [{\citenamefont {Parameswaran}\ \emph {et~al.}(2013)\citenamefont
  {Parameswaran}, \citenamefont {Roy},\ and\ \citenamefont
  {Sondhi}}]{Parameswaran13}%
  \BibitemOpen
  \bibfield  {author} {\bibinfo {author} {\bibfnamefont {S.~A.}\ \bibnamefont
  {Parameswaran}}, \bibinfo {author} {\bibfnamefont {R.}~\bibnamefont {Roy}}, \
  and\ \bibinfo {author} {\bibfnamefont {S.~L.}\ \bibnamefont {Sondhi}},\
  }\href {\doibase https://doi.org/10.1016/j.crhy.2013.04.003} {\bibfield
  {journal} {\bibinfo  {journal} {Comptes Rendus Physique}\ }\textbf {\bibinfo
  {volume} {14}},\ \bibinfo {pages} {816} (\bibinfo {year} {2013})},\ \bibinfo
  {note} {topological insulators / Isolants topologiques}\BibitemShut {NoStop}%
\bibitem [{\citenamefont {S{\o}rensen}\ \emph {et~al.}(2005)\citenamefont
  {S{\o}rensen}, \citenamefont {Demler},\ and\ \citenamefont
  {Lukin}}]{Sorensen05}%
  \BibitemOpen
  \bibfield  {author} {\bibinfo {author} {\bibfnamefont {A.~S.}\ \bibnamefont
  {S{\o}rensen}}, \bibinfo {author} {\bibfnamefont {E.}~\bibnamefont {Demler}},
  \ and\ \bibinfo {author} {\bibfnamefont {M.~D.}\ \bibnamefont {Lukin}},\
  }\href@noop {} {\bibfield  {journal} {\bibinfo  {journal} {Phys. Rev. Lett.}\
  }\textbf {\bibinfo {volume} {94}},\ \bibinfo {pages} {086803} (\bibinfo
  {year} {2005})}\BibitemShut {NoStop}%
\bibitem [{\citenamefont {Hafezi}\ \emph {et~al.}(2007)\citenamefont {Hafezi},
  \citenamefont {S\o{}rensen}, \citenamefont {Demler},\ and\ \citenamefont
  {Lukin}}]{Hafezi07}%
  \BibitemOpen
  \bibfield  {author} {\bibinfo {author} {\bibfnamefont {M.}~\bibnamefont
  {Hafezi}}, \bibinfo {author} {\bibfnamefont {A.~S.}\ \bibnamefont
  {S\o{}rensen}}, \bibinfo {author} {\bibfnamefont {E.}~\bibnamefont {Demler}},
  \ and\ \bibinfo {author} {\bibfnamefont {M.~D.}\ \bibnamefont {Lukin}},\
  }\href {\doibase 10.1103/PhysRevA.76.023613} {\bibfield  {journal} {\bibinfo
  {journal} {Phys. Rev. A}\ }\textbf {\bibinfo {volume} {76}},\ \bibinfo
  {pages} {023613} (\bibinfo {year} {2007})}\BibitemShut {NoStop}%
\bibitem [{\citenamefont {M\"oller}\ and\ \citenamefont
  {Cooper}(2009)}]{Moller09}%
  \BibitemOpen
  \bibfield  {author} {\bibinfo {author} {\bibfnamefont {G.}~\bibnamefont
  {M\"oller}}\ and\ \bibinfo {author} {\bibfnamefont {N.~R.}\ \bibnamefont
  {Cooper}},\ }\href {\doibase 10.1103/PhysRevLett.103.105303} {\bibfield
  {journal} {\bibinfo  {journal} {Phys. Rev. Lett.}\ }\textbf {\bibinfo
  {volume} {103}},\ \bibinfo {pages} {105303} (\bibinfo {year}
  {2009})}\BibitemShut {NoStop}%
\bibitem [{\citenamefont {Scaffidi}\ and\ \citenamefont
  {M\"oller}(2012)}]{Scaffidi12}%
  \BibitemOpen
  \bibfield  {author} {\bibinfo {author} {\bibfnamefont {T.}~\bibnamefont
  {Scaffidi}}\ and\ \bibinfo {author} {\bibfnamefont {G.}~\bibnamefont
  {M\"oller}},\ }\href {\doibase 10.1103/PhysRevLett.109.246805} {\bibfield
  {journal} {\bibinfo  {journal} {Phys. Rev. Lett.}\ }\textbf {\bibinfo
  {volume} {109}},\ \bibinfo {pages} {246805} (\bibinfo {year}
  {2012})}\BibitemShut {NoStop}%
\bibitem [{\citenamefont {Zhang}\ and\ \citenamefont {Shi}(2016)}]{Zhang16}%
  \BibitemOpen
  \bibfield  {author} {\bibinfo {author} {\bibfnamefont {Y.}~\bibnamefont
  {Zhang}}\ and\ \bibinfo {author} {\bibfnamefont {J.}~\bibnamefont {Shi}},\
  }\href {\doibase 10.1103/PhysRevB.93.165129} {\bibfield  {journal} {\bibinfo
  {journal} {Phys. Rev. B}\ }\textbf {\bibinfo {volume} {93}},\ \bibinfo
  {pages} {165129} (\bibinfo {year} {2016})}\BibitemShut {NoStop}%
\bibitem [{\citenamefont {Parameswaran}\ \emph {et~al.}(2012)\citenamefont
  {Parameswaran}, \citenamefont {Roy},\ and\ \citenamefont
  {Sondhi}}]{Parameswaran12}%
  \BibitemOpen
  \bibfield  {author} {\bibinfo {author} {\bibfnamefont {S.~A.}\ \bibnamefont
  {Parameswaran}}, \bibinfo {author} {\bibfnamefont {R.}~\bibnamefont {Roy}}, \
  and\ \bibinfo {author} {\bibfnamefont {S.~L.}\ \bibnamefont {Sondhi}},\
  }\href {\doibase 10.1103/PhysRevB.85.241308} {\bibfield  {journal} {\bibinfo
  {journal} {Phys. Rev. B}\ }\textbf {\bibinfo {volume} {85}},\ \bibinfo
  {pages} {241308(R)} (\bibinfo {year} {2012})}\BibitemShut {NoStop}%
\bibitem [{\citenamefont {Roy}(2014)}]{Roy14}%
  \BibitemOpen
  \bibfield  {author} {\bibinfo {author} {\bibfnamefont {R.}~\bibnamefont
  {Roy}},\ }\href {\doibase 10.1103/PhysRevB.90.165139} {\bibfield  {journal}
  {\bibinfo  {journal} {Phys. Rev. B}\ }\textbf {\bibinfo {volume} {90}},\
  \bibinfo {pages} {165139} (\bibinfo {year} {2014})}\BibitemShut {NoStop}%
\bibitem [{\citenamefont {Jackson}\ \emph {et~al.}(2015)\citenamefont
  {Jackson}, \citenamefont {M{\"o}ller},\ and\ \citenamefont
  {Roy}}]{Jackson15}%
  \BibitemOpen
  \bibfield  {author} {\bibinfo {author} {\bibfnamefont {T.~S.}\ \bibnamefont
  {Jackson}}, \bibinfo {author} {\bibfnamefont {G.}~\bibnamefont {M{\"o}ller}},
  \ and\ \bibinfo {author} {\bibfnamefont {R.}~\bibnamefont {Roy}},\ }\href
  {\doibase 10.1038/ncomms9629} {\bibfield  {journal} {\bibinfo  {journal}
  {Nature Communications}\ }\textbf {\bibinfo {volume} {6}},\ \bibinfo {pages}
  {8629} (\bibinfo {year} {2015})}\BibitemShut {NoStop}%
\bibitem [{\citenamefont {Lee}\ \emph {et~al.}(2017)\citenamefont {Lee},
  \citenamefont {Claassen},\ and\ \citenamefont {Thomale}}]{Lee17}%
  \BibitemOpen
  \bibfield  {author} {\bibinfo {author} {\bibfnamefont {C.~H.}\ \bibnamefont
  {Lee}}, \bibinfo {author} {\bibfnamefont {M.}~\bibnamefont {Claassen}}, \
  and\ \bibinfo {author} {\bibfnamefont {R.}~\bibnamefont {Thomale}},\ }\href
  {\doibase 10.1103/PhysRevB.96.165150} {\bibfield  {journal} {\bibinfo
  {journal} {Phys. Rev. B}\ }\textbf {\bibinfo {volume} {96}},\ \bibinfo
  {pages} {165150} (\bibinfo {year} {2017})}\BibitemShut {NoStop}%
\bibitem [{\citenamefont {Kol}\ and\ \citenamefont {Read}(1993)}]{Kol93}%
  \BibitemOpen
  \bibfield  {author} {\bibinfo {author} {\bibfnamefont {A.}~\bibnamefont
  {Kol}}\ and\ \bibinfo {author} {\bibfnamefont {N.}~\bibnamefont {Read}},\
  }\href@noop {} {\bibfield  {journal} {\bibinfo  {journal} {Phys. Rev. B}\
  }\textbf {\bibinfo {volume} {48}},\ \bibinfo {pages} {8890} (\bibinfo {year}
  {1993})}\BibitemShut {NoStop}%
\bibitem [{\citenamefont {Liu}\ \emph {et~al.}(2012)\citenamefont {Liu},
  \citenamefont {Bergholtz}, \citenamefont {Fan},\ and\ \citenamefont
  {L\"auchli}}]{Liu12}%
  \BibitemOpen
  \bibfield  {author} {\bibinfo {author} {\bibfnamefont {Z.}~\bibnamefont
  {Liu}}, \bibinfo {author} {\bibfnamefont {E.~J.}\ \bibnamefont {Bergholtz}},
  \bibinfo {author} {\bibfnamefont {H.}~\bibnamefont {Fan}}, \ and\ \bibinfo
  {author} {\bibfnamefont {A.~M.}\ \bibnamefont {L\"auchli}},\ }\href {\doibase
  10.1103/PhysRevLett.109.186805} {\bibfield  {journal} {\bibinfo  {journal}
  {Phys. Rev. Lett.}\ }\textbf {\bibinfo {volume} {109}},\ \bibinfo {pages}
  {186805} (\bibinfo {year} {2012})}\BibitemShut {NoStop}%
\bibitem [{\citenamefont {Udagawa}\ and\ \citenamefont
  {Bergholtz}(2014)}]{Udagawa14}%
  \BibitemOpen
  \bibfield  {author} {\bibinfo {author} {\bibfnamefont {M.}~\bibnamefont
  {Udagawa}}\ and\ \bibinfo {author} {\bibfnamefont {E.~J.}\ \bibnamefont
  {Bergholtz}},\ }\href {\doibase 10.1088/1742-5468/2014/10/p10012} {\bibfield
  {journal} {\bibinfo  {journal} {Journal of Statistical Mechanics: Theory and
  Experiment}\ }\textbf {\bibinfo {volume} {2014}},\ \bibinfo {pages} {P10012}
  (\bibinfo {year} {2014})}\BibitemShut {NoStop}%
\bibitem [{\citenamefont {Wu}\ \emph {et~al.}(2015)\citenamefont {Wu},
  \citenamefont {Jain},\ and\ \citenamefont {Sun}}]{Wu15}%
  \BibitemOpen
  \bibfield  {author} {\bibinfo {author} {\bibfnamefont {Y.-H.}\ \bibnamefont
  {Wu}}, \bibinfo {author} {\bibfnamefont {J.~K.}\ \bibnamefont {Jain}}, \ and\
  \bibinfo {author} {\bibfnamefont {K.}~\bibnamefont {Sun}},\ }\href {\doibase
  10.1103/PhysRevB.91.041119} {\bibfield  {journal} {\bibinfo  {journal} {Phys.
  Rev. B}\ }\textbf {\bibinfo {volume} {91}},\ \bibinfo {pages} {041119(R)}
  (\bibinfo {year} {2015})}\BibitemShut {NoStop}%
\bibitem [{\citenamefont {M\"oller}\ and\ \citenamefont
  {Cooper}(2015)}]{Moller15}%
  \BibitemOpen
  \bibfield  {author} {\bibinfo {author} {\bibfnamefont {G.}~\bibnamefont
  {M\"oller}}\ and\ \bibinfo {author} {\bibfnamefont {N.~R.}\ \bibnamefont
  {Cooper}},\ }\href {\doibase 10.1103/PhysRevLett.115.126401} {\bibfield
  {journal} {\bibinfo  {journal} {Phys. Rev. Lett.}\ }\textbf {\bibinfo
  {volume} {115}},\ \bibinfo {pages} {126401} (\bibinfo {year}
  {2015})}\BibitemShut {NoStop}%
\bibitem [{\citenamefont {Andrews}\ and\ \citenamefont
  {M\"oller}(2018)}]{Andrews18}%
  \BibitemOpen
  \bibfield  {author} {\bibinfo {author} {\bibfnamefont {B.}~\bibnamefont
  {Andrews}}\ and\ \bibinfo {author} {\bibfnamefont {G.}~\bibnamefont
  {M\"oller}},\ }\href {\doibase 10.1103/PhysRevB.97.035159} {\bibfield
  {journal} {\bibinfo  {journal} {Phys. Rev. B}\ }\textbf {\bibinfo {volume}
  {97}},\ \bibinfo {pages} {035159} (\bibinfo {year} {2018})}\BibitemShut
  {NoStop}%
\bibitem [{\citenamefont {Nayak}\ \emph {et~al.}(2008)\citenamefont {Nayak},
  \citenamefont {Simon}, \citenamefont {Stern}, \citenamefont {Freedman},\ and\
  \citenamefont {Das~Sarma}}]{Nayak08}%
  \BibitemOpen
  \bibfield  {author} {\bibinfo {author} {\bibfnamefont {C.}~\bibnamefont
  {Nayak}}, \bibinfo {author} {\bibfnamefont {S.~H.}\ \bibnamefont {Simon}},
  \bibinfo {author} {\bibfnamefont {A.}~\bibnamefont {Stern}}, \bibinfo
  {author} {\bibfnamefont {M.}~\bibnamefont {Freedman}}, \ and\ \bibinfo
  {author} {\bibfnamefont {S.}~\bibnamefont {Das~Sarma}},\ }\href {\doibase
  10.1103/RevModPhys.80.1083} {\bibfield  {journal} {\bibinfo  {journal} {Rev.
  Mod. Phys.}\ }\textbf {\bibinfo {volume} {80}},\ \bibinfo {pages} {1083}
  (\bibinfo {year} {2008})}\BibitemShut {NoStop}%
\bibitem [{\citenamefont {Barkeshli}\ and\ \citenamefont
  {Qi}(2012)}]{Barkeshli12}%
  \BibitemOpen
  \bibfield  {author} {\bibinfo {author} {\bibfnamefont {M.}~\bibnamefont
  {Barkeshli}}\ and\ \bibinfo {author} {\bibfnamefont {X.-L.}\ \bibnamefont
  {Qi}},\ }\href {\doibase 10.1103/PhysRevX.2.031013} {\bibfield  {journal}
  {\bibinfo  {journal} {Phys. Rev. X}\ }\textbf {\bibinfo {volume} {2}},\
  \bibinfo {pages} {031013} (\bibinfo {year} {2012})}\BibitemShut {NoStop}%
\bibitem [{\citenamefont {Barkeshli}\ \emph {et~al.}(2013)\citenamefont
  {Barkeshli}, \citenamefont {Jian},\ and\ \citenamefont {Qi}}]{Barkeshli13}%
  \BibitemOpen
  \bibfield  {author} {\bibinfo {author} {\bibfnamefont {M.}~\bibnamefont
  {Barkeshli}}, \bibinfo {author} {\bibfnamefont {C.-M.}\ \bibnamefont {Jian}},
  \ and\ \bibinfo {author} {\bibfnamefont {X.-L.}\ \bibnamefont {Qi}},\ }\href
  {\doibase 10.1103/PhysRevB.87.045130} {\bibfield  {journal} {\bibinfo
  {journal} {Phys. Rev. B}\ }\textbf {\bibinfo {volume} {87}},\ \bibinfo
  {pages} {045130} (\bibinfo {year} {2013})}\BibitemShut {NoStop}%
\bibitem [{\citenamefont {Jaworowski}\ \emph {et~al.}(2019)\citenamefont
  {Jaworowski}, \citenamefont {Regnault},\ and\ \citenamefont
  {Liu}}]{Jaworowski19}%
  \BibitemOpen
  \bibfield  {author} {\bibinfo {author} {\bibfnamefont {B.}~\bibnamefont
  {Jaworowski}}, \bibinfo {author} {\bibfnamefont {N.}~\bibnamefont
  {Regnault}}, \ and\ \bibinfo {author} {\bibfnamefont {Z.}~\bibnamefont
  {Liu}},\ }\href {\doibase 10.1103/PhysRevB.99.045136} {\bibfield  {journal}
  {\bibinfo  {journal} {Phys. Rev. B}\ }\textbf {\bibinfo {volume} {99}},\
  \bibinfo {pages} {045136} (\bibinfo {year} {2019})}\BibitemShut {NoStop}%
\bibitem [{\citenamefont {Sterdyniak}\ \emph {et~al.}(2013)\citenamefont
  {Sterdyniak}, \citenamefont {Repellin}, \citenamefont {Bernevig},\ and\
  \citenamefont {Regnault}}]{Sterdyniak13}%
  \BibitemOpen
  \bibfield  {author} {\bibinfo {author} {\bibfnamefont {A.}~\bibnamefont
  {Sterdyniak}}, \bibinfo {author} {\bibfnamefont {C.}~\bibnamefont
  {Repellin}}, \bibinfo {author} {\bibfnamefont {B.~A.}\ \bibnamefont
  {Bernevig}}, \ and\ \bibinfo {author} {\bibfnamefont {N.}~\bibnamefont
  {Regnault}},\ }\href {\doibase 10.1103/PhysRevB.87.205137} {\bibfield
  {journal} {\bibinfo  {journal} {Phys. Rev. B}\ }\textbf {\bibinfo {volume}
  {87}},\ \bibinfo {pages} {205137} (\bibinfo {year} {2013})}\BibitemShut
  {NoStop}%
\bibitem [{\citenamefont {Bergholtz}\ \emph {et~al.}(2015)\citenamefont
  {Bergholtz}, \citenamefont {Liu}, \citenamefont {Trescher}, \citenamefont
  {Moessner},\ and\ \citenamefont {Udagawa}}]{Bergholtz15}%
  \BibitemOpen
  \bibfield  {author} {\bibinfo {author} {\bibfnamefont {E.~J.}\ \bibnamefont
  {Bergholtz}}, \bibinfo {author} {\bibfnamefont {Z.}~\bibnamefont {Liu}},
  \bibinfo {author} {\bibfnamefont {M.}~\bibnamefont {Trescher}}, \bibinfo
  {author} {\bibfnamefont {R.}~\bibnamefont {Moessner}}, \ and\ \bibinfo
  {author} {\bibfnamefont {M.}~\bibnamefont {Udagawa}},\ }\href {\doibase
  10.1103/PhysRevLett.114.016806} {\bibfield  {journal} {\bibinfo  {journal}
  {Phys. Rev. Lett.}\ }\textbf {\bibinfo {volume} {114}},\ \bibinfo {pages}
  {016806} (\bibinfo {year} {2015})}\BibitemShut {NoStop}%
\bibitem [{\citenamefont {Liu}\ \emph {et~al.}(2017)\citenamefont {Liu},
  \citenamefont {M\"oller},\ and\ \citenamefont {Bergholtz}}]{Liu17}%
  \BibitemOpen
  \bibfield  {author} {\bibinfo {author} {\bibfnamefont {Z.}~\bibnamefont
  {Liu}}, \bibinfo {author} {\bibfnamefont {G.}~\bibnamefont {M\"oller}}, \
  and\ \bibinfo {author} {\bibfnamefont {E.~J.}\ \bibnamefont {Bergholtz}},\
  }\href {\doibase 10.1103/PhysRevLett.119.106801} {\bibfield  {journal}
  {\bibinfo  {journal} {Phys. Rev. Lett.}\ }\textbf {\bibinfo {volume} {119}},\
  \bibinfo {pages} {106801} (\bibinfo {year} {2017})}\BibitemShut {NoStop}%
\bibitem [{\citenamefont {He}\ \emph {et~al.}(2015)\citenamefont {He},
  \citenamefont {Bhattacharjee}, \citenamefont {Moessner},\ and\ \citenamefont
  {Pollmann}}]{He15}%
  \BibitemOpen
  \bibfield  {author} {\bibinfo {author} {\bibfnamefont {Y.-C.}\ \bibnamefont
  {He}}, \bibinfo {author} {\bibfnamefont {S.}~\bibnamefont {Bhattacharjee}},
  \bibinfo {author} {\bibfnamefont {R.}~\bibnamefont {Moessner}}, \ and\
  \bibinfo {author} {\bibfnamefont {F.}~\bibnamefont {Pollmann}},\ }\href
  {\doibase 10.1103/PhysRevLett.115.116803} {\bibfield  {journal} {\bibinfo
  {journal} {Phys. Rev. Lett.}\ }\textbf {\bibinfo {volume} {115}},\ \bibinfo
  {pages} {116803} (\bibinfo {year} {2015})}\BibitemShut {NoStop}%
\bibitem [{\citenamefont {Zeng}\ \emph {et~al.}(2016)\citenamefont {Zeng},
  \citenamefont {Zhu},\ and\ \citenamefont {Sheng}}]{Zeng16}%
  \BibitemOpen
  \bibfield  {author} {\bibinfo {author} {\bibfnamefont {T.-S.}\ \bibnamefont
  {Zeng}}, \bibinfo {author} {\bibfnamefont {W.}~\bibnamefont {Zhu}}, \ and\
  \bibinfo {author} {\bibfnamefont {D.~N.}\ \bibnamefont {Sheng}},\ }\href
  {\doibase 10.1103/PhysRevB.93.195121} {\bibfield  {journal} {\bibinfo
  {journal} {Phys. Rev. B}\ }\textbf {\bibinfo {volume} {93}},\ \bibinfo
  {pages} {195121} (\bibinfo {year} {2016})}\BibitemShut {NoStop}%
\bibitem [{\citenamefont {He}\ \emph {et~al.}(2017)\citenamefont {He},
  \citenamefont {Grusdt}, \citenamefont {Kaufman}, \citenamefont {Greiner},\
  and\ \citenamefont {Vishwanath}}]{He17}%
  \BibitemOpen
  \bibfield  {author} {\bibinfo {author} {\bibfnamefont {Y.-C.}\ \bibnamefont
  {He}}, \bibinfo {author} {\bibfnamefont {F.}~\bibnamefont {Grusdt}}, \bibinfo
  {author} {\bibfnamefont {A.}~\bibnamefont {Kaufman}}, \bibinfo {author}
  {\bibfnamefont {M.}~\bibnamefont {Greiner}}, \ and\ \bibinfo {author}
  {\bibfnamefont {A.}~\bibnamefont {Vishwanath}},\ }\href {\doibase
  10.1103/PhysRevB.96.201103} {\bibfield  {journal} {\bibinfo  {journal} {Phys.
  Rev. B}\ }\textbf {\bibinfo {volume} {96}},\ \bibinfo {pages} {201103(R)}
  (\bibinfo {year} {2017})}\BibitemShut {NoStop}%
\bibitem [{\citenamefont {Spanton}\ \emph {et~al.}(2018)\citenamefont
  {Spanton}, \citenamefont {Zibrov}, \citenamefont {Zhou}, \citenamefont
  {Taniguchi}, \citenamefont {Watanabe}, \citenamefont {Zaletel},\ and\
  \citenamefont {Young}}]{Spanton18}%
  \BibitemOpen
  \bibfield  {author} {\bibinfo {author} {\bibfnamefont {E.~M.}\ \bibnamefont
  {Spanton}}, \bibinfo {author} {\bibfnamefont {A.~A.}\ \bibnamefont {Zibrov}},
  \bibinfo {author} {\bibfnamefont {H.}~\bibnamefont {Zhou}}, \bibinfo {author}
  {\bibfnamefont {T.}~\bibnamefont {Taniguchi}}, \bibinfo {author}
  {\bibfnamefont {K.}~\bibnamefont {Watanabe}}, \bibinfo {author}
  {\bibfnamefont {M.~P.}\ \bibnamefont {Zaletel}}, \ and\ \bibinfo {author}
  {\bibfnamefont {A.~F.}\ \bibnamefont {Young}},\ }\href {\doibase
  10.1126/science.aan8458} {\bibfield  {journal} {\bibinfo  {journal}
  {Science}\ }\textbf {\bibinfo {volume} {360}},\ \bibinfo {pages} {62}
  (\bibinfo {year} {2018})},\ \Eprint
  {http://arxiv.org/abs/https://science.sciencemag.org/content/360/6384/62.full.pdf}
  {https://science.sciencemag.org/content/360/6384/62.full.pdf} \BibitemShut
  {NoStop}%
\bibitem [{\citenamefont {Chen}\ \emph
  {et~al.}(2020{\natexlab{a}})\citenamefont {Chen}, \citenamefont {Sharpe},
  \citenamefont {Fox}, \citenamefont {Zhang}, \citenamefont {Wang},
  \citenamefont {Jiang}, \citenamefont {Lyu}, \citenamefont {Li}, \citenamefont
  {Watanabe}, \citenamefont {Taniguchi}, \citenamefont {Shi}, \citenamefont
  {Senthil}, \citenamefont {Goldhaber-Gordon}, \citenamefont {Zhang},\ and\
  \citenamefont {Wang}}]{Chen20}%
  \BibitemOpen
  \bibfield  {author} {\bibinfo {author} {\bibfnamefont {G.}~\bibnamefont
  {Chen}}, \bibinfo {author} {\bibfnamefont {A.~L.}\ \bibnamefont {Sharpe}},
  \bibinfo {author} {\bibfnamefont {E.~J.}\ \bibnamefont {Fox}}, \bibinfo
  {author} {\bibfnamefont {Y.-H.}\ \bibnamefont {Zhang}}, \bibinfo {author}
  {\bibfnamefont {S.}~\bibnamefont {Wang}}, \bibinfo {author} {\bibfnamefont
  {L.}~\bibnamefont {Jiang}}, \bibinfo {author} {\bibfnamefont
  {B.}~\bibnamefont {Lyu}}, \bibinfo {author} {\bibfnamefont {H.}~\bibnamefont
  {Li}}, \bibinfo {author} {\bibfnamefont {K.}~\bibnamefont {Watanabe}},
  \bibinfo {author} {\bibfnamefont {T.}~\bibnamefont {Taniguchi}}, \bibinfo
  {author} {\bibfnamefont {Z.}~\bibnamefont {Shi}}, \bibinfo {author}
  {\bibfnamefont {T.}~\bibnamefont {Senthil}}, \bibinfo {author} {\bibfnamefont
  {D.}~\bibnamefont {Goldhaber-Gordon}}, \bibinfo {author} {\bibfnamefont
  {Y.}~\bibnamefont {Zhang}}, \ and\ \bibinfo {author} {\bibfnamefont
  {F.}~\bibnamefont {Wang}},\ }\href {\doibase 10.1038/s41586-020-2049-7}
  {\bibfield  {journal} {\bibinfo  {journal} {Nature}\ }\textbf {\bibinfo
  {volume} {579}},\ \bibinfo {pages} {56} (\bibinfo {year}
  {2020}{\natexlab{a}})}\BibitemShut {NoStop}%
\bibitem [{\citenamefont {Xie}\ \emph {et~al.}(2021)\citenamefont {Xie},
  \citenamefont {Pierce}, \citenamefont {Park}, \citenamefont {Parker},
  \citenamefont {Khalaf}, \citenamefont {Ledwith}, \citenamefont {Cao},
  \citenamefont {Lee}, \citenamefont {Chen}, \citenamefont {Forrester},
  \citenamefont {Watanabe}, \citenamefont {Taniguchi}, \citenamefont
  {Vishwanath}, \citenamefont {Jarillo-Herrero},\ and\ \citenamefont
  {Yacoby}}]{Xie21}%
  \BibitemOpen
  \bibfield  {author} {\bibinfo {author} {\bibfnamefont {Y.}~\bibnamefont
  {Xie}}, \bibinfo {author} {\bibfnamefont {A.~T.}\ \bibnamefont {Pierce}},
  \bibinfo {author} {\bibfnamefont {J.~M.}\ \bibnamefont {Park}}, \bibinfo
  {author} {\bibfnamefont {D.~E.}\ \bibnamefont {Parker}}, \bibinfo {author}
  {\bibfnamefont {E.}~\bibnamefont {Khalaf}}, \bibinfo {author} {\bibfnamefont
  {P.}~\bibnamefont {Ledwith}}, \bibinfo {author} {\bibfnamefont
  {Y.}~\bibnamefont {Cao}}, \bibinfo {author} {\bibfnamefont {S.~H.}\
  \bibnamefont {Lee}}, \bibinfo {author} {\bibfnamefont {S.}~\bibnamefont
  {Chen}}, \bibinfo {author} {\bibfnamefont {P.~R.}\ \bibnamefont {Forrester}},
  \bibinfo {author} {\bibfnamefont {K.}~\bibnamefont {Watanabe}}, \bibinfo
  {author} {\bibfnamefont {T.}~\bibnamefont {Taniguchi}}, \bibinfo {author}
  {\bibfnamefont {A.}~\bibnamefont {Vishwanath}}, \bibinfo {author}
  {\bibfnamefont {P.}~\bibnamefont {Jarillo-Herrero}}, \ and\ \bibinfo {author}
  {\bibfnamefont {A.}~\bibnamefont {Yacoby}},\ }\href@noop {} {\enquote
  {\bibinfo {title} {Fractional chern insulators in magic-angle twisted bilayer
  graphene},}\ } (\bibinfo {year} {2021}),\ \Eprint
  {http://arxiv.org/abs/2107.10854} {arXiv:2107.10854 [cond-mat.mes-hall]}
  \BibitemShut {NoStop}%
\bibitem [{\citenamefont {Cooper}\ and\ \citenamefont
  {Dalibard}(2013)}]{Cooper13}%
  \BibitemOpen
  \bibfield  {author} {\bibinfo {author} {\bibfnamefont {N.~R.}\ \bibnamefont
  {Cooper}}\ and\ \bibinfo {author} {\bibfnamefont {J.}~\bibnamefont
  {Dalibard}},\ }\href {\doibase 10.1103/PhysRevLett.110.185301} {\bibfield
  {journal} {\bibinfo  {journal} {Phys. Rev. Lett.}\ }\textbf {\bibinfo
  {volume} {110}},\ \bibinfo {pages} {185301} (\bibinfo {year}
  {2013})}\BibitemShut {NoStop}%
\bibitem [{\citenamefont {Aidelsburger}\ \emph {et~al.}(2015)\citenamefont
  {Aidelsburger}, \citenamefont {Lohse}, \citenamefont {Schweizer},
  \citenamefont {Atala}, \citenamefont {Barreiro}, \citenamefont
  {Nascimb{\`e}ne}, \citenamefont {Cooper}, \citenamefont {Bloch},\ and\
  \citenamefont {Goldman}}]{Aidelsburger15}%
  \BibitemOpen
  \bibfield  {author} {\bibinfo {author} {\bibfnamefont {M.}~\bibnamefont
  {Aidelsburger}}, \bibinfo {author} {\bibfnamefont {M.}~\bibnamefont {Lohse}},
  \bibinfo {author} {\bibfnamefont {C.}~\bibnamefont {Schweizer}}, \bibinfo
  {author} {\bibfnamefont {M.}~\bibnamefont {Atala}}, \bibinfo {author}
  {\bibfnamefont {J.~T.}\ \bibnamefont {Barreiro}}, \bibinfo {author}
  {\bibfnamefont {S.}~\bibnamefont {Nascimb{\`e}ne}}, \bibinfo {author}
  {\bibfnamefont {N.~R.}\ \bibnamefont {Cooper}}, \bibinfo {author}
  {\bibfnamefont {I.}~\bibnamefont {Bloch}}, \ and\ \bibinfo {author}
  {\bibfnamefont {N.}~\bibnamefont {Goldman}},\ }\href {\doibase
  10.1038/nphys3171} {\bibfield  {journal} {\bibinfo  {journal} {Nature
  Physics}\ }\textbf {\bibinfo {volume} {11}},\ \bibinfo {pages} {162}
  (\bibinfo {year} {2015})}\BibitemShut {NoStop}%
\bibitem [{\citenamefont {Cooper}\ \emph {et~al.}(2019)\citenamefont {Cooper},
  \citenamefont {Dalibard},\ and\ \citenamefont {Spielman}}]{Cooper19}%
  \BibitemOpen
  \bibfield  {author} {\bibinfo {author} {\bibfnamefont {N.~R.}\ \bibnamefont
  {Cooper}}, \bibinfo {author} {\bibfnamefont {J.}~\bibnamefont {Dalibard}}, \
  and\ \bibinfo {author} {\bibfnamefont {I.~B.}\ \bibnamefont {Spielman}},\
  }\href {\doibase 10.1103/RevModPhys.91.015005} {\bibfield  {journal}
  {\bibinfo  {journal} {Rev. Mod. Phys.}\ }\textbf {\bibinfo {volume} {91}},\
  \bibinfo {pages} {015005} (\bibinfo {year} {2019})}\BibitemShut {NoStop}%
\bibitem [{\citenamefont {Xiong}\ \emph {et~al.}(2016)\citenamefont {Xiong},
  \citenamefont {Gong},\ and\ \citenamefont {An}}]{Xiong16}%
  \BibitemOpen
  \bibfield  {author} {\bibinfo {author} {\bibfnamefont {T.-S.}\ \bibnamefont
  {Xiong}}, \bibinfo {author} {\bibfnamefont {J.}~\bibnamefont {Gong}}, \ and\
  \bibinfo {author} {\bibfnamefont {J.-H.}\ \bibnamefont {An}},\ }\href
  {\doibase 10.1103/PhysRevB.93.184306} {\bibfield  {journal} {\bibinfo
  {journal} {Phys. Rev. B}\ }\textbf {\bibinfo {volume} {93}},\ \bibinfo
  {pages} {184306} (\bibinfo {year} {2016})}\BibitemShut {NoStop}%
\bibitem [{\citenamefont {Yao}\ \emph {et~al.}(2013)\citenamefont {Yao},
  \citenamefont {Gorshkov}, \citenamefont {Laumann}, \citenamefont {L\"auchli},
  \citenamefont {Ye},\ and\ \citenamefont {Lukin}}]{Yao13}%
  \BibitemOpen
  \bibfield  {author} {\bibinfo {author} {\bibfnamefont {N.~Y.}\ \bibnamefont
  {Yao}}, \bibinfo {author} {\bibfnamefont {A.~V.}\ \bibnamefont {Gorshkov}},
  \bibinfo {author} {\bibfnamefont {C.~R.}\ \bibnamefont {Laumann}}, \bibinfo
  {author} {\bibfnamefont {A.~M.}\ \bibnamefont {L\"auchli}}, \bibinfo {author}
  {\bibfnamefont {J.}~\bibnamefont {Ye}}, \ and\ \bibinfo {author}
  {\bibfnamefont {M.~D.}\ \bibnamefont {Lukin}},\ }\href {\doibase
  10.1103/PhysRevLett.110.185302} {\bibfield  {journal} {\bibinfo  {journal}
  {Phys. Rev. Lett.}\ }\textbf {\bibinfo {volume} {110}},\ \bibinfo {pages}
  {185302} (\bibinfo {year} {2013})}\BibitemShut {NoStop}%
\bibitem [{\citenamefont {Trescher}\ and\ \citenamefont
  {Bergholtz}(2012)}]{Trescher12}%
  \BibitemOpen
  \bibfield  {author} {\bibinfo {author} {\bibfnamefont {M.}~\bibnamefont
  {Trescher}}\ and\ \bibinfo {author} {\bibfnamefont {E.~J.}\ \bibnamefont
  {Bergholtz}},\ }\href {\doibase 10.1103/PhysRevB.86.241111} {\bibfield
  {journal} {\bibinfo  {journal} {Phys. Rev. B}\ }\textbf {\bibinfo {volume}
  {86}},\ \bibinfo {pages} {241111(R)} (\bibinfo {year} {2012})}\BibitemShut
  {NoStop}%
\bibitem [{\citenamefont {Cook}\ and\ \citenamefont
  {Paramekanti}(2014)}]{Cook14}%
  \BibitemOpen
  \bibfield  {author} {\bibinfo {author} {\bibfnamefont {A.~M.}\ \bibnamefont
  {Cook}}\ and\ \bibinfo {author} {\bibfnamefont {A.}~\bibnamefont
  {Paramekanti}},\ }\href {\doibase 10.1103/PhysRevLett.113.077203} {\bibfield
  {journal} {\bibinfo  {journal} {Phys. Rev. Lett.}\ }\textbf {\bibinfo
  {volume} {113}},\ \bibinfo {pages} {077203} (\bibinfo {year}
  {2014})}\BibitemShut {NoStop}%
\bibitem [{\citenamefont {Wu}\ \emph {et~al.}(2012{\natexlab{a}})\citenamefont
  {Wu}, \citenamefont {Bernevig},\ and\ \citenamefont {Regnault}}]{Wu12_2}%
  \BibitemOpen
  \bibfield  {author} {\bibinfo {author} {\bibfnamefont {Y.-L.}\ \bibnamefont
  {Wu}}, \bibinfo {author} {\bibfnamefont {B.~A.}\ \bibnamefont {Bernevig}}, \
  and\ \bibinfo {author} {\bibfnamefont {N.}~\bibnamefont {Regnault}},\ }\href
  {\doibase 10.1103/PhysRevB.85.075116} {\bibfield  {journal} {\bibinfo
  {journal} {Phys. Rev. B}\ }\textbf {\bibinfo {volume} {85}},\ \bibinfo
  {pages} {075116} (\bibinfo {year} {2012}{\natexlab{a}})}\BibitemShut
  {NoStop}%
\bibitem [{\citenamefont {Yang}\ \emph {et~al.}(2012)\citenamefont {Yang},
  \citenamefont {Gu}, \citenamefont {Sun},\ and\ \citenamefont
  {Das~Sarma}}]{Yang12}%
  \BibitemOpen
  \bibfield  {author} {\bibinfo {author} {\bibfnamefont {S.}~\bibnamefont
  {Yang}}, \bibinfo {author} {\bibfnamefont {Z.-C.}\ \bibnamefont {Gu}},
  \bibinfo {author} {\bibfnamefont {K.}~\bibnamefont {Sun}}, \ and\ \bibinfo
  {author} {\bibfnamefont {S.}~\bibnamefont {Das~Sarma}},\ }\href {\doibase
  10.1103/PhysRevB.86.241112} {\bibfield  {journal} {\bibinfo  {journal} {Phys.
  Rev. B}\ }\textbf {\bibinfo {volume} {86}},\ \bibinfo {pages} {241112(R)}
  (\bibinfo {year} {2012})}\BibitemShut {NoStop}%
\bibitem [{\citenamefont {Kourtis}\ \emph {et~al.}(2012)\citenamefont
  {Kourtis}, \citenamefont {Venderbos},\ and\ \citenamefont
  {Daghofer}}]{Kourtis12}%
  \BibitemOpen
  \bibfield  {author} {\bibinfo {author} {\bibfnamefont {S.}~\bibnamefont
  {Kourtis}}, \bibinfo {author} {\bibfnamefont {J.~W.~F.}\ \bibnamefont
  {Venderbos}}, \ and\ \bibinfo {author} {\bibfnamefont {M.}~\bibnamefont
  {Daghofer}},\ }\href {\doibase 10.1103/PhysRevB.86.235118} {\bibfield
  {journal} {\bibinfo  {journal} {Phys. Rev. B}\ }\textbf {\bibinfo {volume}
  {86}},\ \bibinfo {pages} {235118} (\bibinfo {year} {2012})}\BibitemShut
  {NoStop}%
\bibitem [{\citenamefont {Motruk}\ and\ \citenamefont
  {Pollmann}(2017)}]{Motruk17}%
  \BibitemOpen
  \bibfield  {author} {\bibinfo {author} {\bibfnamefont {J.}~\bibnamefont
  {Motruk}}\ and\ \bibinfo {author} {\bibfnamefont {F.}~\bibnamefont
  {Pollmann}},\ }\href {\doibase 10.1103/PhysRevB.96.165107} {\bibfield
  {journal} {\bibinfo  {journal} {Phys. Rev. B}\ }\textbf {\bibinfo {volume}
  {96}},\ \bibinfo {pages} {165107} (\bibinfo {year} {2017})}\BibitemShut
  {NoStop}%
\bibitem [{\citenamefont {Motruk}\ and\ \citenamefont {Na}(2020)}]{Motruk20}%
  \BibitemOpen
  \bibfield  {author} {\bibinfo {author} {\bibfnamefont {J.}~\bibnamefont
  {Motruk}}\ and\ \bibinfo {author} {\bibfnamefont {I.}~\bibnamefont {Na}},\
  }\href {\doibase 10.1103/PhysRevLett.125.236401} {\bibfield  {journal}
  {\bibinfo  {journal} {Phys. Rev. Lett.}\ }\textbf {\bibinfo {volume} {125}},\
  \bibinfo {pages} {236401} (\bibinfo {year} {2020})}\BibitemShut {NoStop}%
\bibitem [{\citenamefont {Peierls}(1933)}]{Peierls33}%
  \BibitemOpen
  \bibfield  {author} {\bibinfo {author} {\bibfnamefont {R.}~\bibnamefont
  {Peierls}},\ }\href {\doibase 10.1007/BF01342591} {\bibfield  {journal}
  {\bibinfo  {journal} {Zeitschrift f{\"u}r Physik}\ }\textbf {\bibinfo
  {volume} {80}},\ \bibinfo {pages} {763} (\bibinfo {year} {1933})}\BibitemShut
  {NoStop}%
\bibitem [{\citenamefont {Zak}(1964)}]{Zak64}%
  \BibitemOpen
  \bibfield  {author} {\bibinfo {author} {\bibfnamefont {J.}~\bibnamefont
  {Zak}},\ }\href {\doibase 10.1103/PhysRev.134.A1602} {\bibfield  {journal}
  {\bibinfo  {journal} {Phys. Rev.}\ }\textbf {\bibinfo {volume} {134}},\
  \bibinfo {pages} {A1602} (\bibinfo {year} {1964})}\BibitemShut {NoStop}%
\bibitem [{\citenamefont {Harper}(1955)}]{Harper55}%
  \BibitemOpen
  \bibfield  {author} {\bibinfo {author} {\bibfnamefont {P.~G.}\ \bibnamefont
  {Harper}},\ }\href {\doibase 10.1088/0370-1298/68/10/304} {\bibfield
  {journal} {\bibinfo  {journal} {Proc. Phys. Soc. A}\ }\textbf {\bibinfo
  {volume} {68}},\ \bibinfo {pages} {874} (\bibinfo {year} {1955})}\BibitemShut
  {NoStop}%
\bibitem [{\citenamefont {Hofstadter}(1976)}]{Hofstadter76}%
  \BibitemOpen
  \bibfield  {author} {\bibinfo {author} {\bibfnamefont {D.~R.}\ \bibnamefont
  {Hofstadter}},\ }\href {\doibase 10.1103/PhysRevB.14.2239} {\bibfield
  {journal} {\bibinfo  {journal} {Phys. Rev. B}\ }\textbf {\bibinfo {volume}
  {14}},\ \bibinfo {pages} {2239} (\bibinfo {year} {1976})}\BibitemShut
  {NoStop}%
\bibitem [{\citenamefont {Liu}\ \emph {et~al.}(2013{\natexlab{a}})\citenamefont
  {Liu}, \citenamefont {Repellin}, \citenamefont {Bernevig},\ and\
  \citenamefont {Regnault}}]{Liu13}%
  \BibitemOpen
  \bibfield  {author} {\bibinfo {author} {\bibfnamefont {T.}~\bibnamefont
  {Liu}}, \bibinfo {author} {\bibfnamefont {C.}~\bibnamefont {Repellin}},
  \bibinfo {author} {\bibfnamefont {B.~A.}\ \bibnamefont {Bernevig}}, \ and\
  \bibinfo {author} {\bibfnamefont {N.}~\bibnamefont {Regnault}},\ }\href
  {\doibase 10.1103/PhysRevB.87.205136} {\bibfield  {journal} {\bibinfo
  {journal} {Phys. Rev. B}\ }\textbf {\bibinfo {volume} {87}},\ \bibinfo
  {pages} {205136} (\bibinfo {year} {2013}{\natexlab{a}})}\BibitemShut
  {NoStop}%
\bibitem [{\citenamefont {Sheng}\ \emph {et~al.}(2011)\citenamefont {Sheng},
  \citenamefont {Gu}, \citenamefont {Sun},\ and\ \citenamefont
  {Sheng}}]{Sheng11}%
  \BibitemOpen
  \bibfield  {author} {\bibinfo {author} {\bibfnamefont {D.~N.}\ \bibnamefont
  {Sheng}}, \bibinfo {author} {\bibfnamefont {Z.-C.}\ \bibnamefont {Gu}},
  \bibinfo {author} {\bibfnamefont {K.}~\bibnamefont {Sun}}, \ and\ \bibinfo
  {author} {\bibfnamefont {L.}~\bibnamefont {Sheng}},\ }\href {\doibase
  10.1038/ncomms1380} {\bibfield  {journal} {\bibinfo  {journal} {Nature
  Communications}\ }\textbf {\bibinfo {volume} {2}},\ \bibinfo {pages} {389}
  (\bibinfo {year} {2011})}\BibitemShut {NoStop}%
\bibitem [{\citenamefont {R\"osner}\ \emph {et~al.}(2015)\citenamefont
  {R\"osner}, \citenamefont {Sasioglu}, \citenamefont {Friedrich},
  \citenamefont {Bl\"ugel},\ and\ \citenamefont {Wehling}}]{Rosner15}%
  \BibitemOpen
  \bibfield  {author} {\bibinfo {author} {\bibfnamefont {M.}~\bibnamefont
  {R\"osner}}, \bibinfo {author} {\bibfnamefont {E.}~\bibnamefont {Sasioglu}},
  \bibinfo {author} {\bibfnamefont {C.}~\bibnamefont {Friedrich}}, \bibinfo
  {author} {\bibfnamefont {S.}~\bibnamefont {Bl\"ugel}}, \ and\ \bibinfo
  {author} {\bibfnamefont {T.~O.}\ \bibnamefont {Wehling}},\ }\href {\doibase
  10.1103/PhysRevB.92.085102} {\bibfield  {journal} {\bibinfo  {journal} {Phys.
  Rev. B}\ }\textbf {\bibinfo {volume} {92}},\ \bibinfo {pages} {085102}
  (\bibinfo {year} {2015})}\BibitemShut {NoStop}%
\bibitem [{\citenamefont {Pizarro}\ \emph {et~al.}(2019)\citenamefont
  {Pizarro}, \citenamefont {R\"osner}, \citenamefont {Thomale}, \citenamefont
  {Valent\'{\i}},\ and\ \citenamefont {Wehling}}]{Pizarro19}%
  \BibitemOpen
  \bibfield  {author} {\bibinfo {author} {\bibfnamefont {J.~M.}\ \bibnamefont
  {Pizarro}}, \bibinfo {author} {\bibfnamefont {M.}~\bibnamefont {R\"osner}},
  \bibinfo {author} {\bibfnamefont {R.}~\bibnamefont {Thomale}}, \bibinfo
  {author} {\bibfnamefont {R.}~\bibnamefont {Valent\'{\i}}}, \ and\ \bibinfo
  {author} {\bibfnamefont {T.~O.}\ \bibnamefont {Wehling}},\ }\href {\doibase
  10.1103/PhysRevB.100.161102} {\bibfield  {journal} {\bibinfo  {journal}
  {Phys. Rev. B}\ }\textbf {\bibinfo {volume} {100}},\ \bibinfo {pages}
  {161102(R)} (\bibinfo {year} {2019})}\BibitemShut {NoStop}%
\bibitem [{\citenamefont {Kim}\ \emph {et~al.}(2017)\citenamefont {Kim},
  \citenamefont {DaSilva}, \citenamefont {Huang}, \citenamefont {Fallahazad},
  \citenamefont {Larentis}, \citenamefont {Taniguchi}, \citenamefont
  {Watanabe}, \citenamefont {LeRoy}, \citenamefont {MacDonald},\ and\
  \citenamefont {Tutuc}}]{Kim17}%
  \BibitemOpen
  \bibfield  {author} {\bibinfo {author} {\bibfnamefont {K.}~\bibnamefont
  {Kim}}, \bibinfo {author} {\bibfnamefont {A.}~\bibnamefont {DaSilva}},
  \bibinfo {author} {\bibfnamefont {S.}~\bibnamefont {Huang}}, \bibinfo
  {author} {\bibfnamefont {B.}~\bibnamefont {Fallahazad}}, \bibinfo {author}
  {\bibfnamefont {S.}~\bibnamefont {Larentis}}, \bibinfo {author}
  {\bibfnamefont {T.}~\bibnamefont {Taniguchi}}, \bibinfo {author}
  {\bibfnamefont {K.}~\bibnamefont {Watanabe}}, \bibinfo {author}
  {\bibfnamefont {B.~J.}\ \bibnamefont {LeRoy}}, \bibinfo {author}
  {\bibfnamefont {A.~H.}\ \bibnamefont {MacDonald}}, \ and\ \bibinfo {author}
  {\bibfnamefont {E.}~\bibnamefont {Tutuc}},\ }\href {\doibase
  10.1073/pnas.1620140114} {\bibfield  {journal} {\bibinfo  {journal}
  {Proceedings of the National Academy of Sciences}\ }\textbf {\bibinfo
  {volume} {114}},\ \bibinfo {pages} {3364} (\bibinfo {year} {2017})},\ \Eprint
  {http://arxiv.org/abs/https://www.pnas.org/content/114/13/3364.full.pdf}
  {https://www.pnas.org/content/114/13/3364.full.pdf} \BibitemShut {NoStop}%
\bibitem [{\citenamefont {Bauer}\ \emph {et~al.}(2016)\citenamefont {Bauer},
  \citenamefont {Jackson},\ and\ \citenamefont {Roy}}]{Bauer16}%
  \BibitemOpen
  \bibfield  {author} {\bibinfo {author} {\bibfnamefont {D.}~\bibnamefont
  {Bauer}}, \bibinfo {author} {\bibfnamefont {T.~S.}\ \bibnamefont {Jackson}},
  \ and\ \bibinfo {author} {\bibfnamefont {R.}~\bibnamefont {Roy}},\ }\href
  {\doibase 10.1103/PhysRevB.93.235133} {\bibfield  {journal} {\bibinfo
  {journal} {Phys. Rev. B}\ }\textbf {\bibinfo {volume} {93}},\ \bibinfo
  {pages} {235133} (\bibinfo {year} {2016})}\BibitemShut {NoStop}%
\bibitem [{\citenamefont {Andrews}\ and\ \citenamefont
  {Soluyanov}(2020)}]{Andrews20}%
  \BibitemOpen
  \bibfield  {author} {\bibinfo {author} {\bibfnamefont {B.}~\bibnamefont
  {Andrews}}\ and\ \bibinfo {author} {\bibfnamefont {A.}~\bibnamefont
  {Soluyanov}},\ }\href {\doibase 10.1103/PhysRevB.101.235312} {\bibfield
  {journal} {\bibinfo  {journal} {Phys. Rev. B}\ }\textbf {\bibinfo {volume}
  {101}},\ \bibinfo {pages} {235312} (\bibinfo {year} {2020})}\BibitemShut
  {NoStop}%
\bibitem [{\citenamefont {Andrews}\ \emph {et~al.}(2021)\citenamefont
  {Andrews}, \citenamefont {Mohan},\ and\ \citenamefont {Neupert}}]{Andrews21}%
  \BibitemOpen
  \bibfield  {author} {\bibinfo {author} {\bibfnamefont {B.}~\bibnamefont
  {Andrews}}, \bibinfo {author} {\bibfnamefont {M.}~\bibnamefont {Mohan}}, \
  and\ \bibinfo {author} {\bibfnamefont {T.}~\bibnamefont {Neupert}},\ }\href
  {\doibase 10.1103/PhysRevB.103.075132} {\bibfield  {journal} {\bibinfo
  {journal} {Phys. Rev. B}\ }\textbf {\bibinfo {volume} {103}},\ \bibinfo
  {pages} {075132} (\bibinfo {year} {2021})}\BibitemShut {NoStop}%
\bibitem [{\citenamefont {Miyake}\ \emph {et~al.}(2013)\citenamefont {Miyake},
  \citenamefont {Siviloglou}, \citenamefont {Kennedy}, \citenamefont {Burton},\
  and\ \citenamefont {Ketterle}}]{Miyake13}%
  \BibitemOpen
  \bibfield  {author} {\bibinfo {author} {\bibfnamefont {H.}~\bibnamefont
  {Miyake}}, \bibinfo {author} {\bibfnamefont {G.~A.}\ \bibnamefont
  {Siviloglou}}, \bibinfo {author} {\bibfnamefont {C.~J.}\ \bibnamefont
  {Kennedy}}, \bibinfo {author} {\bibfnamefont {W.~C.}\ \bibnamefont {Burton}},
  \ and\ \bibinfo {author} {\bibfnamefont {W.}~\bibnamefont {Ketterle}},\
  }\href {\doibase 10.1103/PhysRevLett.111.185302} {\bibfield  {journal}
  {\bibinfo  {journal} {Phys. Rev. Lett.}\ }\textbf {\bibinfo {volume} {111}},\
  \bibinfo {pages} {185302} (\bibinfo {year} {2013})}\BibitemShut {NoStop}%
\bibitem [{\citenamefont {Aidelsburger}\ \emph {et~al.}(2013)\citenamefont
  {Aidelsburger}, \citenamefont {Atala}, \citenamefont {Lohse}, \citenamefont
  {Barreiro}, \citenamefont {Paredes},\ and\ \citenamefont
  {Bloch}}]{Aidelsburger13}%
  \BibitemOpen
  \bibfield  {author} {\bibinfo {author} {\bibfnamefont {M.}~\bibnamefont
  {Aidelsburger}}, \bibinfo {author} {\bibfnamefont {M.}~\bibnamefont {Atala}},
  \bibinfo {author} {\bibfnamefont {M.}~\bibnamefont {Lohse}}, \bibinfo
  {author} {\bibfnamefont {J.~T.}\ \bibnamefont {Barreiro}}, \bibinfo {author}
  {\bibfnamefont {B.}~\bibnamefont {Paredes}}, \ and\ \bibinfo {author}
  {\bibfnamefont {I.}~\bibnamefont {Bloch}},\ }\href {\doibase
  10.1103/PhysRevLett.111.185301} {\bibfield  {journal} {\bibinfo  {journal}
  {Phys. Rev. Lett.}\ }\textbf {\bibinfo {volume} {111}},\ \bibinfo {pages}
  {185301} (\bibinfo {year} {2013})}\BibitemShut {NoStop}%
\bibitem [{\citenamefont {Hafezi}\ \emph {et~al.}(2013)\citenamefont {Hafezi},
  \citenamefont {Mittal}, \citenamefont {Fan}, \citenamefont {Migdall},\ and\
  \citenamefont {Taylor}}]{Hafezi13}%
  \BibitemOpen
  \bibfield  {author} {\bibinfo {author} {\bibfnamefont {M.}~\bibnamefont
  {Hafezi}}, \bibinfo {author} {\bibfnamefont {S.}~\bibnamefont {Mittal}},
  \bibinfo {author} {\bibfnamefont {J.}~\bibnamefont {Fan}}, \bibinfo {author}
  {\bibfnamefont {A.}~\bibnamefont {Migdall}}, \ and\ \bibinfo {author}
  {\bibfnamefont {J.~M.}\ \bibnamefont {Taylor}},\ }\href {\doibase
  10.1038/nphoton.2013.274} {\bibfield  {journal} {\bibinfo  {journal} {Nature
  Photonics}\ }\textbf {\bibinfo {volume} {7}},\ \bibinfo {pages} {1001}
  (\bibinfo {year} {2013})}\BibitemShut {NoStop}%
\bibitem [{\citenamefont {Dean}\ \emph {et~al.}(2013)\citenamefont {Dean},
  \citenamefont {Wang}, \citenamefont {Maher}, \citenamefont {Forsythe},
  \citenamefont {Ghahari}, \citenamefont {Gao}, \citenamefont {Katoch},
  \citenamefont {Ishigami}, \citenamefont {Moon}, \citenamefont {Koshino},
  \citenamefont {Taniguchi}, \citenamefont {Watanabe}, \citenamefont {Shepard},
  \citenamefont {Hone},\ and\ \citenamefont {Kim}}]{Dean13}%
  \BibitemOpen
  \bibfield  {author} {\bibinfo {author} {\bibfnamefont {C.~R.}\ \bibnamefont
  {Dean}}, \bibinfo {author} {\bibfnamefont {L.}~\bibnamefont {Wang}}, \bibinfo
  {author} {\bibfnamefont {P.}~\bibnamefont {Maher}}, \bibinfo {author}
  {\bibfnamefont {C.}~\bibnamefont {Forsythe}}, \bibinfo {author}
  {\bibfnamefont {F.}~\bibnamefont {Ghahari}}, \bibinfo {author} {\bibfnamefont
  {Y.}~\bibnamefont {Gao}}, \bibinfo {author} {\bibfnamefont {J.}~\bibnamefont
  {Katoch}}, \bibinfo {author} {\bibfnamefont {M.}~\bibnamefont {Ishigami}},
  \bibinfo {author} {\bibfnamefont {P.}~\bibnamefont {Moon}}, \bibinfo {author}
  {\bibfnamefont {M.}~\bibnamefont {Koshino}}, \bibinfo {author} {\bibfnamefont
  {T.}~\bibnamefont {Taniguchi}}, \bibinfo {author} {\bibfnamefont
  {K.}~\bibnamefont {Watanabe}}, \bibinfo {author} {\bibfnamefont {K.~L.}\
  \bibnamefont {Shepard}}, \bibinfo {author} {\bibfnamefont {J.}~\bibnamefont
  {Hone}}, \ and\ \bibinfo {author} {\bibfnamefont {P.}~\bibnamefont {Kim}},\
  }\href {\doibase 10.1038/nature12186} {\bibfield  {journal} {\bibinfo
  {journal} {Nature}\ }\textbf {\bibinfo {volume} {497}},\ \bibinfo {pages}
  {598} (\bibinfo {year} {2013})}\BibitemShut {NoStop}%
\bibitem [{\citenamefont {Roushan}\ \emph {et~al.}(2017)\citenamefont
  {Roushan}, \citenamefont {Neill}, \citenamefont {Tangpanitanon},
  \citenamefont {Bastidas}, \citenamefont {Megrant}, \citenamefont {Barends},
  \citenamefont {Chen}, \citenamefont {Chen}, \citenamefont {Chiaro},
  \citenamefont {Dunsworth}, \citenamefont {Fowler}, \citenamefont {Foxen},
  \citenamefont {Giustina}, \citenamefont {Jeffrey}, \citenamefont {Kelly},
  \citenamefont {Lucero}, \citenamefont {Mutus}, \citenamefont {Neeley},
  \citenamefont {Quintana}, \citenamefont {Sank}, \citenamefont {Vainsencher},
  \citenamefont {Wenner}, \citenamefont {White}, \citenamefont {Neven},
  \citenamefont {Angelakis},\ and\ \citenamefont {Martinis}}]{Roushan17}%
  \BibitemOpen
  \bibfield  {author} {\bibinfo {author} {\bibfnamefont {P.}~\bibnamefont
  {Roushan}}, \bibinfo {author} {\bibfnamefont {C.}~\bibnamefont {Neill}},
  \bibinfo {author} {\bibfnamefont {J.}~\bibnamefont {Tangpanitanon}}, \bibinfo
  {author} {\bibfnamefont {V.~M.}\ \bibnamefont {Bastidas}}, \bibinfo {author}
  {\bibfnamefont {A.}~\bibnamefont {Megrant}}, \bibinfo {author} {\bibfnamefont
  {R.}~\bibnamefont {Barends}}, \bibinfo {author} {\bibfnamefont
  {Y.}~\bibnamefont {Chen}}, \bibinfo {author} {\bibfnamefont {Z.}~\bibnamefont
  {Chen}}, \bibinfo {author} {\bibfnamefont {B.}~\bibnamefont {Chiaro}},
  \bibinfo {author} {\bibfnamefont {A.}~\bibnamefont {Dunsworth}}, \bibinfo
  {author} {\bibfnamefont {A.}~\bibnamefont {Fowler}}, \bibinfo {author}
  {\bibfnamefont {B.}~\bibnamefont {Foxen}}, \bibinfo {author} {\bibfnamefont
  {M.}~\bibnamefont {Giustina}}, \bibinfo {author} {\bibfnamefont
  {E.}~\bibnamefont {Jeffrey}}, \bibinfo {author} {\bibfnamefont
  {J.}~\bibnamefont {Kelly}}, \bibinfo {author} {\bibfnamefont
  {E.}~\bibnamefont {Lucero}}, \bibinfo {author} {\bibfnamefont
  {J.}~\bibnamefont {Mutus}}, \bibinfo {author} {\bibfnamefont
  {M.}~\bibnamefont {Neeley}}, \bibinfo {author} {\bibfnamefont
  {C.}~\bibnamefont {Quintana}}, \bibinfo {author} {\bibfnamefont
  {D.}~\bibnamefont {Sank}}, \bibinfo {author} {\bibfnamefont {A.}~\bibnamefont
  {Vainsencher}}, \bibinfo {author} {\bibfnamefont {J.}~\bibnamefont {Wenner}},
  \bibinfo {author} {\bibfnamefont {T.}~\bibnamefont {White}}, \bibinfo
  {author} {\bibfnamefont {H.}~\bibnamefont {Neven}}, \bibinfo {author}
  {\bibfnamefont {D.~G.}\ \bibnamefont {Angelakis}}, \ and\ \bibinfo {author}
  {\bibfnamefont {J.}~\bibnamefont {Martinis}},\ }\href {\doibase
  10.1126/science.aao1401} {\bibfield  {journal} {\bibinfo  {journal}
  {Science}\ }\textbf {\bibinfo {volume} {358}},\ \bibinfo {pages} {1175}
  (\bibinfo {year} {2017})},\ \Eprint
  {http://arxiv.org/abs/https://science.sciencemag.org/content/358/6367/1175.full.pdf}
  {https://science.sciencemag.org/content/358/6367/1175.full.pdf} \BibitemShut
  {NoStop}%
\bibitem [{\citenamefont {Ni}\ \emph {et~al.}(2019)\citenamefont {Ni},
  \citenamefont {Chen}, \citenamefont {Weiner}, \citenamefont {Apigo},
  \citenamefont {Prodan}, \citenamefont {Al{\`u}}, \citenamefont {Prodan},\
  and\ \citenamefont {Khanikaev}}]{Ni19}%
  \BibitemOpen
  \bibfield  {author} {\bibinfo {author} {\bibfnamefont {X.}~\bibnamefont
  {Ni}}, \bibinfo {author} {\bibfnamefont {K.}~\bibnamefont {Chen}}, \bibinfo
  {author} {\bibfnamefont {M.}~\bibnamefont {Weiner}}, \bibinfo {author}
  {\bibfnamefont {D.~J.}\ \bibnamefont {Apigo}}, \bibinfo {author}
  {\bibfnamefont {C.}~\bibnamefont {Prodan}}, \bibinfo {author} {\bibfnamefont
  {A.}~\bibnamefont {Al{\`u}}}, \bibinfo {author} {\bibfnamefont
  {E.}~\bibnamefont {Prodan}}, \ and\ \bibinfo {author} {\bibfnamefont {A.~B.}\
  \bibnamefont {Khanikaev}},\ }\href {\doibase 10.1038/s42005-019-0151-7}
  {\bibfield  {journal} {\bibinfo  {journal} {Communications Physics}\ }\textbf
  {\bibinfo {volume} {2}},\ \bibinfo {pages} {55} (\bibinfo {year}
  {2019})}\BibitemShut {NoStop}%
\bibitem [{\citenamefont {Dutt}\ \emph {et~al.}(2020)\citenamefont {Dutt},
  \citenamefont {Lin}, \citenamefont {Yuan}, \citenamefont {Minkov},
  \citenamefont {Xiao},\ and\ \citenamefont {Fan}}]{Dutt20}%
  \BibitemOpen
  \bibfield  {author} {\bibinfo {author} {\bibfnamefont {A.}~\bibnamefont
  {Dutt}}, \bibinfo {author} {\bibfnamefont {Q.}~\bibnamefont {Lin}}, \bibinfo
  {author} {\bibfnamefont {L.}~\bibnamefont {Yuan}}, \bibinfo {author}
  {\bibfnamefont {M.}~\bibnamefont {Minkov}}, \bibinfo {author} {\bibfnamefont
  {M.}~\bibnamefont {Xiao}}, \ and\ \bibinfo {author} {\bibfnamefont
  {S.}~\bibnamefont {Fan}},\ }\href {\doibase 10.1126/science.aaz3071}
  {\bibfield  {journal} {\bibinfo  {journal} {Science}\ }\textbf {\bibinfo
  {volume} {367}},\ \bibinfo {pages} {59} (\bibinfo {year} {2020})},\ \Eprint
  {http://arxiv.org/abs/https://science.sciencemag.org/content/367/6473/59.full.pdf}
  {https://science.sciencemag.org/content/367/6473/59.full.pdf} \BibitemShut
  {NoStop}%
\bibitem [{\citenamefont {Jain}(1989)}]{Jain89}%
  \BibitemOpen
  \bibfield  {author} {\bibinfo {author} {\bibfnamefont {J.~K.}\ \bibnamefont
  {Jain}},\ }\href {\doibase 10.1103/PhysRevLett.63.199} {\bibfield  {journal}
  {\bibinfo  {journal} {Phys. Rev. Lett.}\ }\textbf {\bibinfo {volume} {63}},\
  \bibinfo {pages} {199} (\bibinfo {year} {1989})}\BibitemShut {NoStop}%
\bibitem [{Note1()}]{Note1}%
  \BibitemOpen
  \bibinfo {note} {This implies that $p$ on the left-hand side of Eq.~\protect
  \textup {\hbox {\mathsurround \z@ \protect \normalfont (\ignorespaces \ref
  {eq:nphi}\unskip \@@italiccorr )}} does not always equal $p'$.}\BibitemShut
  {Stop}%
\bibitem [{\citenamefont {Harper}\ \emph {et~al.}(2014)\citenamefont {Harper},
  \citenamefont {Simon},\ and\ \citenamefont {Roy}}]{Harper14}%
  \BibitemOpen
  \bibfield  {author} {\bibinfo {author} {\bibfnamefont {F.}~\bibnamefont
  {Harper}}, \bibinfo {author} {\bibfnamefont {S.~H.}\ \bibnamefont {Simon}}, \
  and\ \bibinfo {author} {\bibfnamefont {R.}~\bibnamefont {Roy}},\ }\href@noop
  {} {\bibfield  {journal} {\bibinfo  {journal} {Phys. Rev. B}\ }\textbf
  {\bibinfo {volume} {90}},\ \bibinfo {pages} {075104} (\bibinfo {year}
  {2014})}\BibitemShut {NoStop}%
\bibitem [{\citenamefont {Wang}\ \emph {et~al.}(2012)\citenamefont {Wang},
  \citenamefont {Yao}, \citenamefont {Gong},\ and\ \citenamefont
  {Sheng}}]{Wang12}%
  \BibitemOpen
  \bibfield  {author} {\bibinfo {author} {\bibfnamefont {Y.-F.}\ \bibnamefont
  {Wang}}, \bibinfo {author} {\bibfnamefont {H.}~\bibnamefont {Yao}}, \bibinfo
  {author} {\bibfnamefont {C.-D.}\ \bibnamefont {Gong}}, \ and\ \bibinfo
  {author} {\bibfnamefont {D.~N.}\ \bibnamefont {Sheng}},\ }\href {\doibase
  10.1103/PhysRevB.86.201101} {\bibfield  {journal} {\bibinfo  {journal} {Phys.
  Rev. B}\ }\textbf {\bibinfo {volume} {86}},\ \bibinfo {pages} {201101(R)}
  (\bibinfo {year} {2012})}\BibitemShut {NoStop}%
\bibitem [{\citenamefont {Haldane}(1983)}]{Haldane83}%
  \BibitemOpen
  \bibfield  {author} {\bibinfo {author} {\bibfnamefont {F.~D.~M.}\
  \bibnamefont {Haldane}},\ }\href {\doibase 10.1103/PhysRevLett.51.605}
  {\bibfield  {journal} {\bibinfo  {journal} {Phys. Rev. Lett.}\ }\textbf
  {\bibinfo {volume} {51}},\ \bibinfo {pages} {605} (\bibinfo {year}
  {1983})}\BibitemShut {NoStop}%
\bibitem [{\citenamefont {White}(1992)}]{White92}%
  \BibitemOpen
  \bibfield  {author} {\bibinfo {author} {\bibfnamefont {S.~R.}\ \bibnamefont
  {White}},\ }\href {\doibase 10.1103/PhysRevLett.69.2863} {\bibfield
  {journal} {\bibinfo  {journal} {Phys. Rev. Lett.}\ }\textbf {\bibinfo
  {volume} {69}},\ \bibinfo {pages} {2863} (\bibinfo {year}
  {1992})}\BibitemShut {NoStop}%
\bibitem [{\citenamefont {Schollwöck}(2011)}]{Schollwock11}%
  \BibitemOpen
  \bibfield  {author} {\bibinfo {author} {\bibfnamefont {U.}~\bibnamefont
  {Schollwöck}},\ }\href {\doibase https://doi.org/10.1016/j.aop.2010.09.012}
  {\bibfield  {journal} {\bibinfo  {journal} {Annals of Physics}\ }\textbf
  {\bibinfo {volume} {326}},\ \bibinfo {pages} {96} (\bibinfo {year} {2011})},\
  \bibinfo {note} {january 2011 Special Issue}\BibitemShut {NoStop}%
\bibitem [{\citenamefont {Stoudenmire}\ and\ \citenamefont
  {White}(2012)}]{Stoudenmire12}%
  \BibitemOpen
  \bibfield  {author} {\bibinfo {author} {\bibfnamefont {E.}~\bibnamefont
  {Stoudenmire}}\ and\ \bibinfo {author} {\bibfnamefont {S.~R.}\ \bibnamefont
  {White}},\ }\href {\doibase 10.1146/annurev-conmatphys-020911-125018}
  {\bibfield  {journal} {\bibinfo  {journal} {Annual Review of Condensed Matter
  Physics}\ }\textbf {\bibinfo {volume} {3}},\ \bibinfo {pages} {111–128}
  (\bibinfo {year} {2012})}\BibitemShut {NoStop}%
\bibitem [{\citenamefont {Zaletel}\ \emph {et~al.}(2013)\citenamefont
  {Zaletel}, \citenamefont {Mong},\ and\ \citenamefont {Pollmann}}]{Zaletel13}%
  \BibitemOpen
  \bibfield  {author} {\bibinfo {author} {\bibfnamefont {M.~P.}\ \bibnamefont
  {Zaletel}}, \bibinfo {author} {\bibfnamefont {R.~S.~K.}\ \bibnamefont
  {Mong}}, \ and\ \bibinfo {author} {\bibfnamefont {F.}~\bibnamefont
  {Pollmann}},\ }\href {\doibase 10.1103/PhysRevLett.110.236801} {\bibfield
  {journal} {\bibinfo  {journal} {Phys. Rev. Lett.}\ }\textbf {\bibinfo
  {volume} {110}},\ \bibinfo {pages} {236801} (\bibinfo {year}
  {2013})}\BibitemShut {NoStop}%
\bibitem [{\citenamefont {Zaletel}\ \emph {et~al.}(2015)\citenamefont
  {Zaletel}, \citenamefont {Mong}, \citenamefont {Pollmann},\ and\
  \citenamefont {Rezayi}}]{Zaletel15}%
  \BibitemOpen
  \bibfield  {author} {\bibinfo {author} {\bibfnamefont {M.~P.}\ \bibnamefont
  {Zaletel}}, \bibinfo {author} {\bibfnamefont {R.~S.~K.}\ \bibnamefont
  {Mong}}, \bibinfo {author} {\bibfnamefont {F.}~\bibnamefont {Pollmann}}, \
  and\ \bibinfo {author} {\bibfnamefont {E.~H.}\ \bibnamefont {Rezayi}},\
  }\href {\doibase 10.1103/PhysRevB.91.045115} {\bibfield  {journal} {\bibinfo
  {journal} {Phys. Rev. B}\ }\textbf {\bibinfo {volume} {91}},\ \bibinfo
  {pages} {045115} (\bibinfo {year} {2015})}\BibitemShut {NoStop}%
\bibitem [{\citenamefont {Grushin}\ \emph {et~al.}(2015)\citenamefont
  {Grushin}, \citenamefont {Motruk}, \citenamefont {Zaletel},\ and\
  \citenamefont {Pollmann}}]{Grushin15}%
  \BibitemOpen
  \bibfield  {author} {\bibinfo {author} {\bibfnamefont {A.~G.}\ \bibnamefont
  {Grushin}}, \bibinfo {author} {\bibfnamefont {J.}~\bibnamefont {Motruk}},
  \bibinfo {author} {\bibfnamefont {M.~P.}\ \bibnamefont {Zaletel}}, \ and\
  \bibinfo {author} {\bibfnamefont {F.}~\bibnamefont {Pollmann}},\ }\href
  {\doibase 10.1103/PhysRevB.91.035136} {\bibfield  {journal} {\bibinfo
  {journal} {Phys. Rev. B}\ }\textbf {\bibinfo {volume} {91}},\ \bibinfo
  {pages} {035136} (\bibinfo {year} {2015})}\BibitemShut {NoStop}%
\bibitem [{\citenamefont {Schoonderwoerd}\ \emph {et~al.}(2019)\citenamefont
  {Schoonderwoerd}, \citenamefont {Pollmann},\ and\ \citenamefont
  {Möller}}]{Schoonderwoerd19}%
  \BibitemOpen
  \bibfield  {author} {\bibinfo {author} {\bibfnamefont {L.}~\bibnamefont
  {Schoonderwoerd}}, \bibinfo {author} {\bibfnamefont {F.}~\bibnamefont
  {Pollmann}}, \ and\ \bibinfo {author} {\bibfnamefont {G.}~\bibnamefont
  {Möller}},\ }\href@noop {} {\enquote {\bibinfo {title} {Interaction-driven
  plateau transition between integer and fractional chern insulators},}\ }
  (\bibinfo {year} {2019}),\ \Eprint {http://arxiv.org/abs/1908.00988}
  {arXiv:1908.00988 [cond-mat.str-el]} \BibitemShut {NoStop}%
\bibitem [{\citenamefont {Andrews}(2019)}]{AndrewsThesis}%
  \BibitemOpen
  \bibfield  {author} {\bibinfo {author} {\bibfnamefont {B.}~\bibnamefont
  {Andrews}},\ }\emph {\bibinfo {title} {Stability of Topological States and
  Crystalline Solids}},\ \href {\doibase https://doi.org/10.17863/CAM.36140}
  {Ph.D. thesis},\ \bibinfo  {school} {University of Cambridge,} (\bibinfo
  {year} {2019})\BibitemShut {NoStop}%
\bibitem [{\citenamefont {Schoonderwoerd}(2021)}]{SchoonderwoerdThesis}%
  \BibitemOpen
  \bibfield  {author} {\bibinfo {author} {\bibfnamefont {L.}~\bibnamefont
  {Schoonderwoerd}},\ }\emph {\bibinfo {title} {Quantum Hall states in the
  Harper-Hofstadter model: Existence, stability and novel phase transitions}},\
  \href {\doibase 10.22024/UniKent/01.02.88063} {Ph.D. thesis},\ \bibinfo
  {school} {University of Kent,} (\bibinfo {year} {2021})\BibitemShut {NoStop}%
\bibitem [{Note2()}]{Note2}%
  \BibitemOpen
  \bibinfo {note} {We note that variants of the DMRG algorithm geared towards
  excited states are currently being explored~\cite {Khemani16}.}\BibitemShut
  {Stop}%
\bibitem [{\citenamefont {Laughlin}(1981)}]{Laughlin81}%
  \BibitemOpen
  \bibfield  {author} {\bibinfo {author} {\bibfnamefont {R.~B.}\ \bibnamefont
  {Laughlin}},\ }\href {\doibase 10.1103/PhysRevB.23.5632} {\bibfield
  {journal} {\bibinfo  {journal} {Phys. Rev. B}\ }\textbf {\bibinfo {volume}
  {23}},\ \bibinfo {pages} {5632} (\bibinfo {year} {1981})}\BibitemShut
  {NoStop}%
\bibitem [{\citenamefont {Cincio}\ and\ \citenamefont
  {Vidal}(2013)}]{Cincio13}%
  \BibitemOpen
  \bibfield  {author} {\bibinfo {author} {\bibfnamefont {L.}~\bibnamefont
  {Cincio}}\ and\ \bibinfo {author} {\bibfnamefont {G.}~\bibnamefont {Vidal}},\
  }\href {\doibase 10.1103/PhysRevLett.110.067208} {\bibfield  {journal}
  {\bibinfo  {journal} {Phys. Rev. Lett.}\ }\textbf {\bibinfo {volume} {110}},\
  \bibinfo {pages} {067208} (\bibinfo {year} {2013})}\BibitemShut {NoStop}%
\bibitem [{\citenamefont {Kourtis}\ \emph {et~al.}(2014)\citenamefont
  {Kourtis}, \citenamefont {Neupert}, \citenamefont {Chamon},\ and\
  \citenamefont {Mudry}}]{Kourtis14}%
  \BibitemOpen
  \bibfield  {author} {\bibinfo {author} {\bibfnamefont {S.}~\bibnamefont
  {Kourtis}}, \bibinfo {author} {\bibfnamefont {T.}~\bibnamefont {Neupert}},
  \bibinfo {author} {\bibfnamefont {C.}~\bibnamefont {Chamon}}, \ and\ \bibinfo
  {author} {\bibfnamefont {C.}~\bibnamefont {Mudry}},\ }\href {\doibase
  10.1103/PhysRevLett.112.126806} {\bibfield  {journal} {\bibinfo  {journal}
  {Phys. Rev. Lett.}\ }\textbf {\bibinfo {volume} {112}},\ \bibinfo {pages}
  {126806} (\bibinfo {year} {2014})}\BibitemShut {NoStop}%
\bibitem [{\citenamefont {Simon}\ and\ \citenamefont {Rudner}(2020)}]{Simon20}%
  \BibitemOpen
  \bibfield  {author} {\bibinfo {author} {\bibfnamefont {S.~H.}\ \bibnamefont
  {Simon}}\ and\ \bibinfo {author} {\bibfnamefont {M.~S.}\ \bibnamefont
  {Rudner}},\ }\href@noop {} {\bibfield  {journal} {\bibinfo  {journal} {Phys.
  Rev. B}\ }\textbf {\bibinfo {volume} {102}},\ \bibinfo {pages} {165148}
  (\bibinfo {year} {2020})}\BibitemShut {NoStop}%
\bibitem [{Note3()}]{Note3}%
  \BibitemOpen
  \bibinfo {note} {It has recently been shown that it is impossible to engineer
  an ideal flat band with constant Berry curvature from a lattice model with a
  finite number of sites per unit cell~\cite {Varjas21, Mera21}.}\BibitemShut
  {Stop}%
\bibitem [{Note4()}]{Note4}%
  \BibitemOpen
  \bibinfo {note} {We did not observe any FCI breakdown transitions as the
  interaction strength is increased to $V=10$}\BibitemShut {NoStop}%
\bibitem [{Note5()}]{Note5}%
  \BibitemOpen
  \bibinfo {note} {The interaction energy $\mathinner {\langle {\protect \hat
  {V}}\rangle }=V\mathinner {\langle {\DOTSB \sum@ \slimits@ _{\mathinner
  {\langle {i,j}\rangle }}\rho _i\rho _j}\rangle }$ is computed for the
  ground-state wavefunction on the MPS unit cell with periodic boundary
  conditions.}\BibitemShut {Stop}%
\bibitem [{Note6()}]{Note6}%
  \BibitemOpen
  \bibinfo {note} {The kinetic energy $\mathinner {\langle {\protect \hat
  {T}}\rangle }=E-\mathinner {\langle {\protect \hat {V}}\rangle }$ provides
  similar insight.}\BibitemShut {Stop}%
\bibitem [{\citenamefont {Wen}(1992)}]{Wen92}%
  \BibitemOpen
  \bibfield  {author} {\bibinfo {author} {\bibfnamefont {X.-G.}\ \bibnamefont
  {Wen}},\ }\href {\doibase 10.1142/S0217979292000840} {\bibfield  {journal}
  {\bibinfo  {journal} {International Journal of Modern Physics B}\ }\textbf
  {\bibinfo {volume} {06}},\ \bibinfo {pages} {1711} (\bibinfo {year}
  {1992})},\ \Eprint
  {http://arxiv.org/abs/https://doi.org/10.1142/S0217979292000840}
  {https://doi.org/10.1142/S0217979292000840} \BibitemShut {NoStop}%
\bibitem [{\citenamefont {Li}\ and\ \citenamefont {Haldane}(2008)}]{Li08}%
  \BibitemOpen
  \bibfield  {author} {\bibinfo {author} {\bibfnamefont {H.}~\bibnamefont
  {Li}}\ and\ \bibinfo {author} {\bibfnamefont {F.~D.~M.}\ \bibnamefont
  {Haldane}},\ }\href {\doibase 10.1103/PhysRevLett.101.010504} {\bibfield
  {journal} {\bibinfo  {journal} {Phys. Rev. Lett.}\ }\textbf {\bibinfo
  {volume} {101}},\ \bibinfo {pages} {010504} (\bibinfo {year}
  {2008})}\BibitemShut {NoStop}%
\bibitem [{Note7()}]{Note7}%
  \BibitemOpen
  \bibinfo {note} {The edge-state counting cannot be resolved.}\BibitemShut
  {Stop}%
\bibitem [{\citenamefont {Liu}\ \emph {et~al.}(2021)\citenamefont {Liu},
  \citenamefont {Abouelkomsan},\ and\ \citenamefont {Bergholtz}}]{Liu21}%
  \BibitemOpen
  \bibfield  {author} {\bibinfo {author} {\bibfnamefont {Z.}~\bibnamefont
  {Liu}}, \bibinfo {author} {\bibfnamefont {A.}~\bibnamefont {Abouelkomsan}}, \
  and\ \bibinfo {author} {\bibfnamefont {E.~J.}\ \bibnamefont {Bergholtz}},\
  }\href {\doibase 10.1103/PhysRevLett.126.026801} {\bibfield  {journal}
  {\bibinfo  {journal} {Phys. Rev. Lett.}\ }\textbf {\bibinfo {volume} {126}},\
  \bibinfo {pages} {026801} (\bibinfo {year} {2021})}\BibitemShut {NoStop}%
\bibitem [{\citenamefont {Repellin}\ \emph {et~al.}(2020)\citenamefont
  {Repellin}, \citenamefont {L\'eonard},\ and\ \citenamefont
  {Goldman}}]{Repellin20}%
  \BibitemOpen
  \bibfield  {author} {\bibinfo {author} {\bibfnamefont {C.}~\bibnamefont
  {Repellin}}, \bibinfo {author} {\bibfnamefont {J.}~\bibnamefont {L\'eonard}},
  \ and\ \bibinfo {author} {\bibfnamefont {N.}~\bibnamefont {Goldman}},\ }\href
  {\doibase 10.1103/PhysRevA.102.063316} {\bibfield  {journal} {\bibinfo
  {journal} {Phys. Rev. A}\ }\textbf {\bibinfo {volume} {102}},\ \bibinfo
  {pages} {063316} (\bibinfo {year} {2020})}\BibitemShut {NoStop}%
\bibitem [{\citenamefont {Qi}(2011)}]{Qi11}%
  \BibitemOpen
  \bibfield  {author} {\bibinfo {author} {\bibfnamefont {X.-L.}\ \bibnamefont
  {Qi}},\ }\href {\doibase 10.1103/PhysRevLett.107.126803} {\bibfield
  {journal} {\bibinfo  {journal} {Phys. Rev. Lett.}\ }\textbf {\bibinfo
  {volume} {107}},\ \bibinfo {pages} {126803} (\bibinfo {year}
  {2011})}\BibitemShut {NoStop}%
\bibitem [{\citenamefont {Wu}\ \emph {et~al.}(2012{\natexlab{b}})\citenamefont
  {Wu}, \citenamefont {Regnault},\ and\ \citenamefont {Bernevig}}]{Wu12}%
  \BibitemOpen
  \bibfield  {author} {\bibinfo {author} {\bibfnamefont {Y.-L.}\ \bibnamefont
  {Wu}}, \bibinfo {author} {\bibfnamefont {N.}~\bibnamefont {Regnault}}, \ and\
  \bibinfo {author} {\bibfnamefont {B.~A.}\ \bibnamefont {Bernevig}},\ }\href
  {\doibase 10.1103/PhysRevB.86.085129} {\bibfield  {journal} {\bibinfo
  {journal} {Phys. Rev. B}\ }\textbf {\bibinfo {volume} {86}},\ \bibinfo
  {pages} {085129} (\bibinfo {year} {2012}{\natexlab{b}})}\BibitemShut
  {NoStop}%
\bibitem [{\citenamefont {Wu}\ \emph {et~al.}(2013)\citenamefont {Wu},
  \citenamefont {Regnault},\ and\ \citenamefont {Bernevig}}]{Wu13}%
  \BibitemOpen
  \bibfield  {author} {\bibinfo {author} {\bibfnamefont {Y.-L.}\ \bibnamefont
  {Wu}}, \bibinfo {author} {\bibfnamefont {N.}~\bibnamefont {Regnault}}, \ and\
  \bibinfo {author} {\bibfnamefont {B.~A.}\ \bibnamefont {Bernevig}},\ }\href
  {\doibase 10.1103/PhysRevLett.110.106802} {\bibfield  {journal} {\bibinfo
  {journal} {Phys. Rev. Lett.}\ }\textbf {\bibinfo {volume} {110}},\ \bibinfo
  {pages} {106802} (\bibinfo {year} {2013})}\BibitemShut {NoStop}%
\bibitem [{\citenamefont {Wu}\ \emph {et~al.}(2014)\citenamefont {Wu},
  \citenamefont {Regnault},\ and\ \citenamefont {Bernevig}}]{Wu14}%
  \BibitemOpen
  \bibfield  {author} {\bibinfo {author} {\bibfnamefont {Y.-L.}\ \bibnamefont
  {Wu}}, \bibinfo {author} {\bibfnamefont {N.}~\bibnamefont {Regnault}}, \ and\
  \bibinfo {author} {\bibfnamefont {B.~A.}\ \bibnamefont {Bernevig}},\ }\href
  {\doibase 10.1103/PhysRevB.89.155113} {\bibfield  {journal} {\bibinfo
  {journal} {Phys. Rev. B}\ }\textbf {\bibinfo {volume} {89}},\ \bibinfo
  {pages} {155113} (\bibinfo {year} {2014})}\BibitemShut {NoStop}%
\bibitem [{\citenamefont {Knapp}\ \emph {et~al.}(2019)\citenamefont {Knapp},
  \citenamefont {Spanton}, \citenamefont {Young}, \citenamefont {Nayak},\ and\
  \citenamefont {Zaletel}}]{Knapp19}%
  \BibitemOpen
  \bibfield  {author} {\bibinfo {author} {\bibfnamefont {C.}~\bibnamefont
  {Knapp}}, \bibinfo {author} {\bibfnamefont {E.~M.}\ \bibnamefont {Spanton}},
  \bibinfo {author} {\bibfnamefont {A.~F.}\ \bibnamefont {Young}}, \bibinfo
  {author} {\bibfnamefont {C.}~\bibnamefont {Nayak}}, \ and\ \bibinfo {author}
  {\bibfnamefont {M.~P.}\ \bibnamefont {Zaletel}},\ }\href {\doibase
  10.1103/PhysRevB.99.081114} {\bibfield  {journal} {\bibinfo  {journal} {Phys.
  Rev. B}\ }\textbf {\bibinfo {volume} {99}},\ \bibinfo {pages} {081114(R)}
  (\bibinfo {year} {2019})}\BibitemShut {NoStop}%
\bibitem [{\citenamefont {Fu}\ \emph {et~al.}(2016)\citenamefont {Fu},
  \citenamefont {Wang}, \citenamefont {Shan}, \citenamefont {Xiong},
  \citenamefont {Pfeiffer}, \citenamefont {West}, \citenamefont {Kastner},\
  and\ \citenamefont {Lin}}]{Fu16}%
  \BibitemOpen
  \bibfield  {author} {\bibinfo {author} {\bibfnamefont {H.}~\bibnamefont
  {Fu}}, \bibinfo {author} {\bibfnamefont {P.}~\bibnamefont {Wang}}, \bibinfo
  {author} {\bibfnamefont {P.}~\bibnamefont {Shan}}, \bibinfo {author}
  {\bibfnamefont {L.}~\bibnamefont {Xiong}}, \bibinfo {author} {\bibfnamefont
  {L.~N.}\ \bibnamefont {Pfeiffer}}, \bibinfo {author} {\bibfnamefont
  {K.}~\bibnamefont {West}}, \bibinfo {author} {\bibfnamefont {M.~A.}\
  \bibnamefont {Kastner}}, \ and\ \bibinfo {author} {\bibfnamefont
  {X.}~\bibnamefont {Lin}},\ }\href {\doibase 10.1073/pnas.1614543113}
  {\bibfield  {journal} {\bibinfo  {journal} {Proceedings of the National
  Academy of Sciences}\ }\textbf {\bibinfo {volume} {113}},\ \bibinfo {pages}
  {12386} (\bibinfo {year} {2016})},\ \Eprint
  {http://arxiv.org/abs/https://www.pnas.org/content/113/44/12386.full.pdf}
  {https://www.pnas.org/content/113/44/12386.full.pdf} \BibitemShut {NoStop}%
\bibitem [{\citenamefont {Liu}\ \emph {et~al.}(2013{\natexlab{b}})\citenamefont
  {Liu}, \citenamefont {Bergholtz},\ and\ \citenamefont {Kapit}}]{Liu13_2}%
  \BibitemOpen
  \bibfield  {author} {\bibinfo {author} {\bibfnamefont {Z.}~\bibnamefont
  {Liu}}, \bibinfo {author} {\bibfnamefont {E.~J.}\ \bibnamefont {Bergholtz}},
  \ and\ \bibinfo {author} {\bibfnamefont {E.}~\bibnamefont {Kapit}},\ }\href
  {\doibase 10.1103/PhysRevB.88.205101} {\bibfield  {journal} {\bibinfo
  {journal} {Phys. Rev. B}\ }\textbf {\bibinfo {volume} {88}},\ \bibinfo
  {pages} {205101} (\bibinfo {year} {2013}{\natexlab{b}})}\BibitemShut
  {NoStop}%
\bibitem [{\citenamefont {Yang}\ \emph {et~al.}(2019)\citenamefont {Yang},
  \citenamefont {Wu},\ and\ \citenamefont {Papi\ifmmode~\acute{c}\else
  \'{c}\fi{}}}]{Yang19}%
  \BibitemOpen
  \bibfield  {author} {\bibinfo {author} {\bibfnamefont {B.}~\bibnamefont
  {Yang}}, \bibinfo {author} {\bibfnamefont {Y.-H.}\ \bibnamefont {Wu}}, \ and\
  \bibinfo {author} {\bibfnamefont {Z.}~\bibnamefont
  {Papi\ifmmode~\acute{c}\else \'{c}\fi{}}},\ }\href {\doibase
  10.1103/PhysRevB.100.245303} {\bibfield  {journal} {\bibinfo  {journal}
  {Phys. Rev. B}\ }\textbf {\bibinfo {volume} {100}},\ \bibinfo {pages}
  {245303} (\bibinfo {year} {2019})}\BibitemShut {NoStop}%
\bibitem [{\citenamefont {Chen}\ \emph
  {et~al.}(2020{\natexlab{b}})\citenamefont {Chen}, \citenamefont {Capponi},
  \citenamefont {Wietek}, \citenamefont {Mambrini}, \citenamefont {Schuch},\
  and\ \citenamefont {Poilblanc}}]{Chen20_2}%
  \BibitemOpen
  \bibfield  {author} {\bibinfo {author} {\bibfnamefont {J.-Y.}\ \bibnamefont
  {Chen}}, \bibinfo {author} {\bibfnamefont {S.}~\bibnamefont {Capponi}},
  \bibinfo {author} {\bibfnamefont {A.}~\bibnamefont {Wietek}}, \bibinfo
  {author} {\bibfnamefont {M.}~\bibnamefont {Mambrini}}, \bibinfo {author}
  {\bibfnamefont {N.}~\bibnamefont {Schuch}}, \ and\ \bibinfo {author}
  {\bibfnamefont {D.}~\bibnamefont {Poilblanc}},\ }\href {\doibase
  10.1103/PhysRevLett.125.017201} {\bibfield  {journal} {\bibinfo  {journal}
  {Phys. Rev. Lett.}\ }\textbf {\bibinfo {volume} {125}},\ \bibinfo {pages}
  {017201} (\bibinfo {year} {2020}{\natexlab{b}})}\BibitemShut {NoStop}%
\bibitem [{\citenamefont {Hauschild}\ and\ \citenamefont
  {Pollmann}(2018)}]{tenpy}%
  \BibitemOpen
  \bibfield  {author} {\bibinfo {author} {\bibfnamefont {J.}~\bibnamefont
  {Hauschild}}\ and\ \bibinfo {author} {\bibfnamefont {F.}~\bibnamefont
  {Pollmann}},\ }\href {\doibase 10.21468/SciPostPhysLectNotes.5} {\bibfield
  {journal} {\bibinfo  {journal} {SciPost Phys. Lect. Notes}\ ,\ \bibinfo
  {pages} {5}} (\bibinfo {year} {2018})},\ \bibinfo {note} {code available from
  \url{https://github.com/tenpy/tenpy}},\ \Eprint
  {http://arxiv.org/abs/1805.00055} {arXiv:1805.00055} \BibitemShut {NoStop}%
\bibitem [{\citenamefont {Tange}(2011)}]{Tange11}%
  \BibitemOpen
  \bibfield  {author} {\bibinfo {author} {\bibfnamefont {O.}~\bibnamefont
  {Tange}},\ }\href {\doibase http://dx.doi.org/10.5281/zenodo.16303}
  {\bibfield  {journal} {\bibinfo  {journal} {;login: The USENIX Magazine}\
  }\textbf {\bibinfo {volume} {36}},\ \bibinfo {pages} {42} (\bibinfo {year}
  {2011})}\BibitemShut {NoStop}%
\bibitem [{\citenamefont {Pu}\ \emph {et~al.}(2017)\citenamefont {Pu},
  \citenamefont {Wu},\ and\ \citenamefont {Jain}}]{Pu17}%
  \BibitemOpen
  \bibfield  {author} {\bibinfo {author} {\bibfnamefont {S.}~\bibnamefont
  {Pu}}, \bibinfo {author} {\bibfnamefont {Y.-H.}\ \bibnamefont {Wu}}, \ and\
  \bibinfo {author} {\bibfnamefont {J.~K.}\ \bibnamefont {Jain}},\ }\href
  {\doibase 10.1103/PhysRevB.96.195302} {\bibfield  {journal} {\bibinfo
  {journal} {Phys. Rev. B}\ }\textbf {\bibinfo {volume} {96}},\ \bibinfo
  {pages} {195302} (\bibinfo {year} {2017})}\BibitemShut {NoStop}%
\bibitem [{\citenamefont {Chakraborty}\ and\ \citenamefont
  {Pietiläinen}(1995)}]{Chakraborty}%
  \BibitemOpen
  \bibfield  {author} {\bibinfo {author} {\bibfnamefont {T.}~\bibnamefont
  {Chakraborty}}\ and\ \bibinfo {author} {\bibfnamefont {P.}~\bibnamefont
  {Pietiläinen}},\ }\href@noop {} {\emph {\bibinfo {title} {The Quantum Hall
  Effects}}},\ \bibinfo {edition} {2nd}\ ed.,\ \bibinfo {series} {Springer
  Series in Solid-State Sciences}, Vol.~\bibinfo {volume} {85}\ (\bibinfo
  {publisher} {Springer},\ \bibinfo {address} {Heidelberg, Germany},\ \bibinfo
  {year} {1995})\BibitemShut {NoStop}%
\bibitem [{\citenamefont {Khemani}\ \emph {et~al.}(2016)\citenamefont
  {Khemani}, \citenamefont {Pollmann},\ and\ \citenamefont
  {Sondhi}}]{Khemani16}%
  \BibitemOpen
  \bibfield  {author} {\bibinfo {author} {\bibfnamefont {V.}~\bibnamefont
  {Khemani}}, \bibinfo {author} {\bibfnamefont {F.}~\bibnamefont {Pollmann}}, \
  and\ \bibinfo {author} {\bibfnamefont {S.~L.}\ \bibnamefont {Sondhi}},\
  }\href {\doibase 10.1103/PhysRevLett.116.247204} {\bibfield  {journal}
  {\bibinfo  {journal} {Phys. Rev. Lett.}\ }\textbf {\bibinfo {volume} {116}},\
  \bibinfo {pages} {247204} (\bibinfo {year} {2016})}\BibitemShut {NoStop}%
\bibitem [{\citenamefont {Varjas}\ \emph {et~al.}(2021)\citenamefont {Varjas},
  \citenamefont {Abouelkomsan}, \citenamefont {Yang},\ and\ \citenamefont
  {Bergholtz}}]{Varjas21}%
  \BibitemOpen
  \bibfield  {author} {\bibinfo {author} {\bibfnamefont {D.}~\bibnamefont
  {Varjas}}, \bibinfo {author} {\bibfnamefont {A.}~\bibnamefont
  {Abouelkomsan}}, \bibinfo {author} {\bibfnamefont {K.}~\bibnamefont {Yang}},
  \ and\ \bibinfo {author} {\bibfnamefont {E.~J.}\ \bibnamefont {Bergholtz}},\
  }\href@noop {} {\enquote {\bibinfo {title} {Topological lattice models with
  constant berry curvature},}\ } (\bibinfo {year} {2021}),\ \Eprint
  {http://arxiv.org/abs/2107.06902} {arXiv:2107.06902 [cond-mat.str-el]}
  \BibitemShut {NoStop}%
\bibitem [{\citenamefont {Mera}\ and\ \citenamefont {Ozawa}(2021)}]{Mera21}%
  \BibitemOpen
  \bibfield  {author} {\bibinfo {author} {\bibfnamefont {B.}~\bibnamefont
  {Mera}}\ and\ \bibinfo {author} {\bibfnamefont {T.}~\bibnamefont {Ozawa}},\
  }\href@noop {} {\enquote {\bibinfo {title} {Engineering geometrically flat
  chern bands with fubini-study k\"ahler structure},}\ } (\bibinfo {year}
  {2021}),\ \Eprint {http://arxiv.org/abs/2107.09039} {arXiv:2107.09039
  [cond-mat.mes-hall]} \BibitemShut {NoStop}%
\end{thebibliography}%

\end{document}